\documentclass[lettersize,journal]{IEEEtran}
\usepackage{amsmath,amsfonts}

\usepackage{algorithm}
\usepackage{array}
\usepackage[caption=false,font=normalsize,labelfont=sf,textfont=sf]{subfig}
\usepackage{textcomp}
\usepackage{stfloats}
\usepackage{url}
\usepackage{verbatim}
\usepackage{graphicx}
\usepackage{cite}
\usepackage{color}
\usepackage{algpseudocode}
\usepackage{subfig}
\usepackage{booktabs}
\usepackage{multirow}
\usepackage{bm}
\usepackage{pifont}       
\usepackage{bbding}       
\usepackage{fontawesome}
\usepackage{ulem}
\usepackage{makecell}
\usepackage{xcolor} 
\usepackage{soul}   
\usepackage{hyperref}
\sethlcolor{yellow} 

\newcommand{\netname}{SimpleUNet}

\begin{document}

\title{Simple is what you need for efficient and accurate medical image segmentation}

\author{
    \IEEEauthorblockN{Xiang Yu\textsuperscript{1},
                      Yayan Chen\textsuperscript{2},
                      Guannan He\textsuperscript{3},
                      Qing Zeng\textsuperscript{1},
                      Yue Qin\textsuperscript{1},
                      Meiling Liang\textsuperscript{1},
                      Dandan Luo\textsuperscript{1},
                      Yimei Liao\textsuperscript{1},
                      Zeyu Ren\textsuperscript{4},
                      Cheng Kang\textsuperscript{5},
                      Delong Yang\textsuperscript{6},
                      Bocheng Liang\textsuperscript{1},
                      Bin Pu\textsuperscript{7},
                      Ying Yuan\textsuperscript{1},
                      Shengli Li\textsuperscript{1*}
                      }
    
    \IEEEauthorblockA{\textsuperscript{1}Department of Ultrasound, Shenzhen Maternity and Child Healthcare Hospital, Women and Children's Medical Center, Southern Medical University, Shenzhen, Guangdong Province, China.}\\
    \textsuperscript{2}Ultrasound Department of Longhua District Maternal and Child Healthcare Hospital, Shenzhen, China.\\
    \textsuperscript{3}Sichuan Provincial Maternity and Child Health Care Hospital, Chengdu, China. \\
    \textsuperscript{4}College of Agronomy, Jilin Agricultural University, Changchun, China.\\
    \textsuperscript{5}Department of Cybernetics and Robotics, Czech Technical University, Prague, Czech Republic.\\
    \textsuperscript{6}Institute for Engineering Medicine, Kunming Medical University, Kunming, China.\\
    \textsuperscript{7}Department of Electronic and Computer Engineering, The Hong Kong University of Science and Technology.\\
}
\markboth{~}%
{Shell \MakeLowercase{\textit{et al.}}: A Sample Article Using IEEEtran.cls for IEEE Journals}

\IEEEpubid{~}

\maketitle

\begin{abstract}
While modern segmentation models often prioritize performance over practicality, we advocate a design philosophy prioritizing simplicity and efficiency, and attempted high-performance segmentation model design. This paper presents \netname{}, a scalable ultra-lightweight medical image segmentation model with three key innovations: (1) A partial feature selection mechanism in skip connections for redundancy reduction while enhancing segmentation performance; (2) A fixed-width architecture that prevents exponential parameter growth across network stages; (3) An adaptive feature fusion module achieving enhanced representation with minimal computational overhead. With a record-breaking \textbf{16 KB} parameter configuration, \netname{} outperforms LBUNet and other lightweight benchmarks across multiple public datasets. The \textbf{0.67 MB} variant achieves superior efficiency (8.60 GFLOPs) and accuracy, attaining a mean DSC/IoU of 85.76$\%$/75.60$\%$ on multi-center breast lesion datasets, surpassing both U-Net and TransUNet. Evaluations on skin lesion datasets (ISIC 2017/2018: mDice 84.86$\%$/88.77$\%$) and endoscopic polyp segmentation (KVASIR-SEG: 86.46$\%$/76.48$\%$ mDice/mIoU) confirm consistent dominance over state-of-the-art models. This work demonstrates that extreme model compression need not compromise performance, providing new insights for efficient and accurate medical image segmentation. Codes can be found at \url{https://github.com/Frankyu5666666/SimpleUNet}.
\end{abstract}

\begin{IEEEkeywords}
Medical Image Segmentation, Lightweight, Efficiency, Feature Selection, Feature Fusion.
\end{IEEEkeywords}

\section{Introduction}
\IEEEPARstart{I}{n} medical image segmentation, U-Net has been acknowledged as a successful and robust framework distinguished by its unique U-shaped architecture comprising an encoder-decoder pathway\cite{ronneberger2015u, liang2025mambasam, 10899855, CHEN2025128749}. The encoder captures contextual information through convolution operations and downsampling, while the decoder facilitates precise localization by upsampling and concatenating feature maps from the encoder. Generally, the skip connections between the encoder and the decoder are considered to concatenate the lower-level features from the decoder to the high-level features from the encoder for hierarchical feature fusion, mitigating issues like gradient vanishing or explosion, and thus leading to higher performance. The modular design has made U-Net a popular choice for semantic segmentation, especially in medical image segmentation scenarios where available datasets are limited. However, the model's progressively increasing width and the feature concatenation mechanism inherently introduce more parameters in the decoder path for information fusion, potentially resulting in information redundancy and reduced efficiency.

Recent advances have sought to enhance segmentation performance by introducing novel computing operations and attention modules\cite{transunet2021,chen2024esknet,moradzadeh2024ukan, chen2024mlmseg}. Although these innovations have improved accuracy, they often come at an expensive cost regarding parameter and computational complexity that challenges practical deployment in resource-constrained environments. In light of these limitations, researchers in the area endeavored to develop lightweight yet high-performance models for medical image segmentation, such as those utilizing depthwise convolution and state-space-based models\cite{pu2022mobileunet, chollet2017xception, ruan2024vm}. Although these approaches are more computationally efficient, substantial room remains for performance improvement. What is more depressing is that these models are usually well-crafted for computation and parameter efficiency and might not be readily intelligible or generable.

Desperately, we wonder if we can improve U-Net with simplicity instead of leveraging obscure modules or techniques, given the simplicity of U-Net itself. To this end, we developed three simple yet effective strategies: First, we reconsidered the necessity of concatenating all of the shortcut features for skip connections in the decoder, which may lead to redundant information and parameter surge. Consequently, we propose to concatenate only partial shortcut features through an efficient feature selection strategy so that only the most representative information is forwarded for feature fusion in the decoding stage. By doing so, the overall number of parameters in the decoder can be drastically minimized due to the reduced features after concatenation. Second, we propose to apply an intuitive width control strategy by fixing the model width in the encoder, as we note that stagewise model widening in conventional U-Net leads to layer-wise exponential parameter increase in U-Net. By combining these two strategies, we can have an unprecedentedly minimized model simply based on U-Net. To enhance the feature fusion process in the developed lightweight model, we further propose an adaptive feature fusion strategy by reassigning linear weights of the selected shortcut features and deep features, which brings in negligible parameter and computation overheads. Straightforwardly, our developed strategies are strongly oriented to the lightweight model, yet effective without harming the model's performance. The contributions of our \netname{} can be summarized as below:
\begin{enumerate}
\item We developed a novel and efficient feature selection and fusion strategy that mitigates information redundancy in U-Net, leading to parameter reduction with boosted segmentation performance. By selecting the most representative features from the encoder and the deep layers from the decoder, the encoder has become more compact and powerful as a result of applying adaptive feature selection. Also, experiments on U-Nets demonstrate the effectiveness and simplicity of the developed feature selection strategy regarding breast lesion segmentation in ultrasound images.
\item We challenge the traditional model width expansion and shrinkage patterns in U-Net and therefore propose to apply an intuitive model width control strategy by fixing the model width instead, which turns out to a straightforward yet rewarding strategy for model parameter reduction, though at the sacrifice of model performance. In traditional U-Net, we note that stagewise increased model width leads to parameter and computational cost surges. Intuitively, this situation can be effectively mitigated if the model width is simply fixed and resulting in linear increases of the parameter and computational cost, only with acceptable performance deterioration.     
\item Based on the developed strategies and conventional U-Net, our developed \netname{} can effectively achieve similar or even better performance, but with a much smaller number of parameters compared to the state-of-the-art models. Extensive experiments on multiple public datasets demonstrate our superiority to both lightweight and parameter-intensive models regarding model size and performance. On a curated public ultrasound breast lesion dataset, our model with only \textbf{16 KB} parameters achieved a mean DSC and IoU of 82.93$\%$ and 71.25$\%$ across 5 runs, outperforming known lightweight benchmarks such as LBUNet\cite{xu2024lb, wu2024ultralight}. On the other hand, our best model gives a mean DSC value of 85.76$\%$ and 75.60$\%$ with only \textbf{0.67 MB} parameters, which is around 140 times fewer than that of TransUNet with a mean DSC and IoU of 85.32$\%$ and 74.89$\%$, respectively. On three other datasets, including KVASIR-SEG, ISIC2017, and ISIC2018, our best model showed consistent and comparable performance against the edge-cutting models with mDSC values of 86.46$\%$, 84.86$\%$, and 88.77$\%$, respectively, which demonstrates the generality of our model.

\end{enumerate}

The remainder of this paper will be arranged as follows: In Section II, we will revisit the latest representative models that have shown powerful segmentation performance in medical image segmentation scenarios from the perspective of lightweight and heavyweight. In Section III, we will move to our methodology, mainly based on the aforementioned three key strategies, which are validated via extensive experimental results in Section IV. We then discuss the key issues in the model design in Section V, and we end this paper in Section VI with a summary and a quick look at future work.


\section{Related Works}
U-Net has established itself as a de facto architecture for medical image segmentation, driving extensive research into its adaptation across diverse clinical applications. This evolution grossly diverges into two distinct paradigms: lightweight variants prioritize computational efficiency through streamlined architectures for edge deployment, while large-scale counterparts leverage parameter-intensive designs to achieve superior segmentation accuracy at the expense of increased computational latency. We will briefly revisit the prevalent lightweight models and parameter-intensive models in semantic medical image segmentation as we aim to develop a novel and generable segmentation framework for medical image segmentation in a lightweight way, while its performance is comparable to large-scale ones.

\subsection{Lightweight models for medical image segmentation} 
The exploration of lightweight segmentation model design dates back to ENet\cite{paszke2016enet}, where it deployed factorized convolutions, bottleneck modules, and aggressive early downsampling, drastically reducing computational complexity while maintaining segmentation capability for real-time edge deployment. By doing so, the overall number of parameters is only 0.4 MB. In LinkNet, the model achieved efficient feature fusion simply by applying element-wise addition of the features from the decoder to the features in the decoder and thus avoided the introduction of extra parameters in U-Net\cite{chaurasia2017linknet}. UNext introduces tokenized MLP(Multiple Layer Perceptron) to process spatially-reduced features while applying convolution to spatial-intensive features\cite{valanarasu2022unext}. The authors claimed that the resultant model had 72 times fewer parameters than models such as TranUNet\cite{transunet2021}. Yet, a more straightforward way to restrain the model's parameters from drastic growth is to control the number of output channels from the very beginning. By leveraging this strategy, LF-UNet starts with only 8 output channels at the first encoding stage and then progressively and linearly increases the output channels to 128\cite{deng2023lfu}. To optimize the gradient flow between the encoder and the decoder, the authors utilized a full connection scheme that connects all layers, which resulted in relatively expensive computational costs. So far, these works can be considered valuable attempts regarding the lightweight architectural streamlines of U-Net.

More recently, researchers in the area have tended to develop novel lightweight modules to achieve model efficiency. In  MALUNet\cite{ruan2022malunet}, Ruan et al. developed four modules by integrating multiple attention modules, such as channel attention, spatial attention, and cross-sample external attention, which results in extremely low computational costs. Similarly, EGE-UNet employed a Group Multi-Axis Hadamard Product Attention module and Group Aggregation Bridge module, where the authors claimed that GHPA extracts pathological information from varied perspectives while GAB fuses multi-scale information by grouping features from different levels. Similar to LF-UNet, EGE-UNet chose a small number as the initial output channel number. To enhance the model's capability to cope with ambiguous boundaries, LB-UNet integrates a Prediction Map Auxiliary module, which can be further decomposed into three submodules for boundary information enhancement\cite{xu2024lb}, leading to parameters of 38 KB only and GFlops of 0.1. Recently, Mamba-based models have raised wide attention from the community, given their high efficiency and decent performance \cite{wu2024ultralight}. Built upon the novel state space model (SSM) architecture with selective scanning mechanisms, these models demonstrate linear-time computational complexity while maintaining global receptive fields \cite{gu2023mamba}. This breakthrough addresses the quadratic scaling limitations of conventional Transformer architectures, particularly in processing long-sequence tasks such as genomic sequence analysis and high-resolution image understanding

\subsection{Parameter-intensive models for medical image segmentation}
Compared to lightweight models, parameter-intensive models are more favourable regarding segmentation performance, as the learning capacity of learnable models tends to increase along with the growth of parameters. Currently, U-Net is improved from varied perspectives, including the integration of novel attention modules, multi-scale feature fusion, and so forth \cite{transunet2021, woo2018cbam, oktay2018attention, pramanik2024dau, iqbal2022mda, zhou2018unet++, wang2024mf}. 

Attention modules are still popular and considered free plugin components for model performance improvement. Attention U-Net integrates spatial attention gates into skip connections to dynamically suppress irrelevant regions and emphasize lesion boundaries\cite{oktay2018attention}. This addresses U-Net’s limitation in handling complex backgrounds, particularly in medical images with low contrast. In another work named ESKNet, the authors proposed to enhance the selective kernel convolution \cite{chen2024esknet} via channel and spatial attention modules, where the authors claimed that the developed attention modules can adaptively recalibrate feature maps in channel and spatial dimensions. The above-mentioned works focus on feature extraction enhancement in either the encoder or decoder. Alternatively, UNetv2 considers feature enhancement in the skip connections via the proposed semantics and detail infusion (SDI) modules, where varied scales of intermediate features are mapped to a specific scale and fused through Hadamard product operations. By doing so, the authors claimed that both the semantics and details are enhanced. 

When considering multi-scale feature fusion, U-Net++ introduces nested dense skip connections to aggregate multi-scale features hierarchically \cite{zhou2018unet++}, resolving the "semantic gap" between encoder and decoder. Further, UNet 3+ takes advantage of full-scale skip connections by integrating low-level details with high-level semantics from feature maps in different scales \cite{huang2020unet}. MDA-Net achieves multiscale feature fusion by fusing features enhanced through the developed channel-based and lesion-based attention modules. TinyU-Net is a lightweight medical image segmentation model (0.48M parameters) with a Cascade Multi-Receptive Fields (CMRF) module that enhances feature representation through cost-effective multi-channel cascading, achieving higher IoU scores than larger models, such as UNet, while drastically reducing computational complexity\cite{chen2024tinyu}. Besides attention module and feature fusion, novel operations such as multiple head attention and state space module also contributed to new U-Net variants with better performance\cite{transunet2021, wu2024ultralight}.

Most of the developed attention modules are carefully designed throughout these models, resulting in abstruse modules and extra heavy numbers of parameters\cite{chen2024esknet, wu2024ultralight}. However, attempts to enhance the shortcut feature are still few and can be further explored. Considering these, we developed our \netname{} based on U-Net with feature selection and adaptive feature fusion, leading to much more parameter-efficient models with even better performance. When considering possible further model minimisation, we propose to apply the intuitive model width control by fixing the model width instead of progressively increasing the width.

\section{Methodology}
In this section, we will have a quick revisit to U-Net and formulate the problems within it. We then move to our methodology in detail, where the key components are the developed selective feature reduction, intuitive model width control, and adaptive feature fusion strategy, followed by the loss function design and overall architecture of \netname{}. Note that all operations or loss functions involved are ubiquitous instead of being meticulously crafted.

\subsection{Problem formulation}
\begin{figure}[!t]
\centering
\includegraphics[width=2.5in]{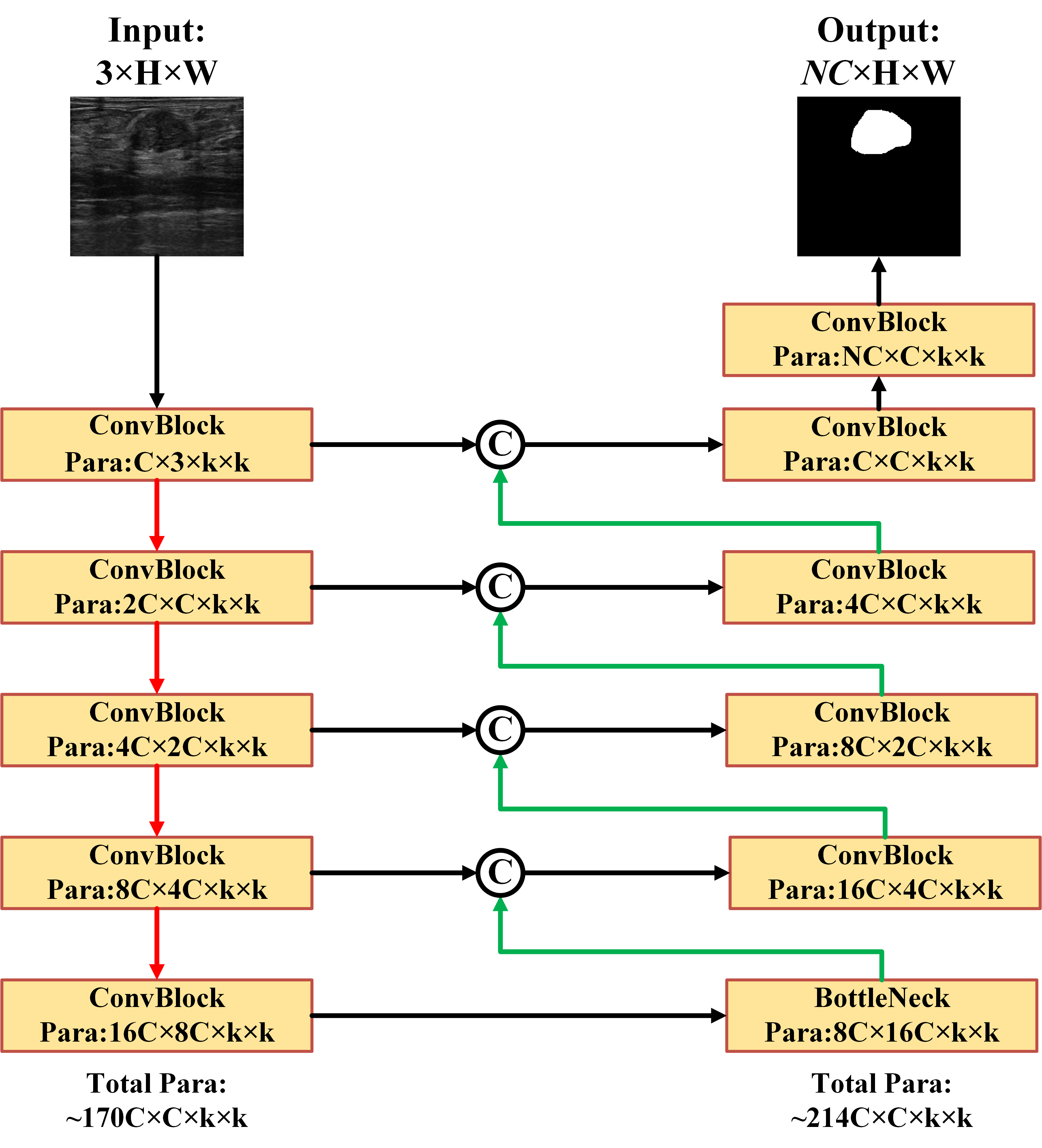}
\caption{A general U-Net. Given the initial model width of $C$, the convolution kernel of $k$, and the number of convolution blocks $N$ as one, the overall number of parameters in the original U-Net is around $384 \ast C^2 \ast k^2$ parameters, whereas the decoder alone contributes around $214 \ast C^2 \ast k^2$ parameters.}
\label{fig1:unet}
\end{figure}

A general architecture of U-Net can be seen in Figure \ref{fig1:unet}, where UNet comprises an encoding path and a decoding path with shortcut connections between them for feature bridging. U-Net's architecture is characterized by its gradually increasing number of output channels, i.e., model width, as the network deepens, a design choice that has been considered a crucial factor in its effectiveness. In the encoder, the number of output channels doubles at each downsampling step, allowing the network to capture increasingly complex and abstract features. The skip connections or shortcuts bridge the encoder and decoder by directly concatenating features from the encoder to the decoder, allowing the model to retain crucial fine spatial details (e.g., edges, textures) for precise pixel-wise segmentation. The encoder gradually receives the features from the bottom deep layer and the shortcut path for feature fusion and spatial information restoration. Given that the kernel sizes throughout the network are  $k\times k$ and the output channel after the first convolution is $C$(where we claim it as initial model width for simplicity), then the overall number of parameters in the network is around $384 \ast C^{2} \ast k^{2}$.

When considering a lightweight model design based on U-Net, we note that there are architectural hindrances to us from doing so. As seen from the Figure \ref{fig1:unet}, the model experiences widening stagewise in the encoder and therefore leads to exponential parameter increases. Moreover, the inclusion of skip connections comes with increased model complexity, computational overhead, and potential information redundancy. More specifically, each skip connection results in doubled channels of the resultant features, and therefore, more parameters are introduced for information processing that the decoder is even slightly parameter-expensive than the encoder. Also, the skip connections don't specifically distinguish the importance of the features, and information redundancy might be a potential issue that compromises the overall performance of U-Net. Therefore, our basic principle regarding lightweight U-Net redesign is to reduce the model size while leveraging the information usage without significant performance compromise. What is even surprising is that experimental results indicate that we successfully reduced the number of parameters to a minimal level while still achieving a higher performance on multiple public datasets. Full of simplicity, our methods only comprise three proposed strategies, including efficient feature selection, intuitive model width control, and adaptive feature fusion. 

\subsection{Efficient feature selection}
\begin{figure}[!t]
\centering
\includegraphics[width=2.5in]{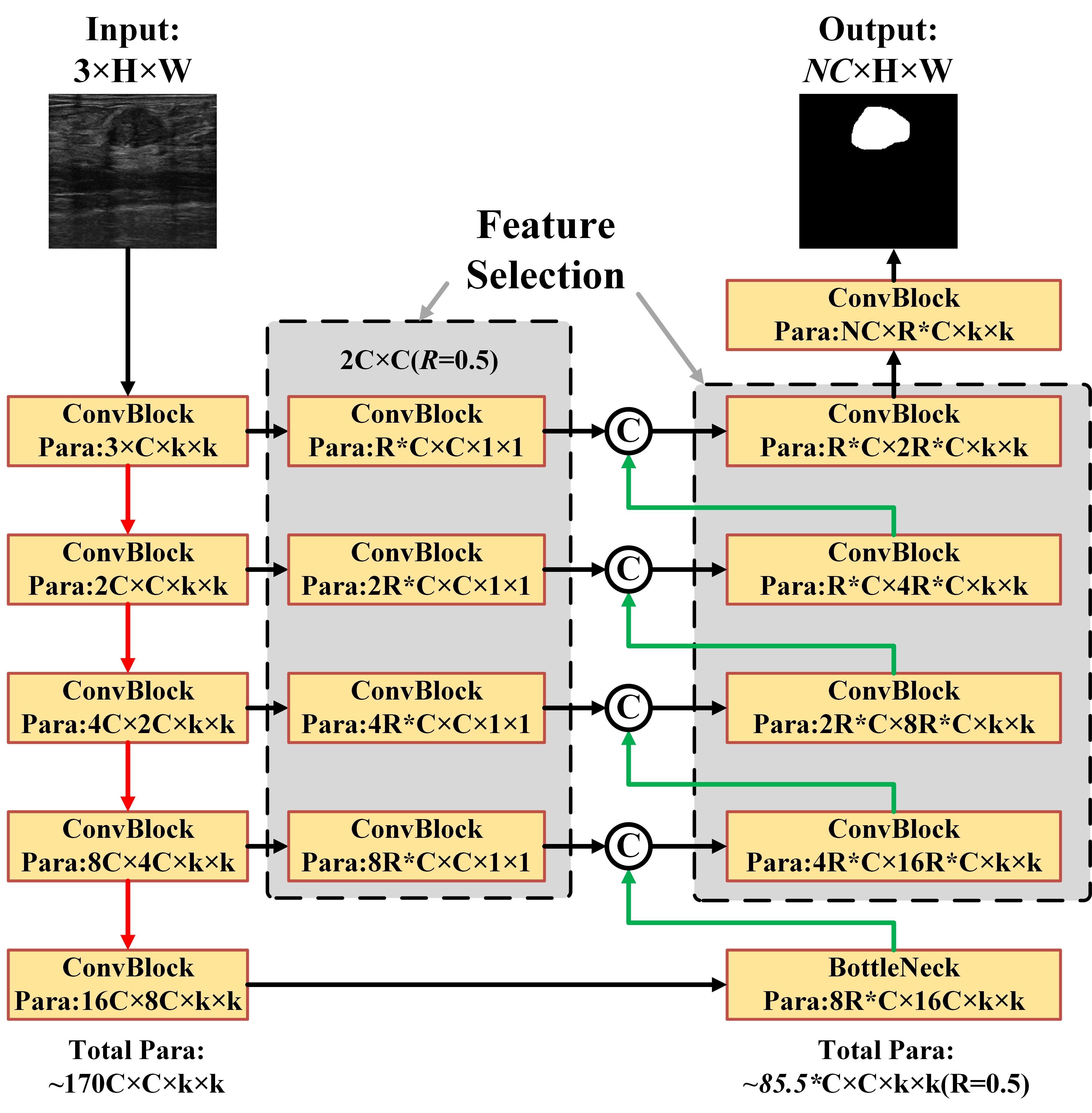}
\caption{The U-Net with feature selection, where the $\bm{R}$ is feature selection rate and is set between 0 and 1. For the shortcut feature selection, $1\times1$ convolution is applied to shrink the input channels to $R$ times. On the other hand, the convolution blocks in the decoder serve to fuse the concatenated features. However, they can also be deemed as feature selection operations, as the output channels of the resultant features are halved.}
\label{fig1:unet_rd}
\end{figure}

The skip connections in the U-Net inspired us to apply feature selection to forward only the most representative features to the decoder. Intuitively, we can significantly reduce the number of model parameters if we selectively pick partial intermediate features as the reduced shortcut features, while the deep features are symmetrically selected to match the reduced shortcut features. Therefore, the conventional U-Net can be turned into the model in Figure \ref{fig1:unet_rd}. More specifically, we apply $1 \times 1$ convolutions to select the representative shortcut features in each stage. For the selection of deep features, we simply adjust the number of output channels to $\bm{R}$ times the input channels when performing convolution, where $\bm{R}$ is the predefined feature selection rate ( $\bm{R>0}$ and $\bm{R \leq 1}$). As a result, the channels of the selected shortcut features and the deep features perfectly match, while the overall number of parameters is reduced to around $85.5\ast\times C^{2} \times k^{2}$ if we have $\bm{R}$ as 0.5, which is a roughly $130\ast\times C^{2} \times k^{2}$ difference to the original decoder.

\subsection{Intuitive model width control}
\begin{figure*}[!t]
\centering
\subfloat[The architecture of U-Net with intuitive model width control.]{\includegraphics[width=2.5in]{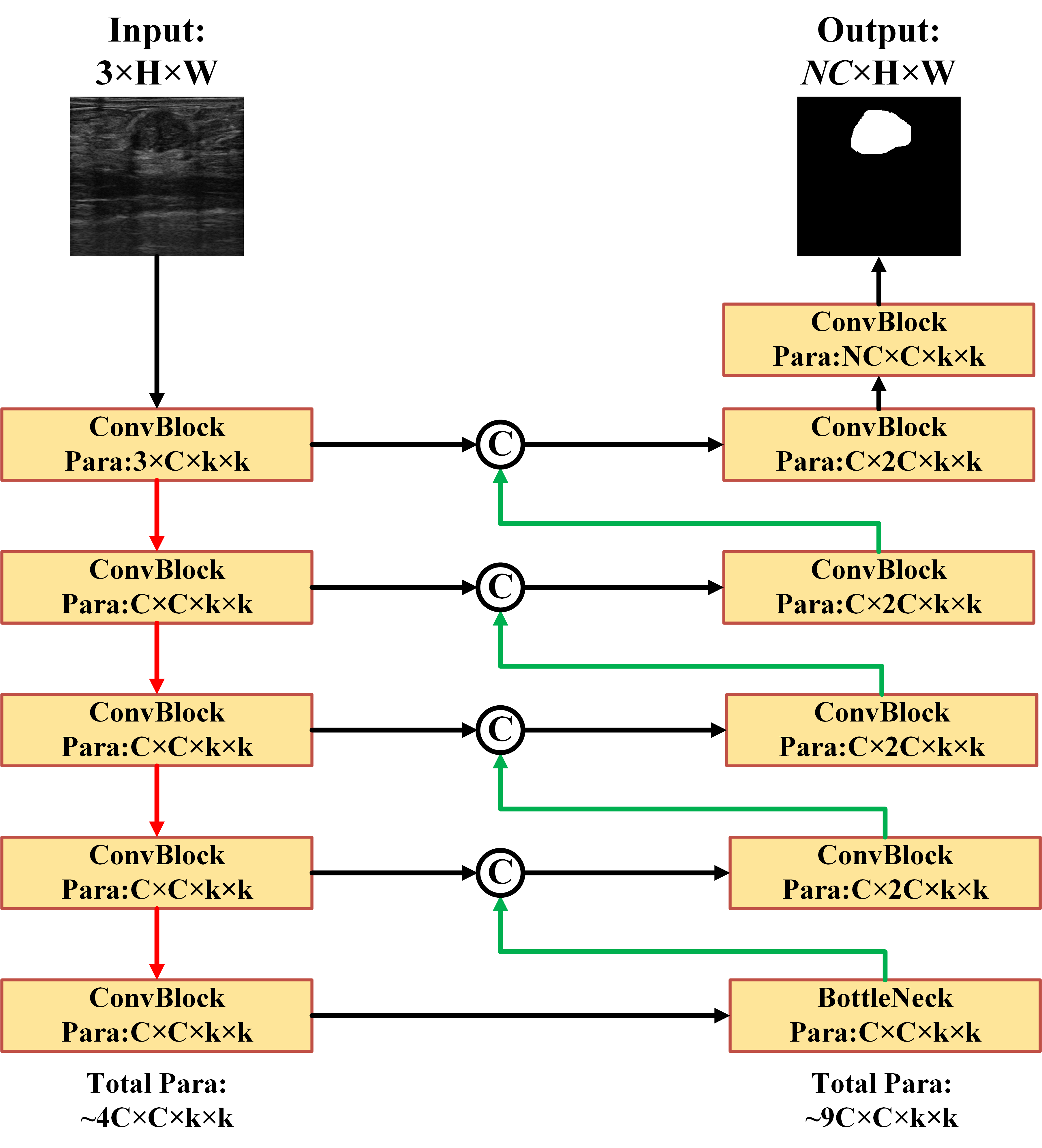}%
\label{fig1_a}}
\hfil
\subfloat[The fixed-width U-Net combining feature selection and intuitive width control.]{\includegraphics[width=2.5in]{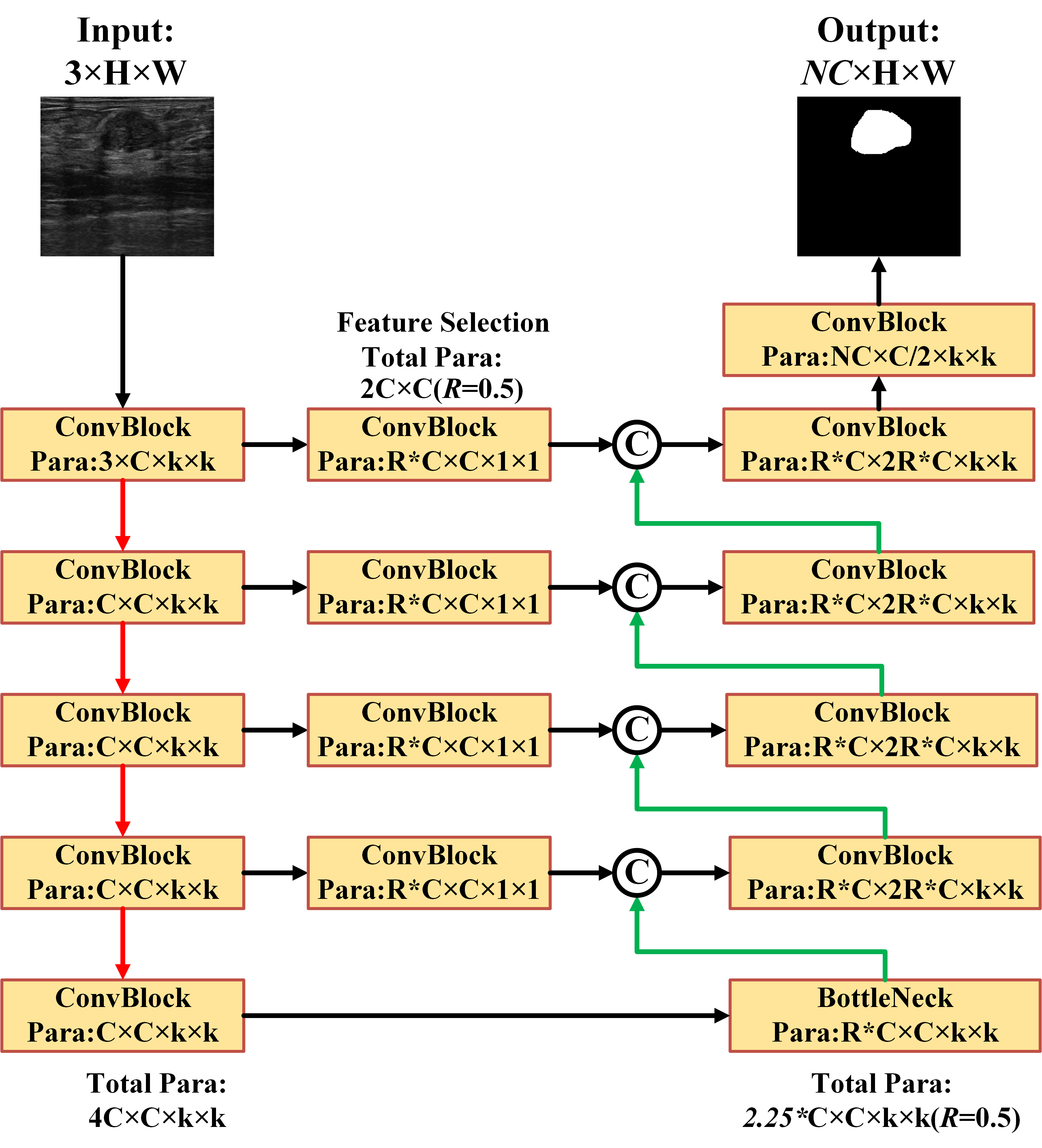}%
\label{fig_second_case}}
\caption{The U-Net variants with our developed strategies. When the model's width is fixed, the overall number of parameter increase linearly along the depth of the model. With the help of feature selection, the overall parameter experiences further reduction, mainly from the encoder part.}
\label{fig1:unet_fwfs}
\end{figure*}

While the proposed feature selection strategy effectively reduces the decoder's parameters, their quantity remains substantial at both the decoder and architectural levels. This necessitates further investigation into more efficient parameter reduction methodologies that optimize computational complexity without compromising model performance. Intuitively, we propose to control the model width by fixing the model width throughout the encoder and thus restraining the model from stagewise exponential parameter increase, as seen in Fig. \ref{fig1_a}. By doing so, the number of parameters in the encoder and the decoder follows a linear increase along with the increase in model depth. While it is a fact that fixed model width is likely to harm the model's performance, however, restraining the model widths from linear increase can be the most intuitive yet straightforward strategy to have a lightweight model when we consider lightweight and simplicity as top priorities. As the remarkable success of U-Net has established it as the prevailing paradigm in medical image segmentation, contemporary works tend to follow the design paradigm instead of challenging the necessity of stagewise increased model width\cite{transunet2021,ruan2024vm}. We reckon that our feature selection and model width control strategy are bold attempts to break the conventions. By combining the feature selection and model width control, we can have an even lighter model based on U-Net shown in Fig. \ref{fig_second_case}, in which the encoder and the decoder have only $4C^{2}\times k^{2}$, and $2.25C^{2}\times k^{2}$, respectively. Unprecedentedly, the model is more than \textbf{60} times lighter than traditional U-Net. Later experiments demonstrate the success of UNet with these two combined strategies. 

\subsection{Adaptive Feature Fusion}
\begin{figure}[!t]
\centering
\includegraphics[width=0.45\textwidth]{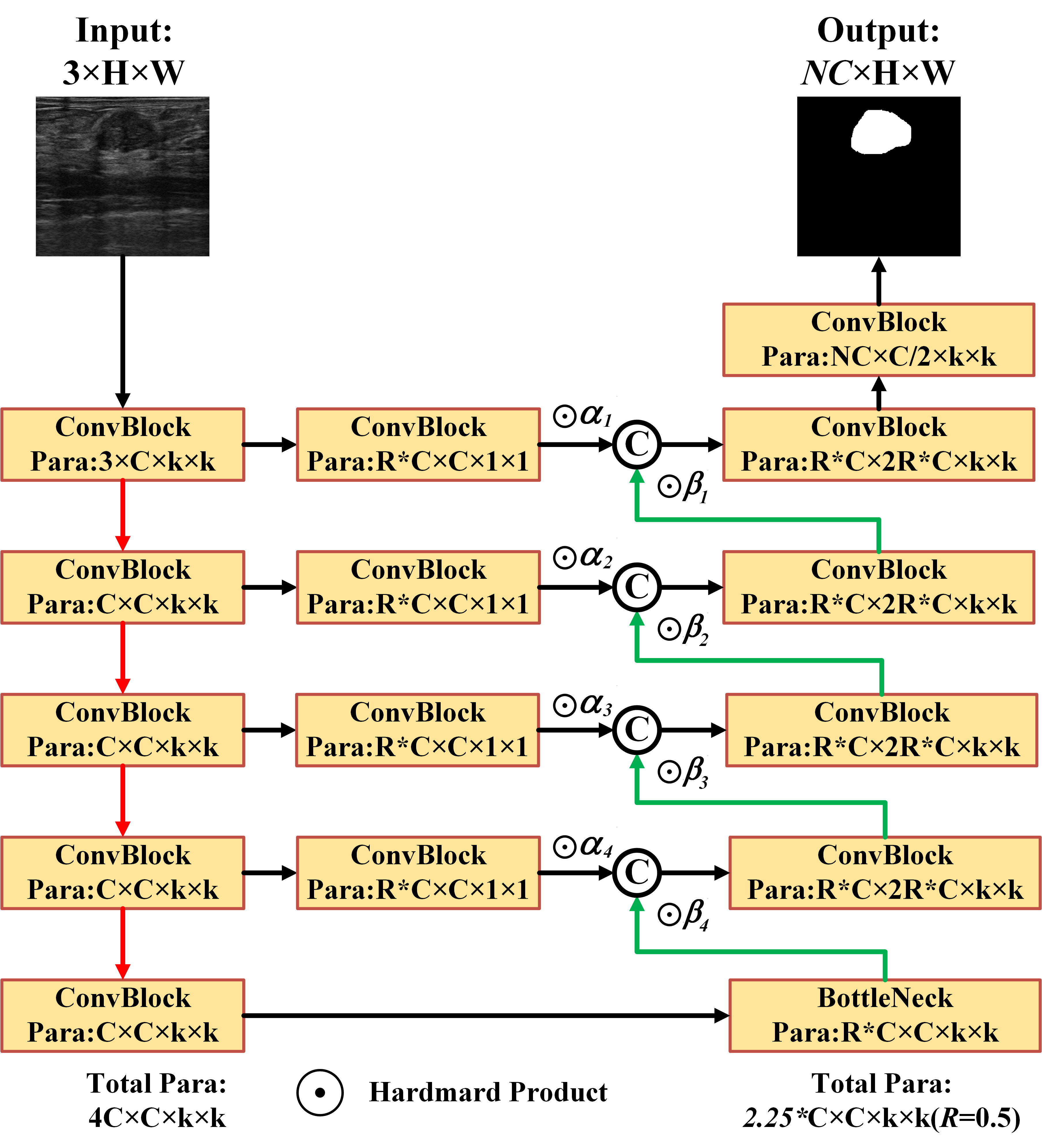}
\caption{The U-Net variant integrated with intuitive width control, feature selection and adaptive feature fusion. The adaptive feature fusion is also a straightforward strategy to reassign channel weights and serves as a cost-minimal channel attention.}
\label{fig:simple}
\end{figure}

Based on the previous strategies, we further propose an adaptive feature fusion strategy for enhanced feature fusion at minimal overhead. Instead of performing a direct concatenation of the reduced shortcut features and deep features, which may treat all feature channels equally regardless of their importance, we propose a simple yet effective feature enhancement strategy using channel-wise adaptive weights. As illustrated in Fig. \ref{fig:simple}, our approach introduces two groups of learnable weights $\bm{\alpha_{i}}$ and $\bm{\beta_{i}}$, each with a dimensionality of $ 2^{(i-1)} \ast \bm{R} \ast C \times 1 \times 1 (i \in \{1, \cdots, 4\})$ throughout stage 1 to 4, where $C$ corresponds to the number of initial model width. The learnable weights dynamically modulate the fusion process by adaptively adjusting the contribution ratios between multi-scale shortcut connections and hierarchical deep features through differentiable attention mechanisms. Specifically, the selected shortcut and deep features with $2^{(i-1)} \ast \bm{R} \ast C$ channels is multiplied with $\bm{\alpha}$ and $\bm{\beta}$ of size $2^{(i-1)} \ast \bm{R} \ast \times C \times 1 \times 1$ through Hardmard product to emphasize important channels and suppress less relevant ones. The advantages of the developed strategy are threefold: (1) it introduces a negligible number of additional parameters, making it highly parameter-efficient; (2) it allows the network to automatically learn the importance of each feature channel, adapting to the specific requirements of the task; and (3) it maintains the simplicity and efficiency of direct feature fusion while improving performance through channel-wise recalibration. This adaptability is particularly useful in tasks like medical image segmentation, where the importance of different feature channels may vary depending on the anatomical structures being analyzed. 
\subsection{Overall Architecture}
\begin{figure*}[!t]
\centering
\includegraphics[width=5in]{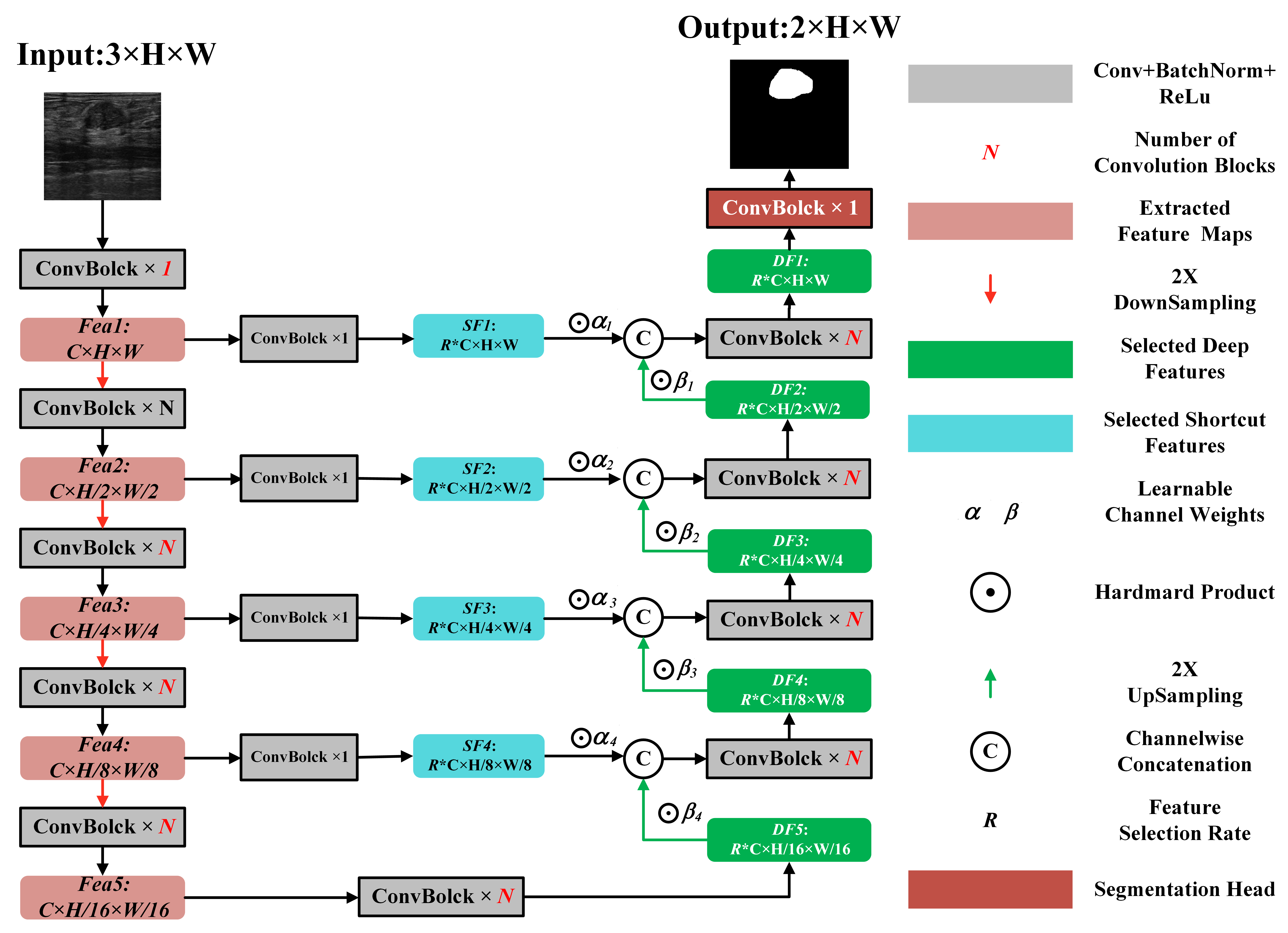}
\caption{The detailed architecture of the proposed \netname{}, where we also explicitly manifest the intermediate features for better understanding. The main difference between our model and U-Net is the shortcut, where we apply $\bm{R}*C$ groups convolution with the kernel size of $1 \times 1$ to shrink the feature channel of C to $\bm{R}*C$ instead. $\alpha$ and $\beta$ are groups of learnable parameters designed for channel weight reassigning. Therefore, U-Net falls into a special case of our \netname{} by configuring $N$, $\bm{R}$, $\alpha$ and $\beta$ to 2, 1, 1 and 1 but having $\alpha$ and $\beta$ fixed.}
\label{fig:arch}
\end{figure*}

Overall, our \netname{} can be seen in Figure \ref{fig:arch}, where multiple $N$ convolution blocks are stacked in each stage for enhanced feature extraction. Then the overall number of parameters is grossly $ 6\ast C^2\ast k^2$ once we have $\bm{R}$ as 0.5 and $N$ as 1, which is around \textbf{65} times parameter efficient than that of U-Net. By combining the above-mentioned strategies into the model design, we can have our \netname{} with extremely low volumes of parameters. As seen, the overall architecture and modules within it are interpretation-friendly though lightweight oriented. The pseudo codes corresponding to the data flow within the developed model can be seen in \textbf{Algorithm} \ref{alg:simple}. Our \netname{} can further be presented in Fig.\ref{fig:arch}, where the original intermediate features are also given for better understanding. 
\begin{algorithm}[!htb]
\caption{Data flow within \netname{} for medical image segmentation.}\label{alg:simple}
\algnewcommand{\ConvolutionBlock}[2]{\text{ConvBlock}_{#1}^{#2}}
\begin{algorithmic}[1] 
\State \# \textcolor{red}{\textit{Data flow in encoding stage}}
\State $\bm{Fea_{0}} \gets \text{Input}$ 
\For{$i=1$ \textbf{to} $5$}
  \State $\bm{Fea_{i}} \gets \ConvolutionBlock{N}{3 \times 3}(\bm{Fea_{i-1}})$
  \If{$i < 5$}
    \State $\bm{RScut}_{i} \gets \ConvolutionBlock{1}{1 \times 1}(\bm{Fea_{i}})$
    \State $\bm{Fea_{i}} \gets \text{DownSample}(\bm{Fea_{i}})$ 
  \EndIf
\EndFor

\State \# \textcolor{red}{\textit{Data flow in bottleneck}} 
\State $\bm{DF_{5}} \gets \ConvolutionBlock{N}{3 \times 3}(\bm{Fea_{5}})$

\State \# \textcolor{red}{\textit{Data flow in decoding stage}} 
\For{$i=4$ \textbf{downto} $1$}
  \State $\bm{DF_{i}} \gets \ConvolutionBlock{N}{3 \times 3} \Big( $ 
  \State \quad $\text{Concat}\big( \textcolor{red}{\alpha_{i}}*\bm{RScut_{i}}, $
  \State \quad $\text{UpSample}(\textcolor{red}{\beta_{i}}*\bm{DF_{i+1}}) \big) \Big) $ 
\EndFor

\State \# \textcolor{red}{\textit{Final output}}
\State \textbf{Output} $\gets \ConvolutionBlock{1}{1 \times 1}(\bm{DF_{1}})$ 
\end{algorithmic}
\end{algorithm}

\section{Experiment}
\subsection{Datasets}
In this study, we examined our proposed strategies and models on seven public medical datasets, including breast lesions, skin lesions, and endoscopic polyps. The numbers of samples in these datasets vary from 1000 to around 3600 and therefore can be taken as representative datasets to measure the performance of deep semantic segmentation models across varied scales. Also, these segmentation tasks vary from lesion types, locations, and shapes, which can be used to measure the generality of the developed models.\\
\textbf{Merged Breast Dataset (MBD)} The developed strategies are validated on four public breast lesion datasets, including BUSI\cite{2020Dataset}, DatasetB\cite{thomas2023bus}, BrEaST\cite{pawlowska2024curated}, and BUS-BRA\cite{gomez2023bus}. Among the datasets considered, BUS-BRA stands out for its substantial size, contributing over 1,800 images on its own. Given that all images across these datasets are meticulously annotated with pixel-level delineations, we then merged them into an enhanced dataset comprising approximately 3,000 images. We name the merged breast dataset as \textbf{MBD} for simplicity. We then randomly partition them into the train set, validation set, and test set in a ratio of 7:1:2. The detailed composition can be seen in Table \ref{tab:breast_data}. 
\begin{table}[!t]
\caption{The Merged Breast Dataset (MBT) for semantic segmentation.\label{tab:breast_data}}
\centering
\begin{tabular}{cccc|c}
\hline
{Dataset} & {Train} & {Validation} & {Test} & {Total}\\
\hline
BUSI\cite{2020Dataset}           &  451 & 65 & 130 & 646\\
DatasetB\cite{thomas2023bus}     &  113 &  17 & 33 & 163\\
BrEaST\cite{pawlowska2024curated} &  175 & 26 & 51 & 252\\
BUS-BRA\cite{gomez2023bus}       & 1311 & 188 & 376 & 1875\\
\hline
Total                             & 2050 & 296 & 590 & 2936\\
\hline
\end{tabular}
\end{table}

\noindent \textbf{ISIC2017 and ISIC2018} ISIC2017 and ISIC2018 are two public dermoscopic lesion datasets released by the International Skin Imaging Collaboration(ISIC) in 2017 and 2018 for developing automated lesion analysis tools \cite{berseth2017isic, wen2018isic}. All images are well-annotated by experts with pixel-level annotations. In ISIC2017, \textbf{2000} images are included in the training set while \textbf{600} images are used for testing, where there are \textbf{2594} and \textbf{1000} images in the counterparts of ISIC2018. The majority of the lesions are neatly centered but vary in shape and size. The details can be seen in Table \ref{tab:isic_data}.
\begin{table}[htb]
  \centering
  \caption{Details of ISIC 2017 and ISIC 2018.}
  \label{tab:isic_data}
  \begin{tabular}{ccc|c}
    \hline
    {Dataset} & {Train} & {Test} & {Overall}\\
    \hline
    ISIC2017 \cite{berseth2017isic} &  2000 & 600 & 2600\\
    ISIC2018 \cite{wen2018isic} &  2594 & 1000 & 3594\\
    \hline
    Total & 4594 & 1600 & 6194\\
    \hline
  \end{tabular}
\end{table}

\noindent \textbf{KVASIR-SEG} KVASIR-SEG is an endoscopic dataset specifically created for pixel-level segmentation of colonic polyps\cite{jha2019kvasir}, and it was featured as part of the MediaEval 2020 benchmarking challenge. The dataset includes 1,000 gastrointestinal polyp images, each paired with corresponding segmentation masks. These masks were carefully annotated and validated by board-certified gastroenterologists to ensure high diagnostic accuracy \cite{jha2019kvasir}. In this study, we randomly divided the dataset into \textbf{700} images for the training set, \textbf{100} images for the validation set, and \textbf{200} images for the testing set, along with their respective labels. For datasets containing validation datasets, we only validate the performance of the model during training after each epoch and iteratively save the weights that give the best performance on the validation set. When model training finishes, we reload the weights from the best-performing model for testing on the unseen test set. Otherwise, the testing set is used as the validation set instead. Some examples and corresponding labels can be seen in Figure \ref{fig:samples}.
\def\myvaluet{0.13}
\begin{figure*}[!t]
\centering
\subfloat[]{\includegraphics[height=\myvaluet\textwidth, width=\myvaluet\textwidth]{Figures/Samples/Images/BUSI_benign_116.png}%
    \label{sample_a}}
\hfil    
\subfloat[]{\includegraphics[height=\myvaluet\textwidth, width=\myvaluet\textwidth]{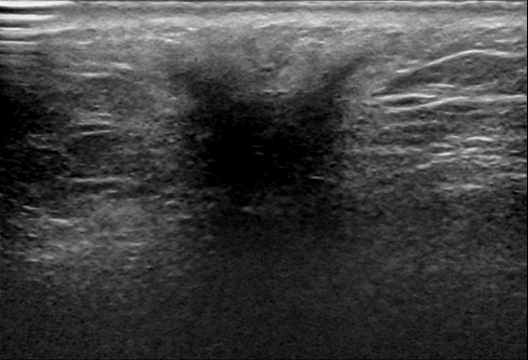}%
\label{sample_b}}
\hfil    
\subfloat[]{\includegraphics[height=\myvaluet\textwidth, width=\myvaluet\textwidth]{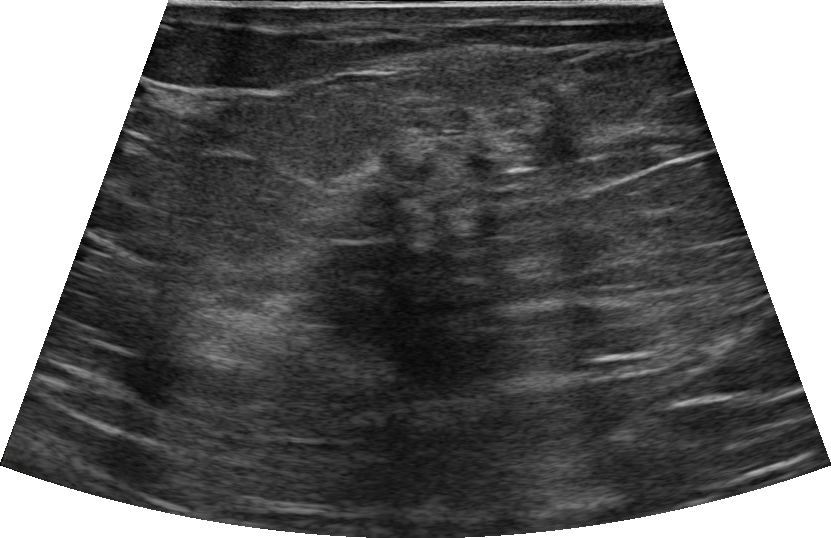}%
\label{sample_c}}
\hfil    
\subfloat[]{\includegraphics[height=\myvaluet\textwidth, width=\myvaluet\textwidth]{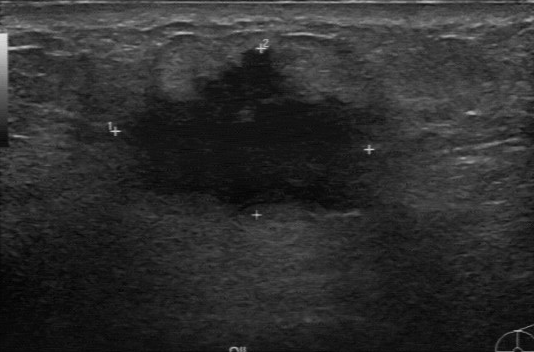}%
\label{sample_d}}
\hfil 
\subfloat[]{\includegraphics[height=\myvaluet\textwidth, width=\myvaluet\textwidth]{Figures/Samples/Images/ISIC_0000052.jpg}%
\label{sample_e}}
\hfil 
\subfloat[]{\includegraphics[height=\myvaluet\textwidth, width=\myvaluet\textwidth]{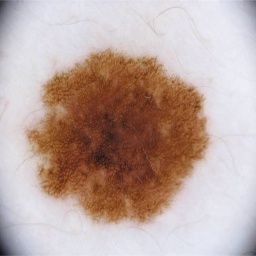}}%
\label{sample_f}
\hfil 
\subfloat[]{\includegraphics[height=\myvaluet\textwidth, width=\myvaluet\textwidth]{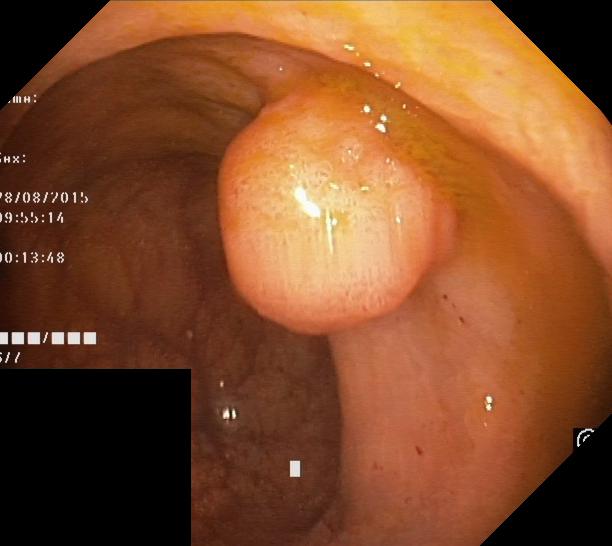}}%
\label{sample_g}
\subfloat[]{\includegraphics[height=\myvaluet\textwidth, width=\myvaluet\textwidth]{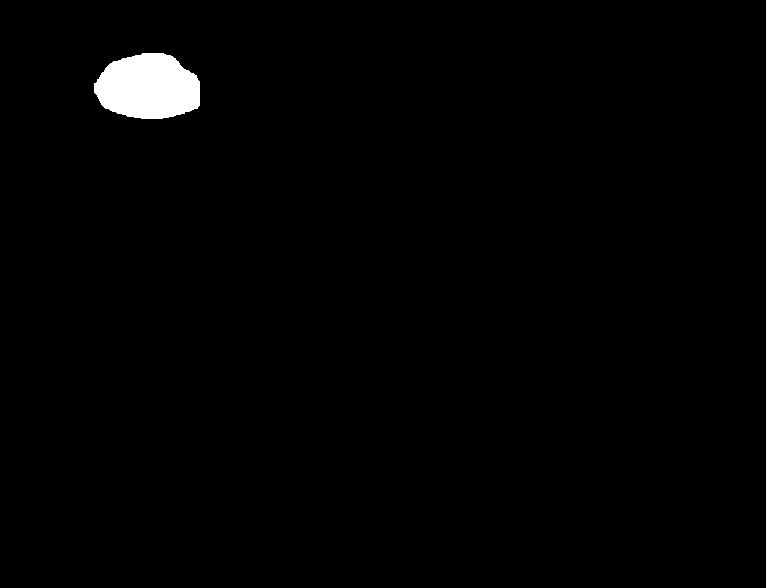}%
    \label{sample_a2}}
\hfil    
\subfloat[]{\includegraphics[height=\myvaluet\textwidth, width=\myvaluet\textwidth]{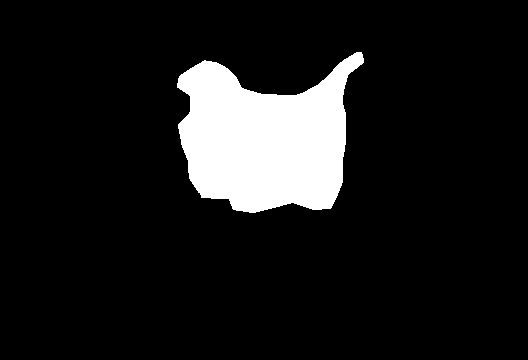}%
\label{sample_b2}}
\hfil    
\subfloat[]{\includegraphics[height=\myvaluet\textwidth, width=\myvaluet\textwidth]{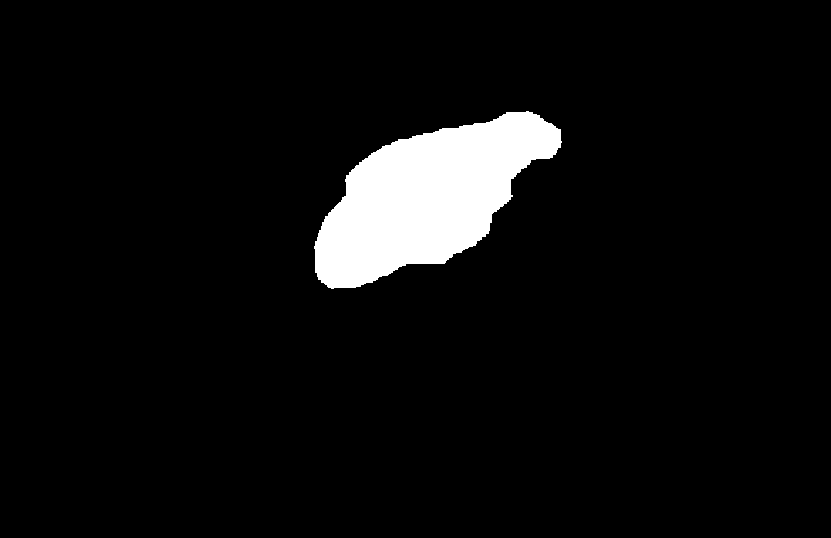}%
\label{sample_c2}}
\hfil    
\subfloat[]{\includegraphics[height=\myvaluet\textwidth, width=\myvaluet\textwidth]{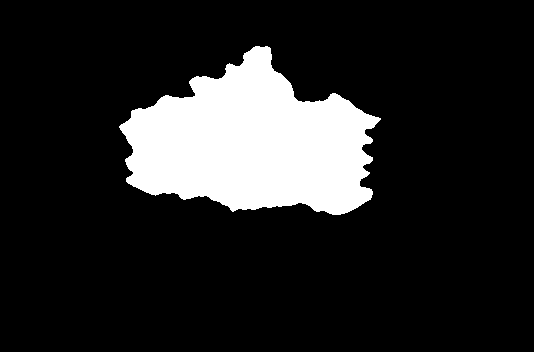}%
\label{sample_d2}}
\hfil 
\subfloat[]{\includegraphics[height=\myvaluet\textwidth, width=\myvaluet\textwidth]{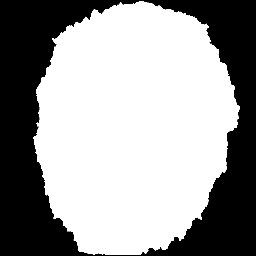}}%
\label{sample_e2}
\hfil 
\subfloat[]{\includegraphics[height=\myvaluet\textwidth, width=\myvaluet\textwidth]{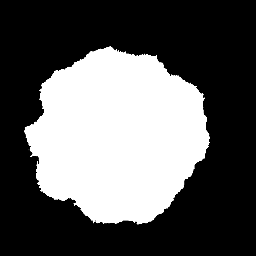}}%
\label{sample_f2}
\hfil 
\subfloat[]{\includegraphics[height=\myvaluet\textwidth, width=\myvaluet\textwidth]{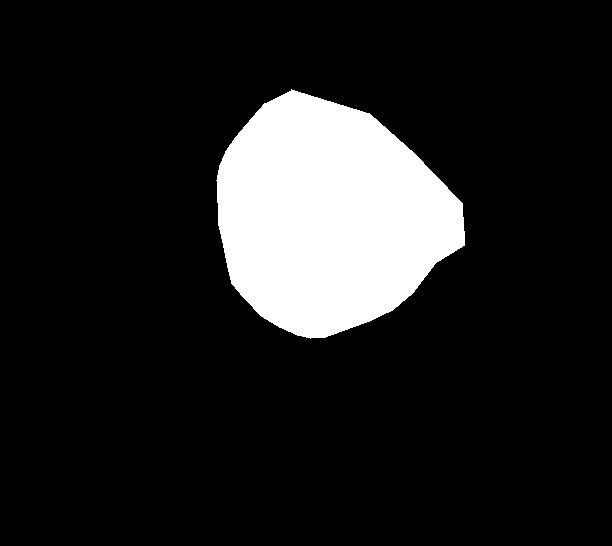}}%
\label{sample_g2}
\caption{From (a) to (g) are images from BUSI, DatasetB, BrEaST, BUS-BRA, ISIC2017, ISIC2018, and KVASIR-SEG, respectively, while (h) to (n) are corresponding labels. The presented multi-source dataset exhibits significant heterogeneity in breast lesion manifestations: samples span diverse scales, anatomical locations, morphological complexity, and dynamic intensity ranges. On the contrary, lesions in ISIC2017 and ISIC2018 images exhibit a centralized tendency near the image centroid. Like the breast lesions, colorectal lesions demonstrate significant heterogeneity in both spatial distribution and morphological characteristics, with greater variability in size and eccentric positioning relative to image boundaries.}\label{fig:samples}
\end{figure*}

\subsection{Evaluation metrics}
we use the Dice Similarity Coefficient (DSC) and Intersection over Union (IoU) as the evaluation metrics in this study, as seen from Equation (\ref{eq:dsc}) and (\ref{eq:miou}), where $TN^{i}$, $TP^{i}$, $FN^{i}$, and $FP^{i}$ for True Negative, True Positive, False Negative and False Positive in $i$th run, respectively, where $M$ is the total number of runs.   
\begin{equation}
\label{eq:dsc}
\mathbf{DSC^{i}} = \frac{2 \times TP^{i}}{2 \times TP^{i} + FP^{i} + FN^{i}}
\end{equation}

\begin{equation}
\mathbf{mDSC} = \frac{1}{M}\sum_{i=1}^{M}\mathbf{DSC^{i}}(i \in \{ 1, \cdots, M \})
\label{eq:mdice}
\end{equation}

\begin{equation}
\label{eq:iou}
\mathbf{IoU^{i}} = \frac{TP^{i}}{TP^{i} + FP^{i} + FN^{i}}
\end{equation}

\begin{equation}
\mathbf{mIoU} = \frac{1}{M}\sum_{i=1}^{M}\mathbf{IoU^{i}}
\label{eq:miou}
\end{equation}

\subsection{Configurations}
\netname{} is developed via the Pytorch framework, while all experiments are conducted on two single NVIDIA L20 GPUs and an NVIDIA GeForce RTX 3090 GPU. By default, we resized the images to $256 \times 256$. We also performed composited data augmentations, including random horizontal flipping, random scale, contrast enhancement via CLAHE\cite{pizer1990contrast}, and color jittering. The optimizer is chosen as AdamW with an initial learning rate of 0.001, while the CosineAnnealingLR is used as the scheduler\cite{loshchilov2017decoupled, loshchilov2016sgdr}. The batch size is set to 16 by default and is halved when out of memory occurs or training stagnates. \textbf{For fair comparisons, all models are individually trained 5 times from scratch ($M=5$ in Equation (\ref{eq:mdice}) and Equation (\ref{eq:miou}))} for 200 epochs without applying early stopping, and we report the averaged evaluation metrics by default. 

\subsection{Ablation studies on module design}
\subsubsection{The effectiveness of the proposed strategies based on U-Net}
Before we move any further, we think examining the developed strategies based on the original U-Net is a necessity. The merged breast lesion dataset MBD is considered here for model training and evaluation because we reckon that the performance of a model on heterogeneous data is more persuasive than the performance on homogeneous and carefully curated data, such as ISIC2017 and ISIC2018. 

\textbf{The vanilla U-Net with varied widths} We then replicate the U-Net with two stacked convolution layers in each block. The kernel size of the convolution is 3 throughout the networks, while the dilation rate is set to 1 by default. We then vary the initial model width from 8 to 64 and stagewise increase the width by a scale factor of 2, where the results can be seen in Table \ref{tab:oriunet}. As seen, all models with varied scales showed competitive performance concerning breast lesion segmentation, in which the model with only 0.54 MB parameters showed an mDSC and mIoU of $83.97\%$ and $72.72\%$. The model with an initial width of 32, however, outperforms its counterparts and gives an mDSC and mIoU of $84.28\$$ and $73.18\%$. Instead, the model of 64 showed slightly declined performance. The results demonstrate that a suitable model should be chosen; otherwise, overly introduced parameters may lead to information redundancy and therefore harm the segmentation performance, which might justify why the model with only 8.64M parameters outperforms the model with 34.53M instead. Based on the results, we reckon that the overall architecture of U-Net can be optimized for possible parameter reduction as well as information redundancy mitigation. 
\begin{table*}[thb]
\caption{The performance of U-Nets with varied scales on the dataset MBD. The output channels of each stage in the segmentation model correspond to the values at their respective index positions in the list $[N, 2*N, 4*N, 8*N, 16*N], N \in \{8, 16, 32, 64\}$. The number of parameters is measured in millions (M). The bold indicates the best, while the underlined indicates the second best.}
\label{tab:oriunet}
\centering
\begin{tabular}{ccccc}
\toprule
Model Width & Para(MB, $\downarrow$) & GFLOPs($\downarrow$)  & mDSC($\%, \uparrow$) & mIoU($\%, \uparrow$) \\
\toprule
$[8, 16,32, 64, 128]$ & \textbf{0.54} & $\textbf{1.06}$ & $83.97\pm0.37$ & $72.72\pm0.54$ \\
$[16, 32, 64, 128, 256]$ & \uline{2.16} & \uline{4.15} & $\uline{84.21\pm0.41}$ & $\uline{73.09\pm0.59}$ \\
$[32, 64, 128, 256, 512]$ &  8.64 & 16.46 & $\bm{84.28\pm0.38}$ & $\bm{73.18\pm0.53}$ \\ 
$[64, 128, 256, 512, 1024]$ & 34.53 & 65.53 & $84.19\pm0.25$ & $73.05\pm0.35$ \\
\bottomrule
\end{tabular}
\end{table*}

\textbf{The U-Nets with efficient feature selection} We then integrate U-Net with our proposed feature selection strategy, and the results can be seen in Table \ref{tab:efsunet}. The UNet with an initial width of 64 has now become the best one with an mDSC and mIoU of $85.15\%$ and $75.52\%$. Besides, the number of parameters has been slashed by nearly half, while the computational cost is reduced by more than 2/3. Moreover, U-Nets across different scales experienced consistent performance improvements, while larger models tend to benefit more from the feature selection strategy. The experiments here show that the proposed efficient feature selection brings multiple benefits to the models, such as reduced model size, performance improvement, and possible shortened latency. The experimental results here again implicitly demonstrate the architectural redundancy within the original U-Nets that improved architecture can lead to better segmentation results with greatly reduced numbers of parameters.
\begin{table*}[thb]
\caption{The performance of U-Nets with efficient feature selection on the dataset MBD.}
\label{tab:efsunet}
\centering
\begin{tabular}{ccccc}
\toprule
 Model Width & Para & GFLOPs  & mDSC & mIoU \\
\toprule
$[8, 16,32, 64, 128]$  & \textbf{0.31} & \textbf{0.34} & $84.17\pm0.31(0.20\uparrow)$ & $73.07\pm0.42(0.35\uparrow)$ \\
$[16,32, 64, 128, 256]$  & \uline{1.24} & \uline{1.28} & $84.65\pm0.57(0.44\uparrow)$ & $73.49\pm0.85(0.40\uparrow)$ \\
$[32, 64, 128, 256, 512]$  &  4.95 & 4.96 & $\uline{85.13\pm0.47}(0.85\uparrow)$ & $\uline{74.50\pm0.69}(1.32\uparrow)$ \\ 
$[64, 128, 256, 512, 1024]$  & 19.80 & 19.56 & $\bm{85.15\pm0.50}(0.96\uparrow)$ & $\bm{74.52\pm0.72}(1.47\uparrow)$ \\
\bottomrule
\end{tabular}
\end{table*}

\textbf{The U-Nets with intuitive model width control} We then fix the model width of U-Nets throughout the models and examine how much the segmentation performance is affected, where the results can be seen in Table \ref{tab：imwcunet}. Overall, the models experienced noticeable performance declines with the mDSC and mIoU dropping by 1.35$\%$ and 1.90$\%$ on average, where the smaller models experienced larger declines. Nevertheless, the model with a fixed width of 64 is the least affected one that gives a mDSC and mIoU of 83.73$\%$ and 72.41$\%$ with only 0.93 MB parameters, which is around 37 times fewer than the original one. From this point and beyond, the intuitive model with control strategy is successful as the overall parameters reduce by around 36 times across varied scales of models without drastically deteriorating performance.
\begin{table*}[thb]
\caption{The performance of U-Nets with fixed model width on the dataset MBD. $5*[N]$ means the model width is fixed at $N$ throughout five stages.}
\label{tab：imwcunet}
\centering
\begin{tabular}{ccccc}
\toprule
 Model Width & Para & GFLOPs  & mDSC & mIoU \\
\toprule
$5*[8]$ & \textbf{0.015} & \textbf{0.29} & $81.39\pm0.41(2.58\downarrow)$ & $69.09\pm0.59(3.63\downarrow)$ \\
$5*[16]$ & \uline{0.059} & \uline{1.11} & $82.76\pm0.40(1.45\downarrow)$ & $71.02\pm0.54(2.07\downarrow)$ \\
$5*[32]$ &  0.23 & 4.33 & \uline{$83.38\pm0.45(0.90\downarrow)$} & \uline{$71.95\pm0.67(1.23\downarrow)$} \\ 
$5*[64]$ & 0.93 & 17.09 & \bm{$83.73\pm0.67(0.46\downarrow)$} & \bm{$72.41\pm0.92(0.64\downarrow)$} \\
\bottomrule
\end{tabular}
\end{table*}

\textbf{The U-Nets with adaptive feature fusion} Based on conventional U-Nets, we apply adaptive feature fusion to U-Nets with varied scales. As seen from Table \ref{tab：affunets}, all models except the one with an initial width of 64 experienced slight declines with negligible parameter and computational costs. In contrast, the model with an initial width of 64 successfully improves the mDSC and mIoU by 0.30$\%$ and 0.44$\%$, respectively, indicating that the adaptive feature fusion can be successful on scaled features. The above experimental results demonstrate the effectiveness of feature selection and intuitive model width control, while the adaptive feature fusion should be carefully applied. Nevertheless, we found that the model with an initial model width of 64 is more likely to show better performance due to a larger number of parameters. 
\begin{table*}[thb]
\caption{The performance of U-Nets with adaptive feature fusion on the dataset MBD.}
\label{tab：affunets}
\centering
\begin{tabular}{cccccc}
\toprule
Model Width & Para & GFLOPs  & mDSC & mIoU \\
\toprule
$[8, 16, 32, 64, 128]$ & \textbf{0.54} & \textbf{1.06} & $83.78\pm0.46(0.19\downarrow)$ & $72.43\pm0.64(0.29\downarrow)$ \\
$[16, 32, 64, 128, 256]$ & \uline{2.16} & \uline{4.15} & $83.80\pm0.54(0.41\downarrow)$ & $72.47\pm0.75(0.62\downarrow)$ \\
$[32, 64, 128, 256, 512]$ &  8.64 & 16.46 & \uline{$84.15\pm0.61(0.13\downarrow)$} & \uline{$73.02\pm0.87(1.32\downarrow)$} \\ 
$[64, 128, 256, 512, 1024]$  & 34.53 & 65.53 & $\bm{84.49\pm0.44(0.30\uparrow)}$ & $\bm{73.49\pm0.65(0.44\uparrow)}$ \\
\bottomrule
\end{tabular}
\end{table*}

\subsubsection{Explorations of \netname{} design}
The main difference between our developed \netname{} and U-Net is that we combine the feature selection and the intuitive width control, while leaving the adaptive feature fusion switched off by default, leading to even more parameter-compact models. Without other specifications, we have the following configurations until specified: The model width is 64 throughout five stages, the kernel size of the convolution is $3\times3$, the dilation rate of the convolution is 2, and the number of convolution blocks is 1. Now, we go through each of these variables to identify their impact on model performance. 
\\
\textbf{Model Width} We investigate the impact of model width on performance with finer-grained intervals, as seen in Table \ref{tab:width}, where we vary the model width from 8 to 128. While wider models tend to show better performance, the models with the narrow widths are also promising, given the model performance and parameter efficiency. Overall, the model with 96 reached an mDSC and mIoU of $85.09\%$ and $74.42\%$, which already surpassed the U-Nets. Moreover, this ablation study shows that wider models don't always guarantee performance, as seen from models with widths of 24 and 128. Notably, the minimal model only has \textbf{4} $\bm{KB}$ parameters but also showed promising results though it might not be suitable for practical deployment. Compared to the original U-Net, our \netname{}s are much lighter and efficient, where the \netname{} with only 0.251M parameters already shows comparable performance to U-Nets but enjoys around \textbf{140 times} parameter efficiency. In conclusion, the results here supported our motivation concerning lightweight model design and led to even better performance.
\begin{table}[htb]
\centering
\caption{Ablation studies of \netname{} concerning model width.}\label{tab:width}
\begin{tabular}{ccccc}
\toprule
Model Width & Para & GFLOPs & mDSC & mIoU \\
\toprule
$5*[8]$  & \textbf{0.004} & \textbf{0.065}  & $80.95\pm0.54$ & $68.43\pm0.72$ \\
$5*[16]$& \uline{0.016} & \uline{0.211}  & $82.93\pm0.20$ & $71.25\pm0.29$ \\
$5*[24]$ & 0.036 & 0.437  & $82.87\pm0.57$ & $71.20\pm0.81$ \\
$5*[32]$ & 0.063 & 0.744  & $83.88\pm0.45$ & $72.64\pm0.64$ \\
$5*[48]$ & 0.142 & 1.602  & $84.12\pm0.17$ & $73.04\pm0.25$\\
$5*[64]$ & 0.251 & 2.783  & $84.22\pm0.39$ & $73.15\pm0.54$ \\
$5*[96]$ & 0.562 & 6.116  & $\mathbf{85.09\pm0.41}$ & $\mathbf{74.42\pm0.62}$ \\
$5*[128]$ & 0.997 & 10.744  & \uline{$84.92\pm0.29$} & \uline{$74.20\pm0.43$} \\
\bottomrule
\end{tabular}
\end{table}

\textbf{Kernel Size} We then vary the kernel sizes of the convolution while fixing the model width as 64 and report the results in Table \ref{tab:kernel}. The model with the kernel size of 5 gives the best results at an mDSC and mIoU of $84.54\%$ and $73.71\%$, indicating that a suitable kernel size should be carefully chosen, while larger kernels lead to an exponential increase in parameters. Aimed at developing a lightweight yet computationally efficient model, we stick with the kernel size of 3 in the remaining experiments.
\begin{table}[htb]
\centering
\caption{Ablation studies of \netname{} concerning kernel size of convolution.}\label{tab:kernel}
\begin{tabular}{ccccc}
\toprule
Kernel Size & Para & GFLOPs & mDSC & mIoU \\
\toprule
3 & \textbf{0.251} & \textbf{2.783} & $84.22\pm0.39$ & $73.15\pm0.54$ \\
5 & \uline{0.680} & \uline{7.271}  & $\mathbf{84.54\pm0.22}$ & $\mathbf{73.71\pm0.33}$ \\
7 & 1.323 & 14.002  & \uline{$84.51\pm0.44$} &\uline{$73.70\pm0.61$} \\
\bottomrule
\end{tabular}
\end{table}

\textbf{Dilation Rate} The results of models with dilation rates ranging from 1 to 3 are presented in Table \ref{tab:dilation}. The results show that larger dilation rates enable models to have larger receptive fields and therefore better performance. As seen from  Table \ref{tab:dilation}, the \netname{} with a dilation rate of 3 turns out the best in terms of mDSC and mIoU with $84.27\%$ and $73.33\%$, respectively. However, large dilation rates can also lead to overly sparse effective weights and significantly increased memory consumption. Therefore, a dilation rate of 2 is recommended over 3.
\begin{table}[htb]
\centering
\caption{Ablation studies concerning dilation rate of convolution $DR$.}\label{tab:dilation}
\begin{tabular}{ccccc}
\toprule
Dilation Rate & Para & GFLOPs & mDSC & mIoU\\
\toprule
1 & 0.251 & 2.783 & $82.54\pm0.24$ & $70.65\pm0.35$ \\
2 &  0.251 & 2.783 & \uline{$84.22\pm0.39$} & \uline{$73.15\pm0.54$} \\
3 & 0.251 & 2.783 & $\mathbf{84.27\pm0.19}$ & $\mathbf{73.33\pm0.26}$ \\
\bottomrule
\end{tabular}
\end{table}

\textbf{Feature Selection Rate $\bm{R}$} We then vary the feature selection rate $\bm{R}$ from 0.25 to 1 and the results can be seen in Table \ref{tab:reduction}, where models with larger rates tend to be more desirable in terms of performance. As a result, the model with the selection rate of 1 provided an mDSC and mIoU of 84.63$\%$ and 73.71$\%$. Nevertheless, the model with a reduction rate of 0.25 also presents promising performance with an mDSC and mIoU of 84.11$\%$ and 73.03$\%$, manifesting the plausibility of applying feature selection in the developed models. The experimental results here again demonstrate the effectiveness of feature selection. Conservatively, we move on with the reduction rate of 0.5 as 0.25 might lead to significant information loss when the model goes deeper and larger.
\begin{table}[htb]
\centering
\caption{Ablation studies concerning feature selection rate $\bm{R}$.}\label{tab:reduction}
\begin{tabular}{ccccc}
\toprule
$\bm{R}$ & Para & GFLOPs & mDSC & mIoU\\
\toprule
0.25 & \textbf{0.182} & \textbf{1.459}  & $84.11\pm0.12$ & $73.03\pm0.16$\\
0.5 &\uline{0.251} & \uline{2.783}  & $84.22\pm0.39$ & $73.15\pm0.54$ \\
0.75 & 0.357 & 4.908  & \uline{$84.32\pm0.42$} & \uline{$73.33\pm0.63$} \\
1 &  0.499 & 7.836  & $\mathbf{84.63\pm0.23}$ & $\mathbf{73.71\pm0.34}$ \\
\bottomrule
\end{tabular}
\end{table}

\textbf{Number of Convolution Blocks} The results of models with varied convolution blocks in each stage are presented in Table \ref{tab:numblk}. The model with 2 convolution blocks in each stage turns out the best with mDSC and mIoU values of $85.44\%$ and $75.06\%$ while the number of parameters is only around 0.45 \textbf{MB}, which again demonstrates the high performance and parameter-efficiency of the developed models.
\begin{table}[htb]
\centering
\caption{Ablation studies concerning the number of convolution blocks $NumBlock$.}\label{tab:numblk}
\begin{tabular}{ccccc}
\toprule
$NumBlock$ & Para & GFLOPs & mDSC & mIoU\\
\toprule
1 &\textbf{ 0.251} & \textbf{2.783}  & $84.22\pm0.39$ & $73.15\pm0.54$ \\
2 & \uline{0.445} & \uline{4.406}  & $\mathbf{85.44\pm0.40}$ & $\mathbf{75.06\pm0.50}$ \\
4 & 0.834 & 7.653  & $\uline{85.36\pm0.50}$ & $\uline{74.99\pm0.67}$ \\
\bottomrule
\end{tabular}
\end{table}

\textbf{Adaptive Feature Fusion} Previous experimental results based on U-Nets showed that adaptive feature fusion should be applied to wider models. We, therefore, examine the performance of our \netname{} of width 64 with or without adaptive feature fusion. For the model without adaptive feature fusion, the values of $\bm{\alpha}$ and $\bm{\beta}$ are simply fixed to constant ones, leading to models with almost the same number of parameters and computational costs. However, adaptive feature fusion enriches feature representation and therefore helps to boost the performance by $0.44\%$ and $0.63\%$ regarding mDSC and mIoU. Therefore, adaptive feature fusion is suggested for the model width of 64 or around. 
\begin{table}[htb]
\centering
\caption{Ablation studies concerning adaptive feature fusion.}\label{tab:ada}
\begin{tabular}{ccccc}
\toprule
AFF & Para & GFLOPs & mDSC & mIoU\\
\toprule
\ding{55} & $\bm{< 0.251}$  & $\bm{<2.783}$  & $84.22\pm0.39$ & $73.15\pm0.54$ \\
\ding{51} & 0.251 & 2.783  & $\mathbf{84.66\pm0.39}$ & $\mathbf{73.78\pm0.55}$ \\
\bottomrule
\end{tabular}
\end{table}

\textbf{Combinations} Based on previous ablation studies, we then combine the settings as we aimed to obtain the most optimal ones. The results can be seen in Table \ref{tab:comb}, where we tweak our model with model widths and number of blocks while switching the adaptive feature fusion on and having the kernel size, dilation rate, and feature selection rate of 3, 2, and 0.5. As seen, the model with a width of 48 gives the highest mDSC and mIoU values of 85.70$\%$ and 75.47$\%$, respectively, suggesting the importance of the stacked blocks. Based on previous experiments, we consider that the developed framework is successful, as our models achieved promising performance while no fancy blocks other than conventional convolution blocks are involved in the model.
\begin{table}[htb]
\centering
\caption{Ablation studies concerning combinations of configurations, where  $[48, 2, AFF]$ signals the model with the width of 48 and number of blocks of 2, while "AFF" means the adaptive feature fusion is applied. So are the other configurations.}\label{tab:comb}
\begin{tabular}{ccccc}
\toprule
Configs & Para & GFLOPs & mDSC & mIoU \\
\toprule
$[48, 2, AFF]$  & \textbf{0.25}  & \textbf{2.52} & $85.02\pm0.38$ & $74.38\pm0.58$ \\
$[48, 4, AFF]$ & \uline{0.47}  & \uline{4.35} & $\mathbf{85.70\pm0.64}$ & $\mathbf{75.47\pm0.87}$\\
$[64, 2, AFF]$ & 0.45  & 4.41 & $85.50\pm0.49$ & $75.09\pm0.68$\\
$[64, 4, AFF]$ & 0.83  & 7.65 & $\uline{85.68\pm0.48}$ & $\uline{75.44\pm0.70}$ \\
\bottomrule
\end{tabular}
\end{table}

\subsubsection{Extendability of \netname{}}
Besides, we would like to demonstrate the high extendability of our models. Based on a high-performance edge-cutting model ESKNet \cite{chen2024esknet}, we simply replace the common convolution blocks with the ones in ESKNet while maintaining the overall architecture, including efficient feature selection, intuitive width control, and adaptive feature fusion. For simplicity, we name the extended model as SimpleESKNet. Deeply bearing in mind that we simply desire models that have fewer than 1 MB parameters, we then restrain the number of parameters by adjusting the model width and number of blocks in each stage. For quick validations, we trained and tested the extended models on the KVASIR-SEG dataset, as it is the smallest dataset among the four datasets, while the results can be seen in Table \ref{tab:mc_simesk_kvasir}. \\
As seen, the adaptive feature fusion helps to improve the minimal model $SimpleESKNet_{16}$ by 0.24$\%$ and 0.40$\%$ in terms of mDice and mIoU, indicating the effectiveness of the adaptive feature fusion. More importantly, all models showed competitive performance with the values of mDSC and mIoU being over 82$\%$ and 70$\%$, where the best one presented an mDice and mIoU of $86.46\%$ and $76.48\%$, respectively. In conclusion, the experimental results here demonstrate the easy integration and the strong transferability of the developed strategies. 
\begin{table}[htb]
\centering
\caption{The quantitative comparison between SimpleESKNets on KVASIR-SEG, where $SimpleESKNet_{32}^{2+AFF}$ indicates that the model width is 32, and the number of convolution blocks is 2 in each stage while the adaptive feature fusion is applied.}\label{tab:mc_simesk_kvasir}
\begin{tabular}{ccccc}
\toprule
\textcolor{red}{SimpleESKNet} &  Para & GFLOPs & mDSC & mIoU \\
\toprule
$[16, 1, -]$ & \textbf{0.10} & \textbf{1.65} & $82.43\pm0.90$ & $70.47\pm1.29$ \\
$[16, 1, AFF]$ & \uline{0.10} & \uline{1.65} & $82.67\pm0.86$ & $70.87\pm1.18$ \\
$[16, 4, AFF]$ & 0.31 & 3.32 & $83.58\pm0.87$ & $72.28\pm1.29$ \\
$[24, 1, AFF]$ & 0.22 & 3.65 & $84.02\pm0.89$ & $72.80\pm1.31$ \\
$[24, 2, AFF]$ & 0.38 & 4.88 & \uline{$85.41\pm0.65$} & \uline{$74.97\pm0.98$} \\
$[32, 1, AFF]$ & 0.40 & 6.42 & $84.76\pm0.60$ & $73.94\pm0.85$ \\
$[32, 2, AFF]$ & 0.67 & 8.60 & \bm{$86.46\pm0.42$} & \bm{$76.48\pm0.60$} \\
$[48, 1, AFF]$ & 0.89 & 14.30 & $84.47\pm0.62$ & $73.56\pm0.93$\\
\bottomrule
\end{tabular}
\end{table}

We then select SimpleESKNet$_{16}^{1+AFF}$ and SimpleESKNet$_{32}^{2+AFF}$ as the representative models and examine their performance on the other three datasets, where the results can be seen in Table \ref{tab:mc_simesk_mbd_isisc}. Surprisingly, the minimal $SimpleESKNet$  with only 0.1 MB parameters still performs well across multiple datasets, especially on the MBD dataset, which could be the intrinsic nature of ESKNet as it was developed for breast lesion segmentation in Ultrasound. Also, we note that the performance difference between the minimal one and the best one has been narrowed, suggesting that a model with only 0.1 MB can also be a powerful substitute for existing well-performed models. 
\begin{table}[htb]
\centering
\caption{The performance of the representative SimpleESKNets on MBD, ISIC2017 and ISIC2018.}\label{tab:mc_simesk_mbd_isisc}
\begin{tabular}{cccc}
\toprule
\textcolor{red}{SimpleESKNet} &  Dataset & mDSC & mIoU \\
\toprule
$[16, 1, AFF]$ & \multirow{2}*{MBD} & $85.48\pm0.35$ & $75.08\pm0.49$ \\
$[32, 2, AFF]$ & ~ & $85.76\pm0.49$ & $75.60\pm0.68$ \\
\midrule
$[16, 1, AFF]$ & \multirow{2}*{ISIC2017} & $84.49\pm0.24$ & $75.99\pm0.37$ \\
$[32, 2, AFF]$ & ~ & $84.86\pm0.34$ & $76.60\pm0.38$ \\
\midrule
$[16, 1, AFF]$ & \multirow{2}*{ISIC2018} & $88.41\pm0.15$ & $81.58\pm0.24$ \\
$[32, 2, AFF]$ & ~ & $88.77\pm0.19$ & $82.13\pm0.23$ \\
\bottomrule
\end{tabular}
\end{table}

\subsubsection{Comparison with the state-of-the-art methods}
We then compare our models with the state-of-the-art lightweight and prevalent methods over different testing sets. The results of representative models on the MBD can be seen in Table \ref{tab:mc_mbd}. In general, all lightweight models barely achieve an mDSC and mIoU over 82$\%$ and 70$\%$. However, our model with only 16 KB parameters obtained an mDSC and mIoU of 82.93$\%$ and 71.25$\%$, respectively. Among other lightweight models, $UNext_S$ provided an mDice of 82.50$\%$, which was only 0.43$\%$ lower than ours. However, the parameters within $UNext_S$ are around 10 times that of our \netname{}, while our model is twice as computationally expensive. While outperforming all lightweight models, the advanced SimpleESKNet also shows superior performance than some heavy models, such as U-Net, TransUNet, and MDA-Net, where only TransUNet showed promising segmentation performance with an mDSC and mIoU of 85.32$\%$ and 74.89$\%$, though with expensive parameters and computational costs. Instead, our best models showed dominating performance while the number of parameters was less than 0.7 MB. Therefore, we may conclude that our models are more advantageous on a heterogeneous dataset regarding breast lesion segmentation. Some segmentation examples can be found in Fig. \ref{fig:ins_cmp_sota}, Fig. \ref{fig:ins_sml_cmp_sota}, and Fig. \ref{fig:ins_med_cmp_sota}, where our paradigm fundamentally resolves the over/under-segmentation trade-off.
\begin{table*}[htb]
\centering
\caption{The quantitative performance comparison between state-of-the-art models on public medical image datasets. We compared our models to both lightweight and representative semantic segmentation models. Best viewed in color. The red indicates the best while the blue signifies the second best.}\label{tab:mc_mbd}
\begin{tabular}{cccccccccccc}
\toprule
\multirow{2}*{Model} & \multirow{2}*{Year} & \multirow{2}*{Para} & \multirow{2}*{GFLOPs} & \multicolumn{2}{c}{MBD}  & \multicolumn{2}{c}{KVASIR-SEG} & \multicolumn{2}{c}{ISIC2017} & \multicolumn{2}{c}{ISIC2018} \\ \cline{5-12}
~ & ~ & ~ & ~ &  mDice & mIoU & mDice & mIoU & mDice & mIoU & mDice & mIoU \\
\toprule
  MALUNet\cite{ruan2022malunet} &  2022  & 0.175 & \uline{0.083} & \makecell{80.87 \\ ± 0.59} & \makecell{68.55 \\ ± 0.87} & \makecell{64.37 \\ ± 1.80} & \makecell{47.76 \\ ± 1.91} & \makecell{83.48\\ ± 0.21} & \makecell{74.57\\ ± 0.25} & \makecell{87.55\\ ± 0.32} & \makecell{80.11\\ ± 0.56}\\
  UNext$_{s}$\cite{valanarasu2022unext}  &  2022 & 0.253 & 0.104 & \makecell{82.50\\ ± 0.46} & \makecell{70.74\\ ± 0.72} & \makecell{72.39 \\ ± 0.72} & \makecell{57.08\\ ± 0.88} & \makecell{\textcolor{blue}{83.79}\\ \textcolor{blue}{± 0.34}} & \makecell{\textcolor{blue}{74.97}\\ \textcolor{blue}{± 0.45}} & \makecell{87.70\\ ± 0.28} & \makecell{80.53\\ ± 0.47}\\
  LF-UNet\cite{deng2023lfu} & 2023 &  0.048  & 0.763 & \makecell{80.92\\ ± 0.45} & \makecell{68.35\\ ± 0.59} & \makecell{70.60 \\ ± 1.40} & \makecell{54.76 \\ ± 1.67} & \makecell{81.30\\ ± 0.21} & \makecell{71.89\\ ± 0.28} & \makecell{86.33\\ ± 0.26} & \makecell{78.17\\ ± 0.37}\\
  LBUNet\cite{xu2024lb}& 2024 & \uline{0.038} & 0.098 & \makecell{81.59\\ ± 0.60} & \makecell{69.44\\ ± 0.84} & \makecell{67.26\\ ± 1.59 } & \makecell{50.97 \\ ± 1.80} & \makecell{83.30\\ ± 0.27} & \makecell{74.54\\ ± 0.33} & \makecell{\textcolor{blue}{87.85}\\ \textcolor{blue}{± 0.50}}& \makecell{\textcolor{blue}{80.69}\\ \textcolor{blue}{± 0.70}}\\
  UltraVMUNet\cite{wu2024ultralight} & 2024 & 0.044 & \textbf{0.065} & \makecell{81.61\\ ± 0.66} & \makecell{69.43\\ ± 0.91} &\makecell{69.93\\ ± 1.19} & \makecell{54.25\\ ± 1.41} & \makecell{83.13\\ ± 0.40} & \makecell{74.33\\ ± 0.44}& \makecell{87.40\\ ± 0.50} & \makecell{79.73\\ ± 0.69}\\
  $\mathbf{SimpleUNet_{16}^{1}(Ours)}$ & - & \textbf{0.016}  & 0.211 & \makecell{\textcolor{blue}{82.93}\\ \textcolor{blue}{± 0.20}} & \makecell{\textcolor{blue}{71.25}\\ \textcolor{blue}{± 0.29}}& \makecell{\textcolor{blue}{73.11}\\ \textcolor{blue}{± 0.62}} & \makecell{\textcolor{blue}{57.96}\\ \textcolor{blue}{± 0.80}} & \makecell{83.47\\ ± 0.24} & \makecell{74.52\\ ± 0.23}& \makecell{87.57\\ ± 0.26} & \makecell{79.98\\ ± 0.38}\\
$\mathbf{SimpleESKNet_{16}^{1+AFF}(Ours)}$ & - & 0.10  & 1.65 & \makecell{\textcolor{red}{85.48}\\ \textcolor{red}{± 0.35}} & \makecell{\textcolor{red}{75.08}\\ \textcolor{red}{± 0.68}} &  \makecell{\textcolor{red}{82.67}\\ \textcolor{red}{± 0.86}} & \makecell{\textcolor{red}{70.87}\\ \textcolor{red}{± 1.18}} & \makecell{\textcolor{red}{84.49}\\ \textcolor{red}{± 0.24}} & \makecell{\textcolor{red}{75.99}\\ \textcolor{red}{± 0.37}} & \makecell{\textcolor{red}{88.41}\\ \textcolor{red}{± 0.15}} & \makecell{\textcolor{red}{81.58}\\\textcolor{red}{ ± 0.23}}\\
  \midrule
  UNet\cite{ronneberger2015u} & 2015 & 8.64 & 16.46 & \makecell{84.28 \\ ± 0.38} & \makecell{73.18\\ ± 0.53} & \makecell{82.63\\ ± 0.48} & \makecell{70.73\\ ± 0.76} & \makecell{83.18\\ ± 0.36} & \makecell{74.25\\ ± 0.41} & \makecell{87.10\\ ± 0.19} & \makecell{79.42\\ ± 0.20}\\ 
  TransUNet\cite{transunet2021} & 2021 & 93.23 & 32.24  &  \makecell{85.32\\ ± 0.48} & \makecell{74.89\\ ± 0.70} &  \makecell{84.21\\ ± 0.75} & \makecell{73.14 \\ ± 1.11} & \makecell{\textcolor{blue}{85.14}\\ \textcolor{blue}{± 0.37}} & \makecell{\textcolor{blue}{76.58}\\ \textcolor{blue}{± 0.45}} & \makecell{\textcolor{red}{89.02}\\ \textcolor{red}{± 0.17}} & \makecell{\textcolor{red}{81.91}\\ \textcolor{red}{± 0.16}}\\
  MDA-Net\cite{iqbal2022mda} & 2022 & 29.84  & 49.56 & \makecell{84.34\\ ± 0.48} & \makecell{73.33\\± 0.72}&  \makecell{80.14\\ ± 1.65} & \makecell{67.20\\ ± 2.25} & \makecell{83.96\\ ± 0.30} & \makecell{75.15\\ ± 0.36} & \makecell{87.52\\ ± 0.14} & \makecell{80.14\\ ± 0.26}\\
  UNetv2\cite{peng2023u} & 2023 & 25.13 & 5.40 & \makecell{83.15 \\ ± 0.36} & \makecell{71.57\\± 0.55}&  \makecell{78.27\\ ± 1.25} & \makecell{64.56\\ ± 1.72} & \makecell{83.49\\ ± 0.25} & \makecell{74.60\\ ± 0.33} & \makecell{87.15\\ ± 0.44} & \makecell{79.62\\ ± 0.53}\\
  Tiny-UNet\cite{chen2024tinyu} & 2024 & \uline{0.48} & \textbf{1.66} & \makecell{84.35\\± 0.31} & \makecell{73.30\\± 0.42} &  \makecell{77.40\\ ± 1.10} & \makecell{63.51\\ ± 1.43} & \makecell{83.05\\ ± 0.21} & \makecell{73.97\\ ± 0.29} & \makecell{87.35\\ ± 0.23} & \makecell{79.76\\ ± 0.31}\\
  ESKNet\cite{chen2024esknet} & 2024 & 26.71 & 45.29 & \makecell{84.98\\± 0.49} & \makecell{74.40\\± 0.69} &  \makecell{\textcolor{red}{86.60}\\ \textcolor{red}{± 0.83}} & \makecell{\textcolor{red}{76.67}\\ \textcolor{red}{± 1.24}} & \makecell{\textcolor{red}{85.24}\\ \textcolor{red}{± 0.35}} & \makecell{\textcolor{red}{77.06}\\ \textcolor{red}{± 0.39}} & \makecell{88.44\\ ± 0.39} & \makecell{81.86\\ ± 0.50}\\
  $\mathbf{SimpleUNet_{48}^{4+AFF}(Ours)}$ & - & \textbf{0.47}  & \uline{4.35} & \makecell{\textcolor{blue}{85.70}\\ \textcolor{blue}{± 0.64}} & \makecell{\textcolor{blue}{75.47}\\ \textcolor{blue}{± 0.87}} &  \makecell{85.72\\ ± 0.55} & \makecell{75.37\\ ± 0.97} & \makecell{84.34\\ ± 0.18} & \makecell{75.75\\ ± 0.24} & \makecell{87.57\\ ± 0.23} & \makecell{80.36\\ ± 0.30}\\
  $\mathbf{SimpleESKNet_{32}^{2+AFF}(Ours)}$ & - & 0.67  & 8.60 & \makecell{\textcolor{red}{85.76}\\ \textcolor{red}{± 0.49}} & \makecell{\textcolor{red}{75.60}\\ \textcolor{red}{± 0.68}} &  \makecell{\textcolor{blue}{86.46}\\ \textcolor{blue}{± 0.42}} & \makecell{\textcolor{blue}{76.48}\\ \textcolor{blue}{± 0.60}} & \makecell{84.86\\ ± 0.34} & \makecell{76.60\\ ± 0.38} & \makecell{\textcolor{blue}{88.77}\\ \textcolor{blue}{± 0.19}} & \makecell{\textcolor{blue}{82.13}\\ \textcolor{blue}{± 0.23}}\\
\bottomrule
\end{tabular}
\end{table*}
\def\myvaluet{0.065}
\vspace{2pt}
\begin{figure*}[htbp]
\centering
\begin{minipage}{0.04\textwidth}
    \centering
    \rotatebox{90}{\scriptsize{BUSI}}
\end{minipage}
\begin{minipage}{0.92\textwidth} 
    \begin{minipage}{\myvaluet\textwidth}
        \centering
        \textbf{\scriptsize Images}\\[1ex]
        \includegraphics[height=\textwidth, width=\textwidth]{Figures/Best/MergedBreast/Challenging/BUSI11/BUSI_benign_311.png}
    \end{minipage}
    \begin{minipage}{\myvaluet\textwidth}
        \centering
        \textbf{\scriptsize GTs}\\[1ex]
        \includegraphics[height=\textwidth, width=\textwidth]{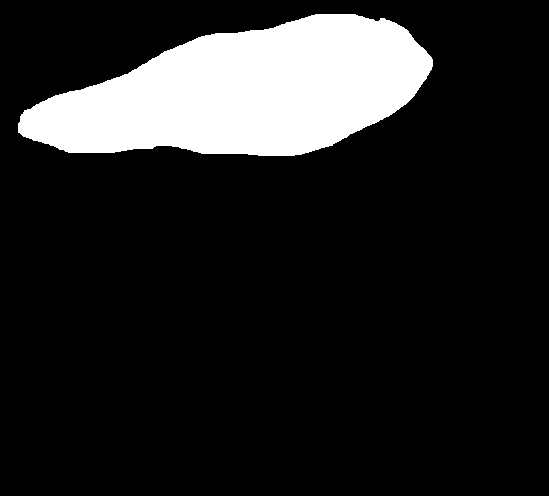}
    \end{minipage}
    \begin{minipage}{\myvaluet\textwidth}
        \centering
        \textbf{\tiny{SimpleESKNet}}\\[1ex]
        \includegraphics[height=\textwidth, width=\textwidth]{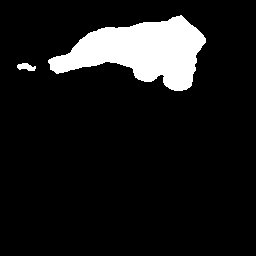}
    \end{minipage}
    \begin{minipage}{\myvaluet\textwidth}
        \centering
        \textbf{\scriptsize MALUNet}\\[1ex]
        \includegraphics[height=\textwidth, width=\textwidth]{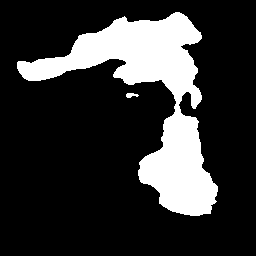}
    \end{minipage}
    \begin{minipage}{\myvaluet\textwidth}
        \centering
        \textbf{\scriptsize UNext$_{s}$}\\[1ex]
        \includegraphics[height=\textwidth, width=\textwidth]{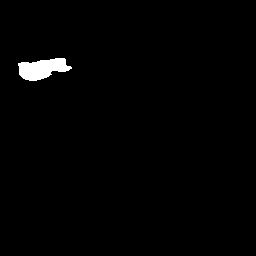}
    \end{minipage}
    \begin{minipage}{\myvaluet\textwidth}
        \centering
        \textbf{\scriptsize LF-UNet}\\[1ex]
        \includegraphics[height=\textwidth, width=\textwidth]{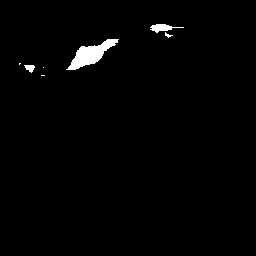}
    \end{minipage}
    \begin{minipage}{\myvaluet\textwidth}
        \centering
        \textbf{\scriptsize LBUNet}\\[1ex]
        \includegraphics[height=\textwidth, width=\textwidth]{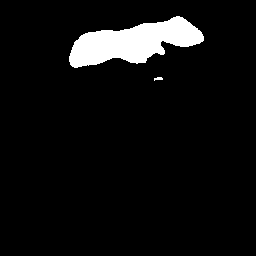}
    \end{minipage}
    \begin{minipage}{\myvaluet\textwidth}
        \centering
        \textbf{\tiny{UltraVMUNet}}\\[1ex]
        \includegraphics[height=\textwidth, width=\textwidth]{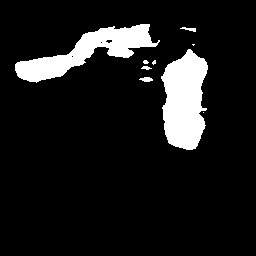}
    \end{minipage}
    \begin{minipage}{\myvaluet\textwidth}
        \centering
        \textbf{\scriptsize UNet}\\[1ex]
        \includegraphics[height=\textwidth, width=\textwidth]{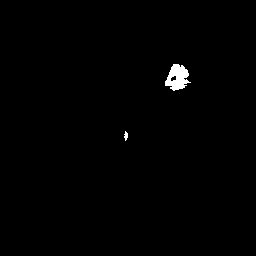}
    \end{minipage}
    \begin{minipage}{\myvaluet\textwidth}
        \centering
        \textbf{\scriptsize TransUNet}\\[1ex]
        \includegraphics[height=\textwidth, width=\textwidth]{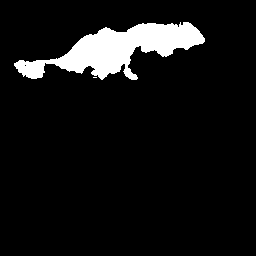}
    \end{minipage}
    \begin{minipage}{\myvaluet\textwidth}
        \centering
        \textbf{\scriptsize MDA-Net}\\[1ex]
        \includegraphics[height=\textwidth, width=\textwidth]{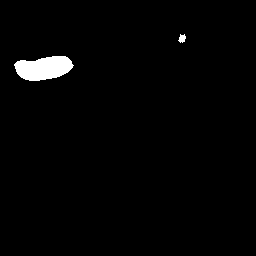}
    \end{minipage}
    \begin{minipage}{\myvaluet\textwidth}
        \centering
        \textbf{\scriptsize UNetv2}\\[1ex]
        \includegraphics[height=\textwidth, width=\textwidth]{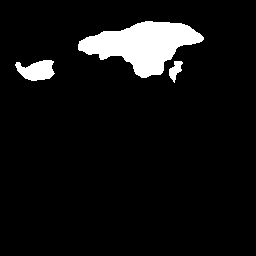}
    \end{minipage}
    \begin{minipage}{\myvaluet\textwidth}
        \centering
        \textbf{\tiny{Tiny-UNet}}\\[1ex]
        \includegraphics[height=\textwidth, width=\textwidth]{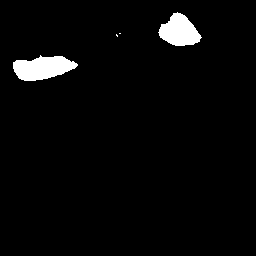}
    \end{minipage}
    \begin{minipage}{\myvaluet\textwidth}
        \centering
        \textbf{\scriptsize ESKNet}\\[1ex]
        \includegraphics[height=\textwidth, width=\textwidth]{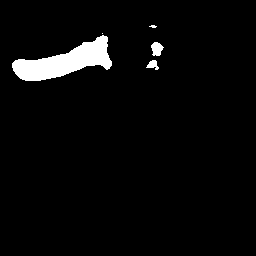}
    \end{minipage}
\end{minipage}
\vspace{2pt}

\begin{minipage}{0.04\textwidth}
    \centering
    \rotatebox{90}{\scriptsize{DatasetB}}
\end{minipage}
\begin{minipage}{0.92\textwidth} 
    \begin{minipage}{\myvaluet\textwidth}
        \centering
        \includegraphics[height=\textwidth, width=\textwidth]{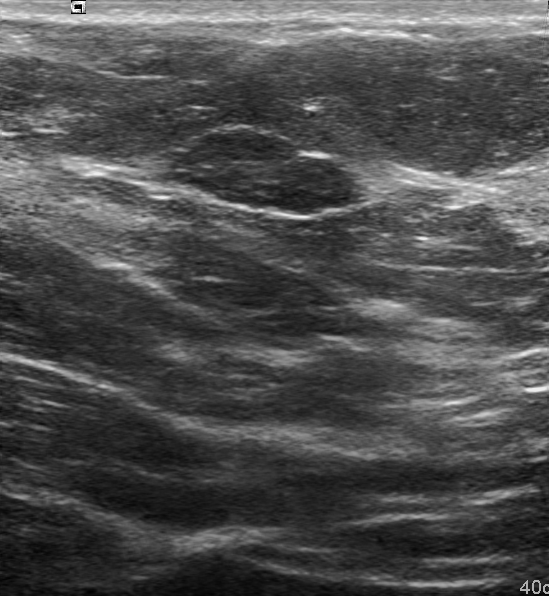}
    \end{minipage}
    \begin{minipage}{\myvaluet\textwidth}
        \centering
        \includegraphics[height=\textwidth, width=\textwidth]{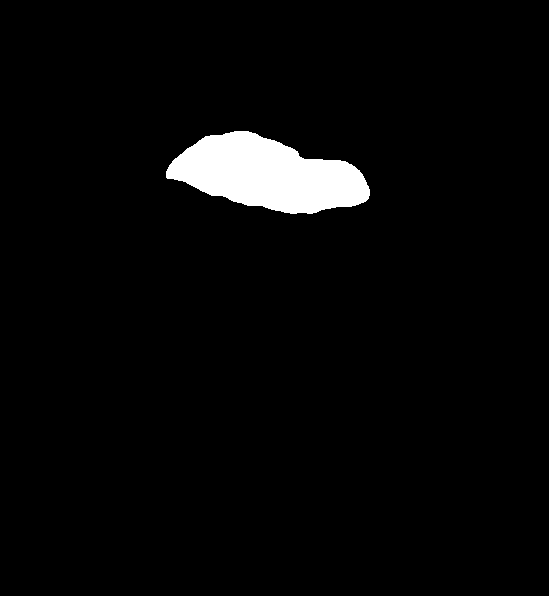}
    \end{minipage}
    \begin{minipage}{\myvaluet\textwidth}
        \centering
        \includegraphics[height=\textwidth, width=\textwidth]{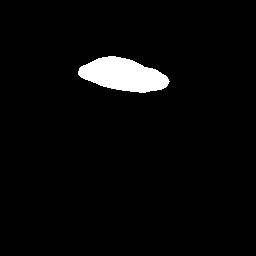}
    \end{minipage}
    \begin{minipage}{\myvaluet\textwidth}
        \centering
        \includegraphics[height=\textwidth, width=\textwidth]{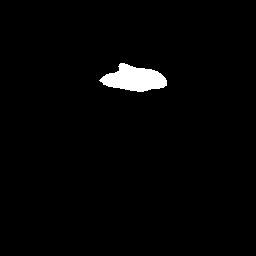}
    \end{minipage}
    \begin{minipage}{\myvaluet\textwidth}
        \centering
        \includegraphics[height=\textwidth, width=\textwidth]{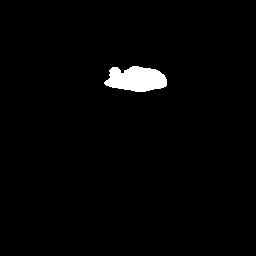}
    \end{minipage}
    \begin{minipage}{\myvaluet\textwidth}
        \centering
        \includegraphics[height=\textwidth, width=\textwidth]{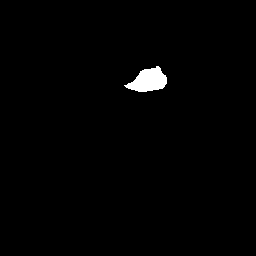}
    \end{minipage}
    \begin{minipage}{\myvaluet\textwidth}
        \centering
        \includegraphics[height=\textwidth, width=\textwidth]{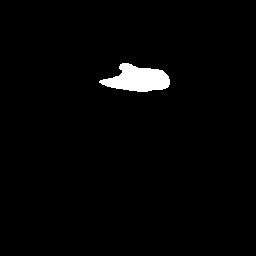}
    \end{minipage}
    \begin{minipage}{\myvaluet\textwidth}
        \centering
        \includegraphics[height=\textwidth, width=\textwidth]{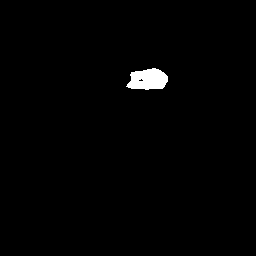}
    \end{minipage}
    \begin{minipage}{\myvaluet\textwidth}
        \centering
        \includegraphics[height=\textwidth, width=\textwidth]{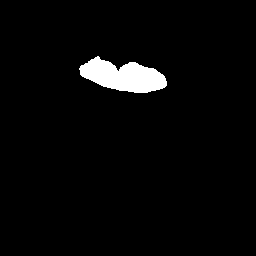}
    \end{minipage}
    \begin{minipage}{\myvaluet\textwidth}
        \centering
        \includegraphics[height=\textwidth, width=\textwidth]{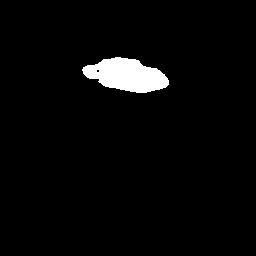}
    \end{minipage}
    \begin{minipage}{\myvaluet\textwidth}
        \centering
        \includegraphics[height=\textwidth, width=\textwidth]{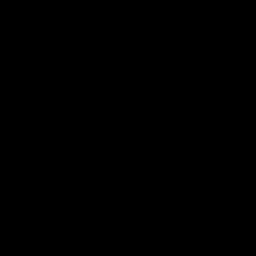}
    \end{minipage}
    \begin{minipage}{\myvaluet\textwidth}
        \centering
        \includegraphics[height=\textwidth, width=\textwidth]{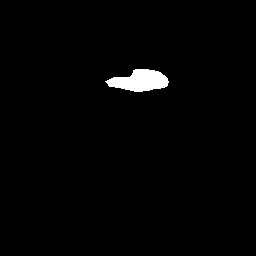}
    \end{minipage}
    \begin{minipage}{\myvaluet\textwidth}
        \centering
        \includegraphics[height=\textwidth, width=\textwidth]{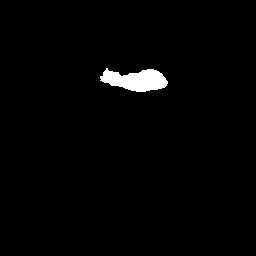}
    \end{minipage}
    \begin{minipage}{\myvaluet\textwidth}
        \centering
        \includegraphics[height=\textwidth, width=\textwidth]{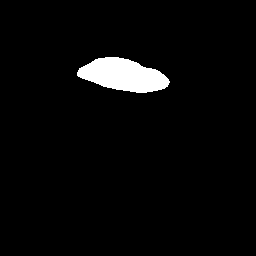}
    \end{minipage}
\end{minipage}
\vspace{2pt}

\begin{minipage}{0.04\textwidth}
    \centering
    \rotatebox{90}{\scriptsize{BrEaST}}
\end{minipage}
\begin{minipage}{0.92\textwidth} 
    \begin{minipage}{\myvaluet\textwidth}
        \centering
        \includegraphics[height=\textwidth, width=\textwidth]{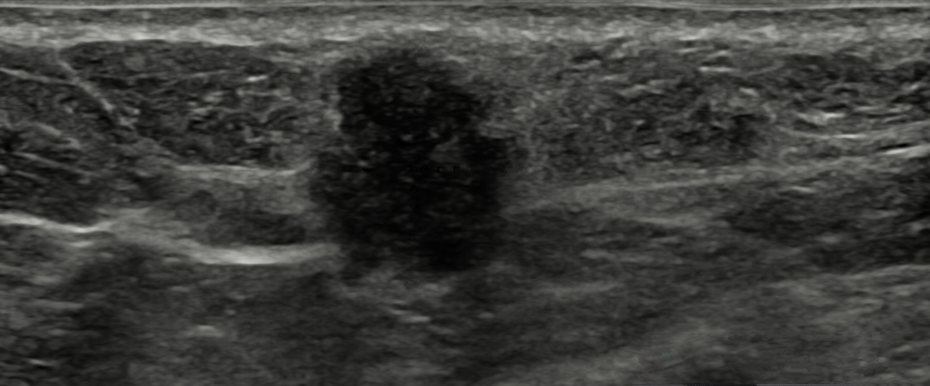}
    \end{minipage}
    \begin{minipage}{\myvaluet\textwidth}
        \centering
        \includegraphics[height=\textwidth, width=\textwidth]{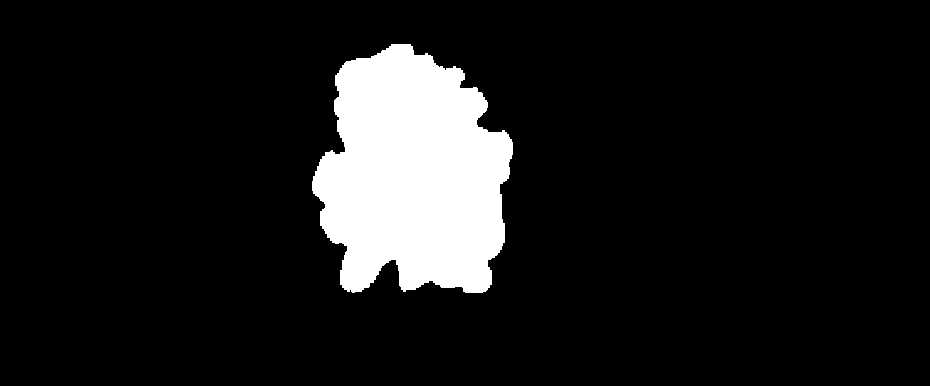}
    \end{minipage}
    \begin{minipage}{\myvaluet\textwidth}
        \centering
        \includegraphics[height=\textwidth, width=\textwidth]{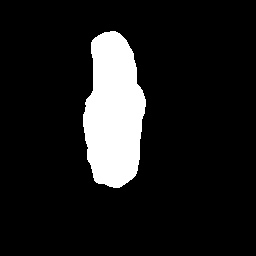}
    \end{minipage}
    \begin{minipage}{\myvaluet\textwidth}
        \centering
        \includegraphics[height=\textwidth, width=\textwidth]{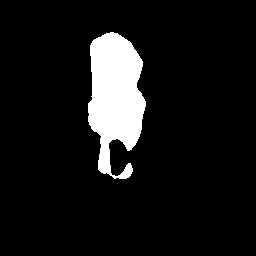}
    \end{minipage}
    \begin{minipage}{\myvaluet\textwidth}
        \centering
        \includegraphics[height=\textwidth, width=\textwidth]{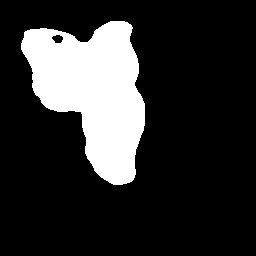}
    \end{minipage}
    \begin{minipage}{\myvaluet\textwidth}
        \centering
        \includegraphics[height=\textwidth, width=\textwidth]{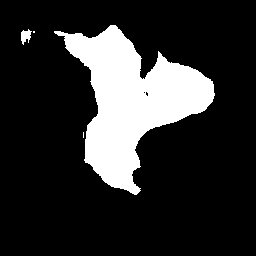}
    \end{minipage}
    \begin{minipage}{\myvaluet\textwidth}
        \centering
        \includegraphics[height=\textwidth, width=\textwidth]{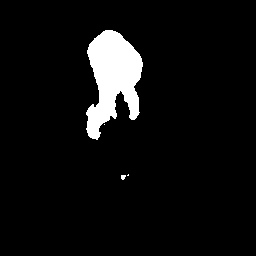}
    \end{minipage}
    \begin{minipage}{\myvaluet\textwidth}
        \centering
        \includegraphics[height=\textwidth, width=\textwidth]{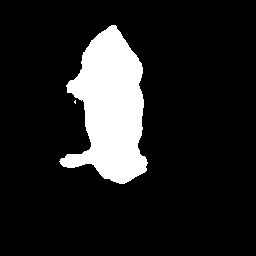}
    \end{minipage}
    \begin{minipage}{\myvaluet\textwidth}
        \centering
        \includegraphics[height=\textwidth, width=\textwidth]{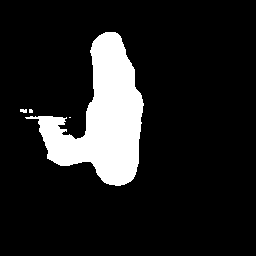}
    \end{minipage}
    \begin{minipage}{\myvaluet\textwidth}
        \centering
        \includegraphics[height=\textwidth, width=\textwidth]{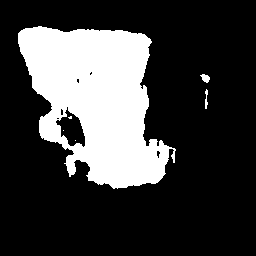}
    \end{minipage}
    \begin{minipage}{\myvaluet\textwidth}
        \centering
        \includegraphics[height=\textwidth, width=\textwidth]{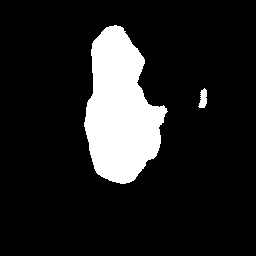}
    \end{minipage}
    \begin{minipage}{\myvaluet\textwidth}
        \centering
        \includegraphics[height=\textwidth, width=\textwidth]{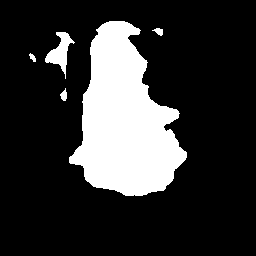}
    \end{minipage}
    \begin{minipage}{\myvaluet\textwidth}
        \centering
        \includegraphics[height=\textwidth, width=\textwidth]{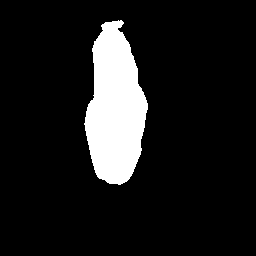}
    \end{minipage}
    \begin{minipage}{\myvaluet\textwidth}
        \centering
        \includegraphics[height=\textwidth, width=\textwidth]{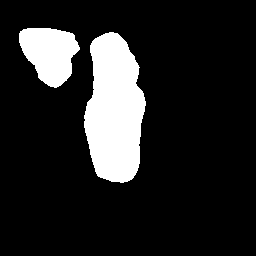}
    \end{minipage}
\end{minipage}
\vspace{2pt}

\begin{minipage}{0.04\textwidth}
    \centering
    \rotatebox{90}{\scriptsize{BUSBRA}}
\end{minipage}
\begin{minipage}{0.92\textwidth} 
    \begin{minipage}{\myvaluet\textwidth}
        \centering
        \includegraphics[height=\textwidth, width=\textwidth]{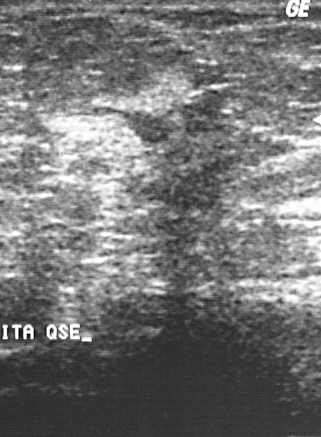}
    \end{minipage}
    \begin{minipage}{\myvaluet\textwidth}
        \centering
        \includegraphics[height=\textwidth, width=\textwidth]{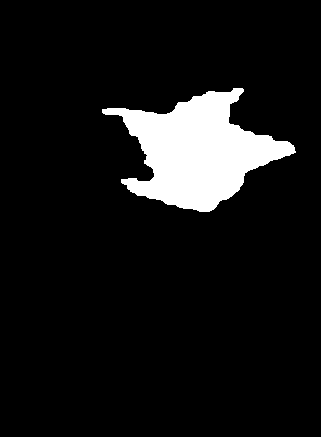}
    \end{minipage}
    \begin{minipage}{\myvaluet\textwidth}
        \centering
        \includegraphics[height=\textwidth, width=\textwidth]{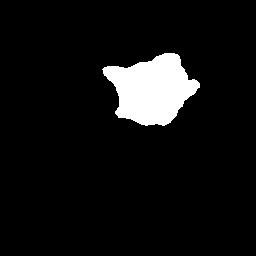}
    \end{minipage}
    \begin{minipage}{\myvaluet\textwidth}
        \centering
        \includegraphics[height=\textwidth, width=\textwidth]{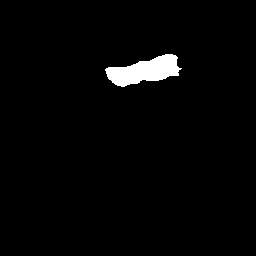}
    \end{minipage}
    \begin{minipage}{\myvaluet\textwidth}
        \centering
        \includegraphics[height=\textwidth, width=\textwidth]{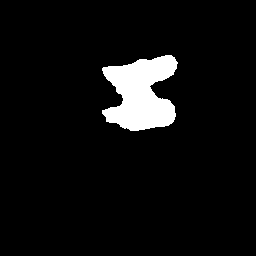}
    \end{minipage}
    \begin{minipage}{\myvaluet\textwidth}
        \centering
        \includegraphics[height=\textwidth, width=\textwidth]{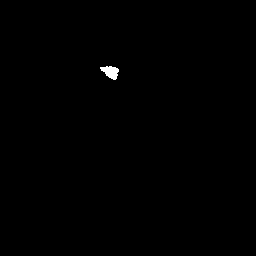}
    \end{minipage}
    \begin{minipage}{\myvaluet\textwidth}
        \centering
        \includegraphics[height=\textwidth, width=\textwidth]{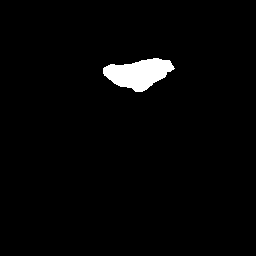}
    \end{minipage}
    \begin{minipage}{\myvaluet\textwidth}
        \centering
        \includegraphics[height=\textwidth, width=\textwidth]{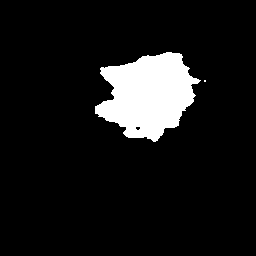}
    \end{minipage}
    \begin{minipage}{\myvaluet\textwidth}
        \centering
        \includegraphics[height=\textwidth, width=\textwidth]{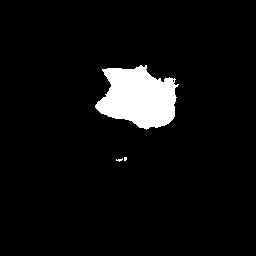}
    \end{minipage}
    \begin{minipage}{\myvaluet\textwidth}
        \centering
        \includegraphics[height=\textwidth, width=\textwidth]{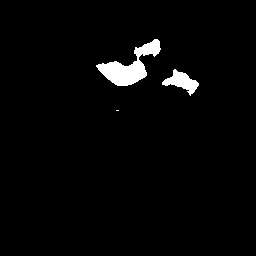}
    \end{minipage}
    \begin{minipage}{\myvaluet\textwidth}
        \centering
        \includegraphics[height=\textwidth, width=\textwidth]{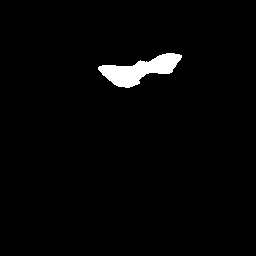}
    \end{minipage}
    \begin{minipage}{\myvaluet\textwidth}
        \centering
        \includegraphics[height=\textwidth, width=\textwidth]{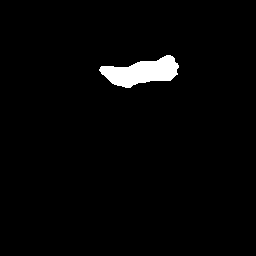}
    \end{minipage}
    \begin{minipage}{\myvaluet\textwidth}
        \centering
        \includegraphics[height=\textwidth, width=\textwidth]{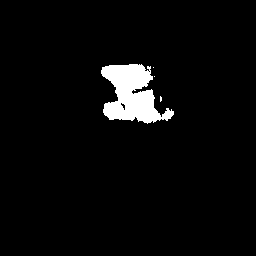}
    \end{minipage}
    \begin{minipage}{\myvaluet\textwidth}
        \centering
        \includegraphics[height=\textwidth, width=\textwidth]{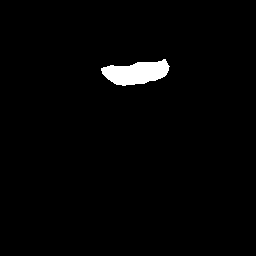}
    \end{minipage}
\end{minipage}
\caption{The qualitative comparison between our model and other models. Images from left to right are input images, corresponding labels, and segmentation results given by the segmentation models. As seen, our model provided the closest prediction results to the ground truths, while other models tend to wrongly or overly segment the lesions.}
\label{fig:ins_cmp_sota}
\end{figure*}

\begin{figure*}[htbp]
\centering
\begin{minipage}{0.04\textwidth}
    \centering
    \rotatebox{90}{\scriptsize{BUSI}}
\end{minipage}
\begin{minipage}{0.92\textwidth}
    \begin{minipage}{\myvaluet\textwidth}
        \centering
        \textbf{\scriptsize Images}\\[1ex]
        \includegraphics[height=\textwidth, width=\textwidth]{Figures/Best/MergedBreast/SmallLesion/BUSI0/BUSI_benign_246.png}
    \end{minipage}
    \begin{minipage}{\myvaluet\textwidth}
        \centering
        \textbf{\scriptsize GTs}\\[1ex]
        \includegraphics[height=\textwidth, width=\textwidth]{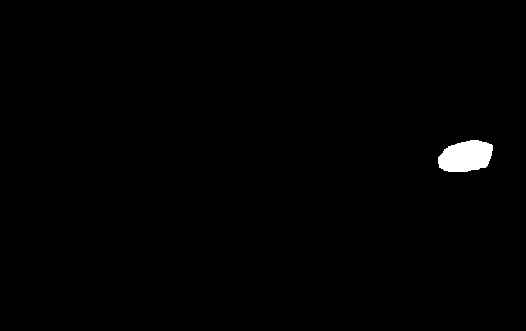}
    \end{minipage}
    \begin{minipage}{\myvaluet\textwidth}
        \centering
        \textbf{\tiny{SimpleESKNet}}\\[1ex]
        \includegraphics[height=\textwidth, width=\textwidth]{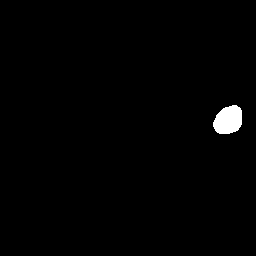}
    \end{minipage}
    \begin{minipage}{\myvaluet\textwidth}
        \centering
        \textbf{\scriptsize MALUNet}\\[1ex]
        \includegraphics[height=\textwidth, width=\textwidth]{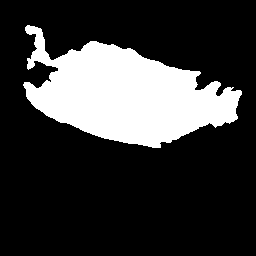}
    \end{minipage}
    \begin{minipage}{\myvaluet\textwidth}
        \centering
        \textbf{\scriptsize UNext$_{s}$}\\[1ex]
        \includegraphics[height=\textwidth, width=\textwidth]{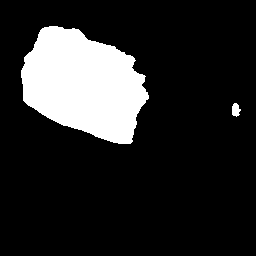}
    \end{minipage}
    \begin{minipage}{\myvaluet\textwidth}
        \centering
        \textbf{\scriptsize LF-UNet}\\[1ex]
        \includegraphics[height=\textwidth, width=\textwidth]{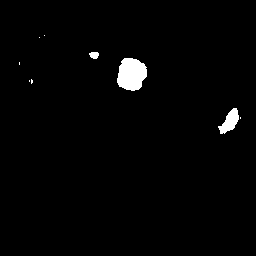}
    \end{minipage}
    \begin{minipage}{\myvaluet\textwidth}
        \centering
        \textbf{\scriptsize LBUNet}\\[1ex]
        \includegraphics[height=\textwidth, width=\textwidth]{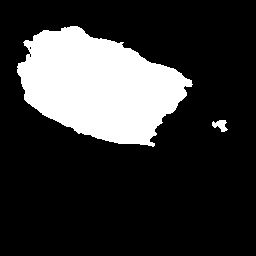}
    \end{minipage}
    \begin{minipage}{\myvaluet\textwidth}
        \centering
        \textbf{\tiny{UltraVMUNet}}\\[1ex]
        \includegraphics[height=\textwidth, width=\textwidth]{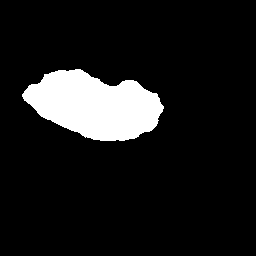}
    \end{minipage}
    \begin{minipage}{\myvaluet\textwidth}
        \centering
        \textbf{\scriptsize UNet}\\[1ex]
        \includegraphics[height=\textwidth, width=\textwidth]{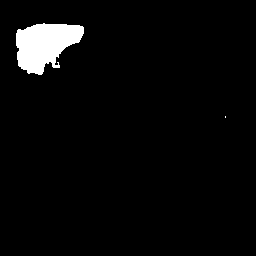}
    \end{minipage}
    \begin{minipage}{\myvaluet\textwidth}
        \centering
        \textbf{\scriptsize TransUNet}\\[1ex]
        \includegraphics[height=\textwidth, width=\textwidth]{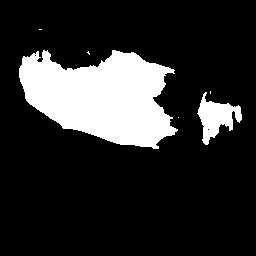}
    \end{minipage}
    \begin{minipage}{\myvaluet\textwidth}
        \centering
        \textbf{\scriptsize MDA-Net}\\[1ex]
        \includegraphics[height=\textwidth, width=\textwidth]{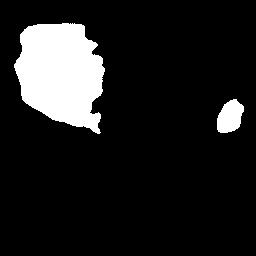}
    \end{minipage}
    \begin{minipage}{\myvaluet\textwidth}
        \centering
        \textbf{\scriptsize UNetv2}\\[1ex]
        \includegraphics[height=\textwidth, width=\textwidth]{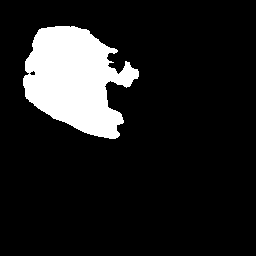}
    \end{minipage}
    \begin{minipage}{\myvaluet\textwidth}
        \centering
        \textbf{\tiny{Tiny-UNet}}\\[1ex]
        \includegraphics[height=\textwidth, width=\textwidth]{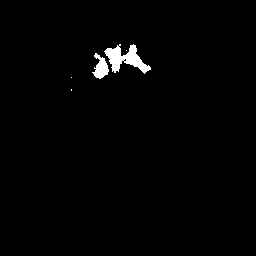}
    \end{minipage}
    \begin{minipage}{\myvaluet\textwidth}
        \centering
        \textbf{\scriptsize ESKNet}\\[1ex]
        \includegraphics[height=\textwidth, width=\textwidth]{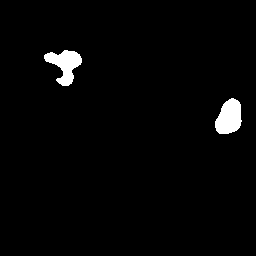}
    \end{minipage}
\end{minipage}

\vspace{2pt}
\begin{minipage}{0.04\textwidth}
    \centering
    \rotatebox{90}{\scriptsize{DatasetB}}
\end{minipage}
\begin{minipage}{0.92\textwidth}
    \begin{minipage}{\myvaluet\textwidth}
        \centering
        \includegraphics[height=\textwidth, width=\textwidth]{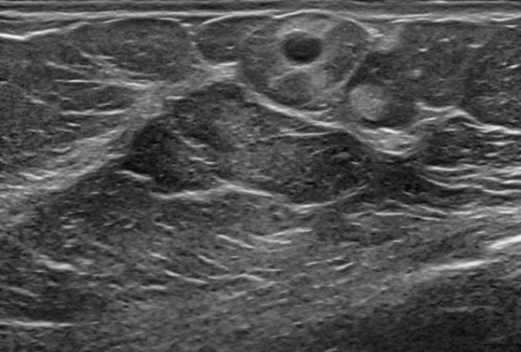}
    \end{minipage}
    \begin{minipage}{\myvaluet\textwidth}
        \centering
        \includegraphics[height=\textwidth, width=\textwidth]{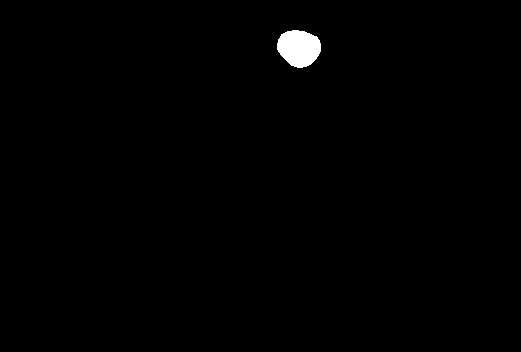}
    \end{minipage}
    \begin{minipage}{\myvaluet\textwidth}
        \centering
        \includegraphics[height=\textwidth, width=\textwidth]{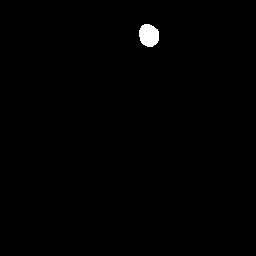}
    \end{minipage}
    \begin{minipage}{\myvaluet\textwidth}
        \centering
        \includegraphics[height=\textwidth, width=\textwidth]{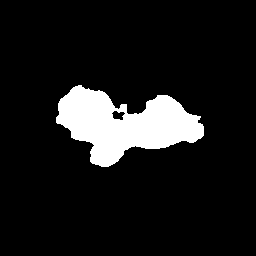}
    \end{minipage}
    \begin{minipage}{\myvaluet\textwidth}
        \centering
        \includegraphics[height=\textwidth, width=\textwidth]{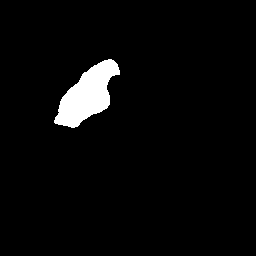}
    \end{minipage}
    \begin{minipage}{\myvaluet\textwidth}
        \centering
        \includegraphics[height=\textwidth, width=\textwidth]{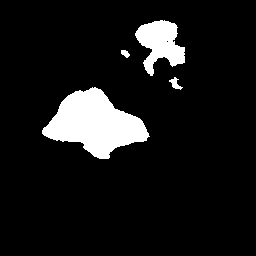}
    \end{minipage}
    \begin{minipage}{\myvaluet\textwidth}
        \centering
        \includegraphics[height=\textwidth, width=\textwidth]{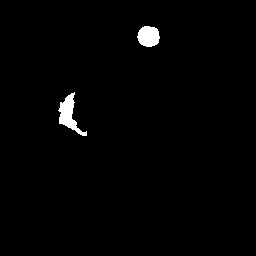}
    \end{minipage}
    \begin{minipage}{\myvaluet\textwidth}
        \centering
        \includegraphics[height=\textwidth, width=\textwidth]{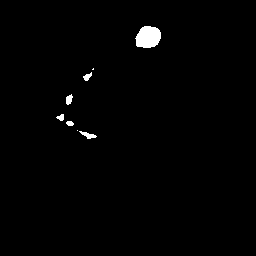}
    \end{minipage}
    \begin{minipage}{\myvaluet\textwidth}
        \centering
        \includegraphics[height=\textwidth, width=\textwidth]{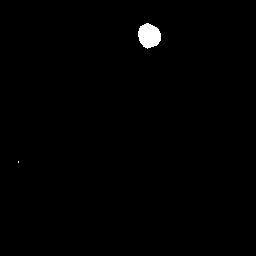}
    \end{minipage}
    \begin{minipage}{\myvaluet\textwidth}
        \centering
        \includegraphics[height=\textwidth, width=\textwidth]{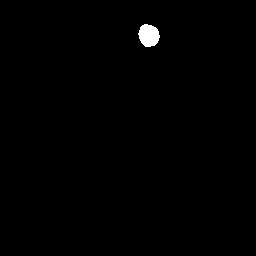}
    \end{minipage}
    \begin{minipage}{\myvaluet\textwidth}
        \centering
        \includegraphics[height=\textwidth, width=\textwidth]{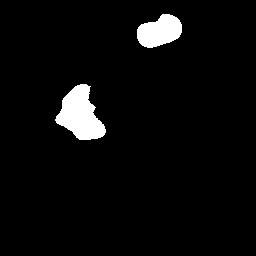}
    \end{minipage}
    \begin{minipage}{\myvaluet\textwidth}
        \centering
        \includegraphics[height=\textwidth, width=\textwidth]{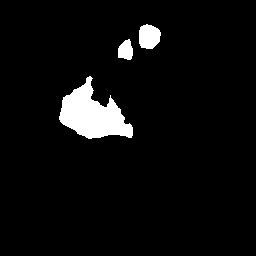}
    \end{minipage}
    \begin{minipage}{\myvaluet\textwidth}
        \centering
        \includegraphics[height=\textwidth, width=\textwidth]{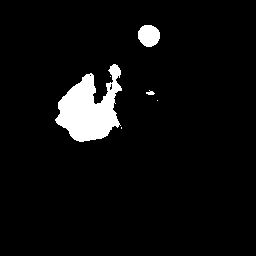}
    \end{minipage}
    \begin{minipage}{\myvaluet\textwidth}
        \centering
        \includegraphics[height=\textwidth, width=\textwidth]{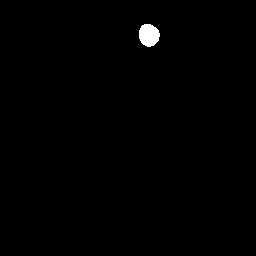}
    \end{minipage}
\end{minipage}

\vspace{2pt}
\begin{minipage}{0.04\textwidth}
    \centering
    \rotatebox{90}{\scriptsize{BrEaST}}
\end{minipage}
\begin{minipage}{0.92\textwidth}
    \begin{minipage}{\myvaluet\textwidth}
        \centering
        \includegraphics[height=\textwidth, width=\textwidth]{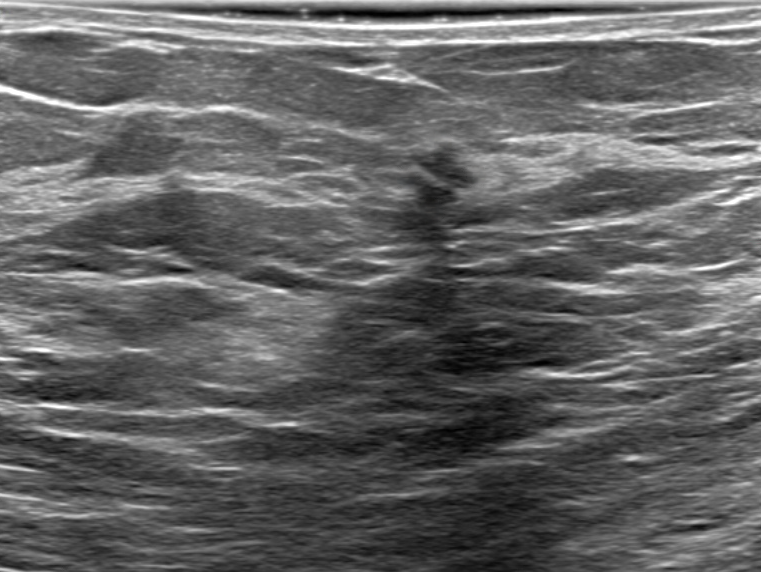}
    \end{minipage}
    \begin{minipage}{\myvaluet\textwidth}
        \centering
        \includegraphics[height=\textwidth, width=\textwidth]{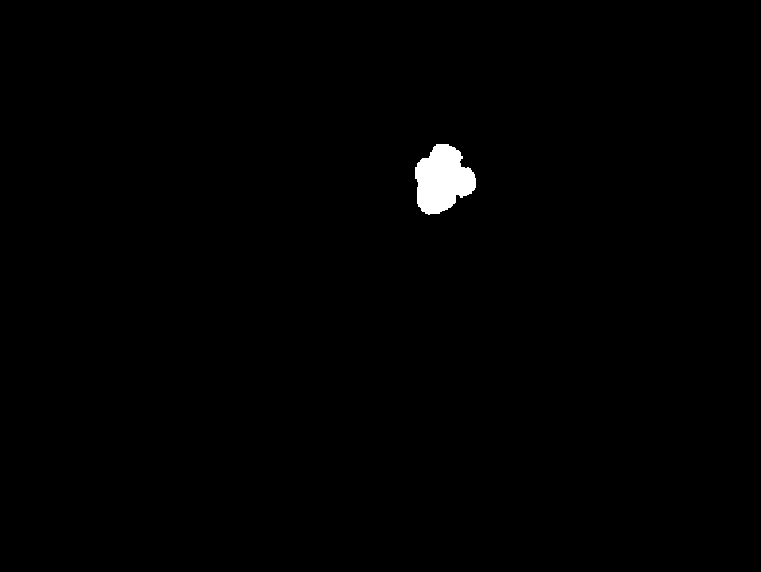}
    \end{minipage}
    \begin{minipage}{\myvaluet\textwidth}
        \centering
        \includegraphics[height=\textwidth, width=\textwidth]{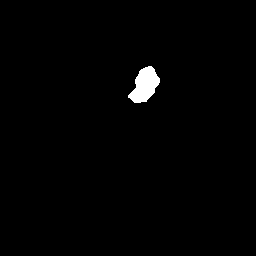}
    \end{minipage}
    \begin{minipage}{\myvaluet\textwidth}
        \centering
        \includegraphics[height=\textwidth, width=\textwidth]{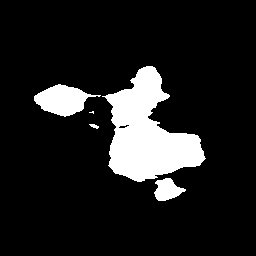}
    \end{minipage}
    \begin{minipage}{\myvaluet\textwidth}
        \centering
        \includegraphics[height=\textwidth, width=\textwidth]{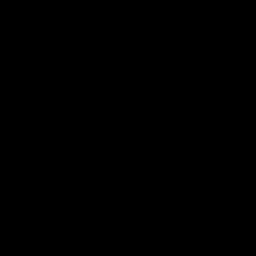}
    \end{minipage}
    \begin{minipage}{\myvaluet\textwidth}
        \centering
        \includegraphics[height=\textwidth, width=\textwidth]{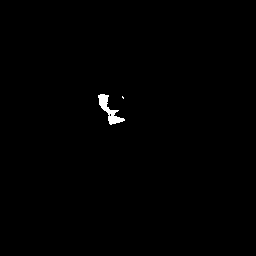}
    \end{minipage}
    \begin{minipage}{\myvaluet\textwidth}
        \centering
        \includegraphics[height=\textwidth, width=\textwidth]{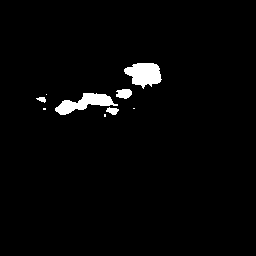}
    \end{minipage}
    \begin{minipage}{\myvaluet\textwidth}
        \centering
        \includegraphics[height=\textwidth, width=\textwidth]{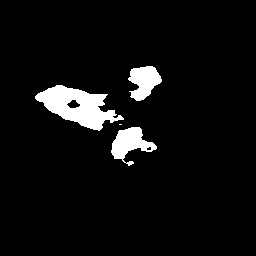}
    \end{minipage}
    \begin{minipage}{\myvaluet\textwidth}
        \centering
        \includegraphics[height=\textwidth, width=\textwidth]{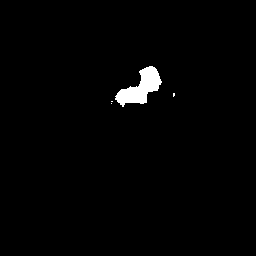}
    \end{minipage}
    \begin{minipage}{\myvaluet\textwidth}
        \centering
        \includegraphics[height=\textwidth, width=\textwidth]{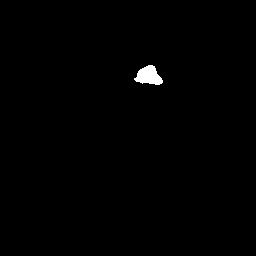}
    \end{minipage}
    \begin{minipage}{\myvaluet\textwidth}
        \centering
        \includegraphics[height=\textwidth, width=\textwidth]{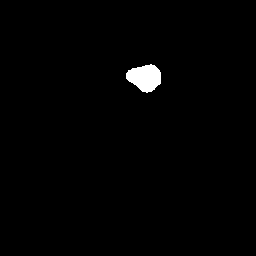}
    \end{minipage}
    \begin{minipage}{\myvaluet\textwidth}
        \centering
        \includegraphics[height=\textwidth, width=\textwidth]{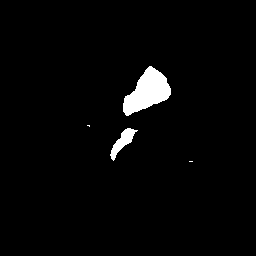}
    \end{minipage}
    \begin{minipage}{\myvaluet\textwidth}
        \centering
        \includegraphics[height=\textwidth, width=\textwidth]{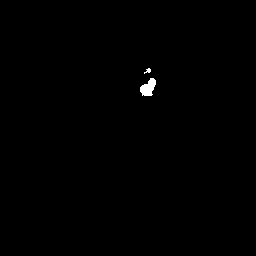}
    \end{minipage}
    \begin{minipage}{\myvaluet\textwidth}
        \centering
        \includegraphics[height=\textwidth, width=\textwidth]{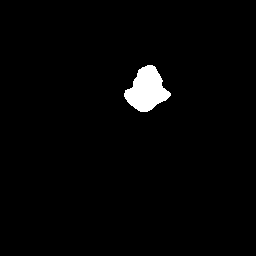}
    \end{minipage}
\end{minipage}

\vspace{2pt}
\begin{minipage}{0.04\textwidth}
    \centering
    \rotatebox{90}{\scriptsize{BUSBRA}}
\end{minipage}
\begin{minipage}{0.92\textwidth}
    \begin{minipage}{\myvaluet\textwidth}
        \centering
        \includegraphics[height=\textwidth, width=\textwidth]{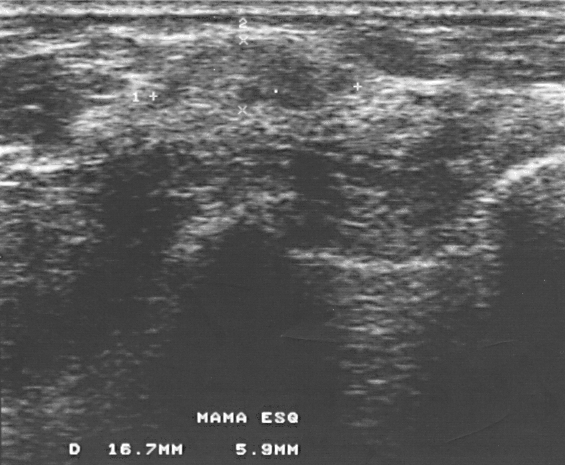}
    \end{minipage}
    \begin{minipage}{\myvaluet\textwidth}
        \centering
        \includegraphics[height=\textwidth, width=\textwidth]{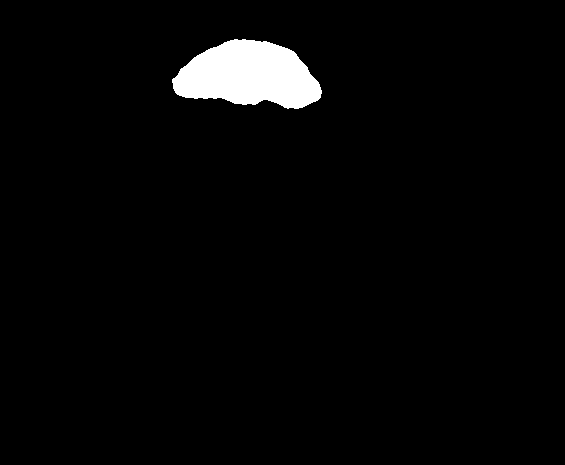}
    \end{minipage}
    \begin{minipage}{\myvaluet\textwidth}
        \centering
        \includegraphics[height=\textwidth, width=\textwidth]{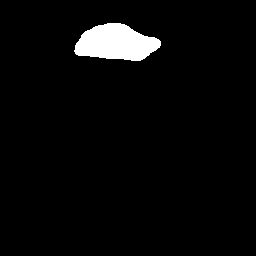}
    \end{minipage}
    \begin{minipage}{\myvaluet\textwidth}
        \centering
        \includegraphics[height=\textwidth, width=\textwidth]{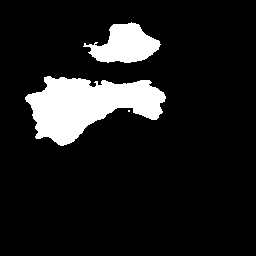}
    \end{minipage}
    \begin{minipage}{\myvaluet\textwidth}
        \centering
        \includegraphics[height=\textwidth, width=\textwidth]{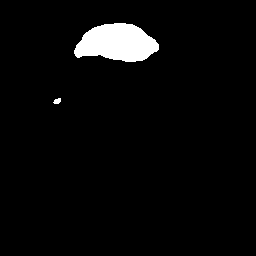}
    \end{minipage}
    \begin{minipage}{\myvaluet\textwidth}
        \centering
        \includegraphics[height=\textwidth, width=\textwidth]{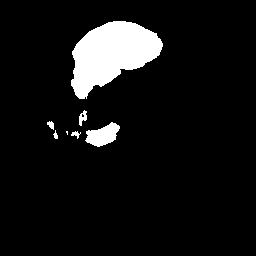}
    \end{minipage}
    \begin{minipage}{\myvaluet\textwidth}
        \centering
        \includegraphics[height=\textwidth, width=\textwidth]{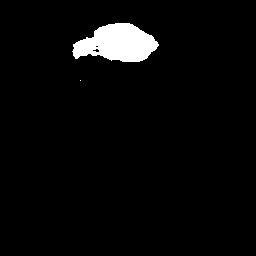}
    \end{minipage}
    \begin{minipage}{\myvaluet\textwidth}
        \centering
        \includegraphics[height=\textwidth, width=\textwidth]{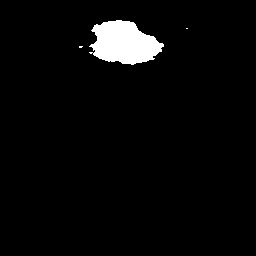}
    \end{minipage}
    \begin{minipage}{\myvaluet\textwidth}
        \centering
        \includegraphics[height=\textwidth, width=\textwidth]{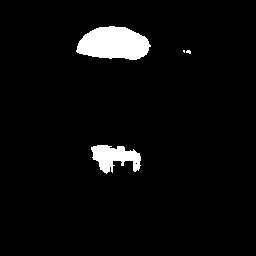}
    \end{minipage}
    \begin{minipage}{\myvaluet\textwidth}
        \centering
        \includegraphics[height=\textwidth, width=\textwidth]{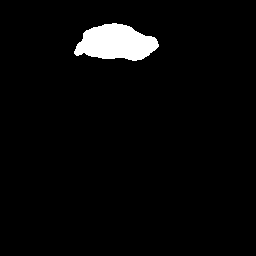}
    \end{minipage}
    \begin{minipage}{\myvaluet\textwidth}
        \centering
        \includegraphics[height=\textwidth, width=\textwidth]{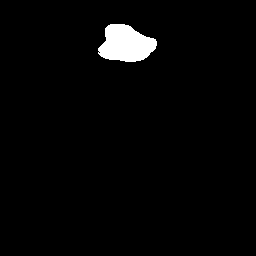}
    \end{minipage}
    \begin{minipage}{\myvaluet\textwidth}
        \centering
        \includegraphics[height=\textwidth, width=\textwidth]{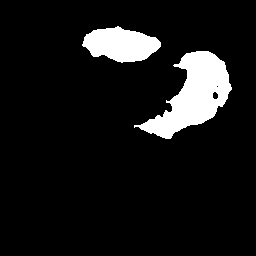}
    \end{minipage}
    \begin{minipage}{\myvaluet\textwidth}
        \centering
        \includegraphics[height=\textwidth, width=\textwidth]{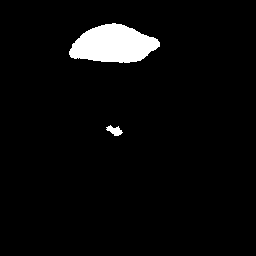}
    \end{minipage}
    \begin{minipage}{\myvaluet\textwidth}
        \centering
        \includegraphics[height=\textwidth, width=\textwidth]{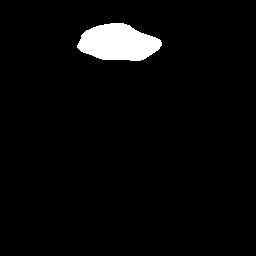}
    \end{minipage}
\end{minipage}
\caption{The qualitative comparison between our model and other models regarding \textcolor{red}{Small Breast Lesions}. Throughout the varied locations of the small-sized lesions, our model consistently shows accurate segmentation results while other models suffer from over-segmentation, indicating the superiority of our model on small lesion segmentation.}
\label{fig:ins_sml_cmp_sota}
\end{figure*}

\begin{figure*}[htbp]
\centering
\begin{minipage}{0.04\textwidth}
    \centering
    \rotatebox{90}{\scriptsize{BUSI}}
\end{minipage}
\begin{minipage}{0.92\textwidth}
    \begin{minipage}{\myvaluet\textwidth}
        \centering
        \textbf{\scriptsize Images}\\[1ex]
        \includegraphics[height=\textwidth, width=\textwidth]{Figures/Best/MergedBreast/MediumLesion/BUSI3/BUSI_benign_418.png}
    \end{minipage}
    \begin{minipage}{\myvaluet\textwidth}
        \centering
        \textbf{\scriptsize GTs}\\[1ex]
        \includegraphics[height=\textwidth, width=\textwidth]{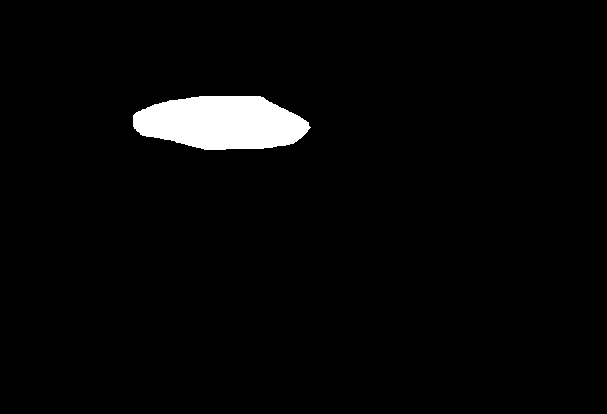}
    \end{minipage}
    \begin{minipage}{\myvaluet\textwidth}
        \centering
        \textbf{\tiny{SimpleESKNet}}\\[1ex]
        \includegraphics[height=\textwidth, width=\textwidth]{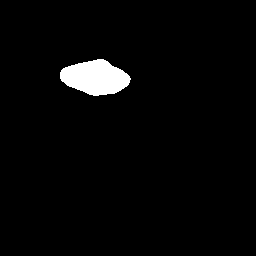}
    \end{minipage}
    \begin{minipage}{\myvaluet\textwidth}
        \centering
        \textbf{\scriptsize MALUNet}\\[1ex]
        \includegraphics[height=\textwidth, width=\textwidth]{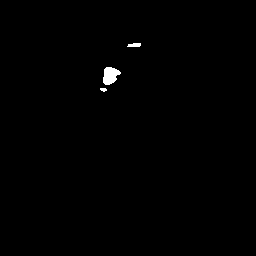}
    \end{minipage}
    \begin{minipage}{\myvaluet\textwidth}
        \centering
        \textbf{\scriptsize UNext$_{s}$}\\[1ex]
        \includegraphics[height=\textwidth, width=\textwidth]{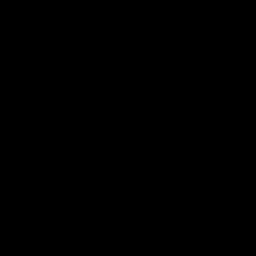}
    \end{minipage}
    \begin{minipage}{\myvaluet\textwidth}
        \centering
        \textbf{\scriptsize LF-UNet}\\[1ex]
        \includegraphics[height=\textwidth, width=\textwidth]{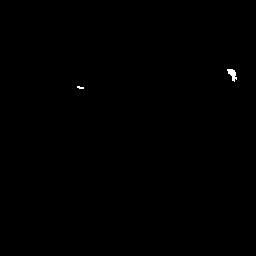}
    \end{minipage}
    \begin{minipage}{\myvaluet\textwidth}
        \centering
        \textbf{\scriptsize LBUNet}\\[1ex]
        \includegraphics[height=\textwidth, width=\textwidth]{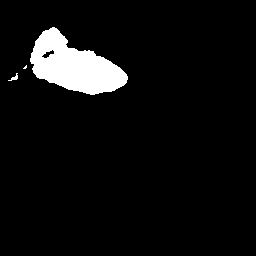}
    \end{minipage}
    \begin{minipage}{\myvaluet\textwidth}
        \centering
        \textbf{\tiny{UltraVMUNet}}\\[1ex]
        \includegraphics[height=\textwidth, width=\textwidth]{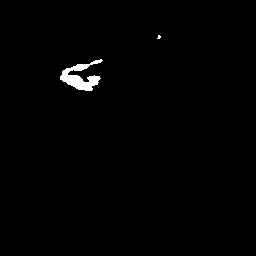}
    \end{minipage}
    \begin{minipage}{\myvaluet\textwidth}
        \centering
        \textbf{\scriptsize UNet}\\[1ex]
        \includegraphics[height=\textwidth, width=\textwidth]{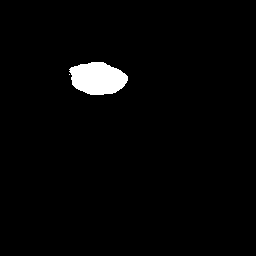}
    \end{minipage}
    \begin{minipage}{\myvaluet\textwidth}
        \centering
        \textbf{\scriptsize TransUNet}\\[1ex]
        \includegraphics[height=\textwidth, width=\textwidth]{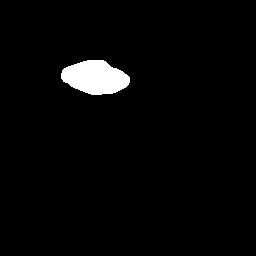}
    \end{minipage}
    \begin{minipage}{\myvaluet\textwidth}
        \centering
        \textbf{\scriptsize MDA-Net}\\[1ex]
        \includegraphics[height=\textwidth, width=\textwidth]{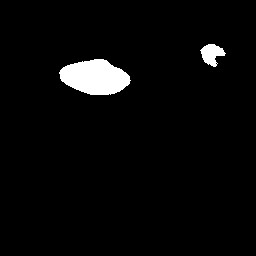}
    \end{minipage}
    \begin{minipage}{\myvaluet\textwidth}
        \centering
        \textbf{\scriptsize UNetv2}\\[1ex]
        \includegraphics[height=\textwidth, width=\textwidth]{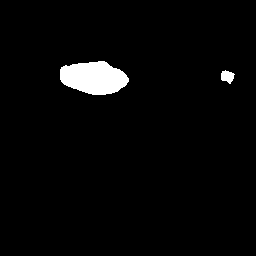}
    \end{minipage}
    \begin{minipage}{\myvaluet\textwidth}
        \centering
        \textbf{\tiny{Tiny-UNet}}\\[1ex]
        \includegraphics[height=\textwidth, width=\textwidth]{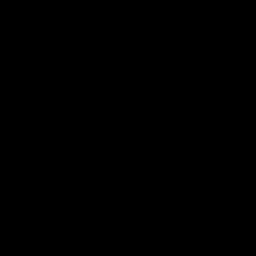}
    \end{minipage}
    \begin{minipage}{\myvaluet\textwidth}
        \centering
        \textbf{\scriptsize ESKNet}\\[1ex]
        \includegraphics[height=\textwidth, width=\textwidth]{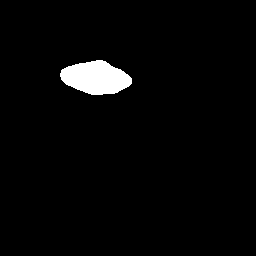}
    \end{minipage}
\end{minipage}

\vspace{2pt}
\begin{minipage}{0.04\textwidth}
    \centering
    \rotatebox{90}{\scriptsize{DatasetB}}
\end{minipage}
\begin{minipage}{0.92\textwidth}
    \begin{minipage}{\myvaluet\textwidth}
        \centering
        \includegraphics[height=\textwidth, width=\textwidth]{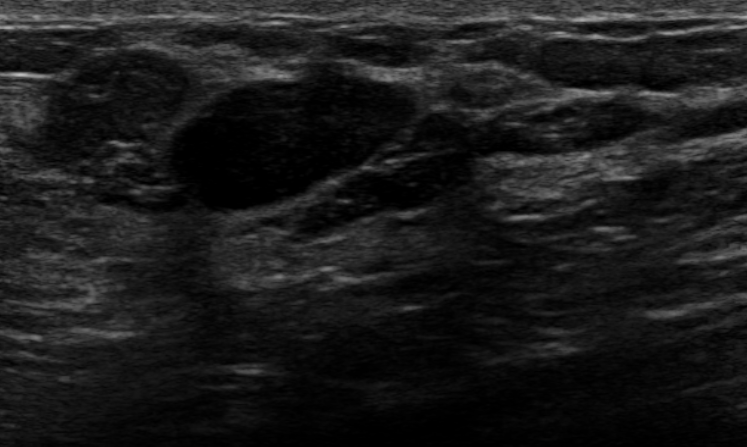}
    \end{minipage}
    \begin{minipage}{\myvaluet\textwidth}
        \centering
        \includegraphics[height=\textwidth, width=\textwidth]{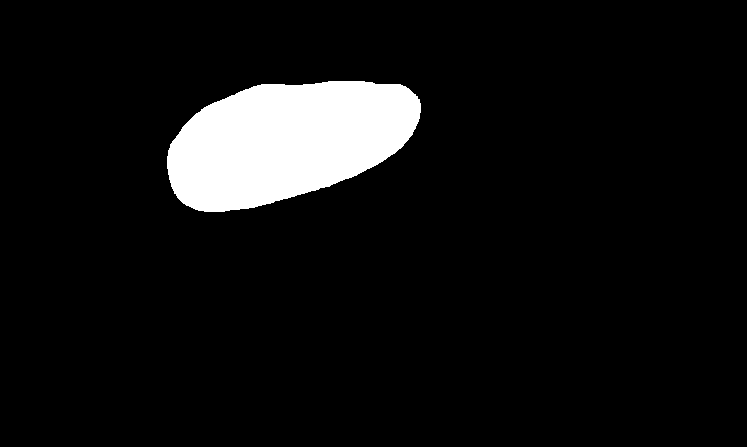}
    \end{minipage}
    \begin{minipage}{\myvaluet\textwidth}
        \centering
        \includegraphics[height=\textwidth, width=\textwidth]{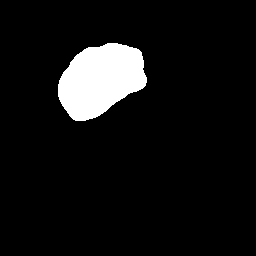}
    \end{minipage}
    \begin{minipage}{\myvaluet\textwidth}
        \centering
        \includegraphics[height=\textwidth, width=\textwidth]{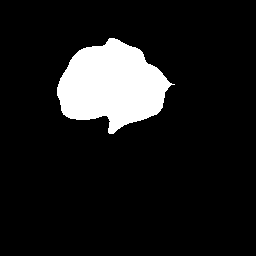}
    \end{minipage}
    \begin{minipage}{\myvaluet\textwidth}
        \centering
        \includegraphics[height=\textwidth, width=\textwidth]{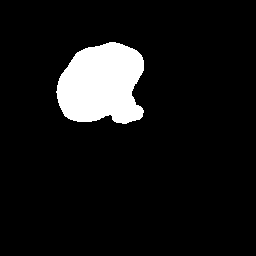}
    \end{minipage}
    \begin{minipage}{\myvaluet\textwidth}
        \centering
        \includegraphics[height=\textwidth, width=\textwidth]{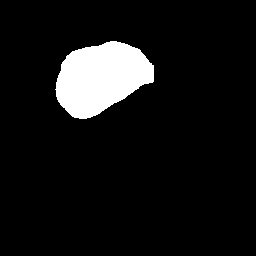}
    \end{minipage}
    \begin{minipage}{\myvaluet\textwidth}
        \centering
        \includegraphics[height=\textwidth, width=\textwidth]{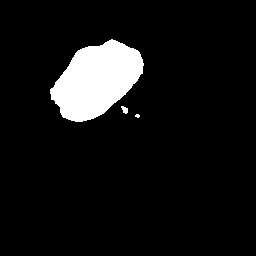}
    \end{minipage}
    \begin{minipage}{\myvaluet\textwidth}
        \centering
        \includegraphics[height=\textwidth, width=\textwidth]{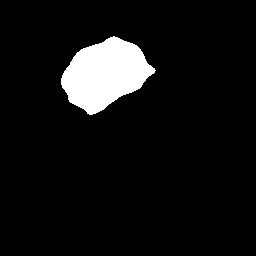}
    \end{minipage}
    \begin{minipage}{\myvaluet\textwidth}
        \centering
        \includegraphics[height=\textwidth, width=\textwidth]{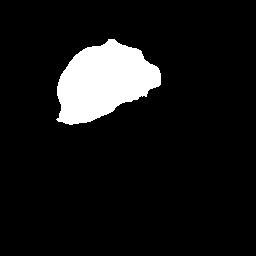}
    \end{minipage}
    \begin{minipage}{\myvaluet\textwidth}
        \centering
        \includegraphics[height=\textwidth, width=\textwidth]{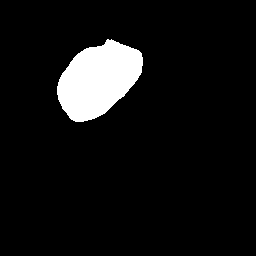}
    \end{minipage}
    \begin{minipage}{\myvaluet\textwidth}
        \centering
        \includegraphics[height=\textwidth, width=\textwidth]{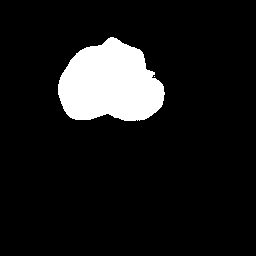}
    \end{minipage}
    \begin{minipage}{\myvaluet\textwidth}
        \centering
        \includegraphics[height=\textwidth, width=\textwidth]{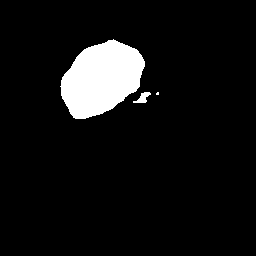}
    \end{minipage}
    \begin{minipage}{\myvaluet\textwidth}
        \centering
        \includegraphics[height=\textwidth, width=\textwidth]{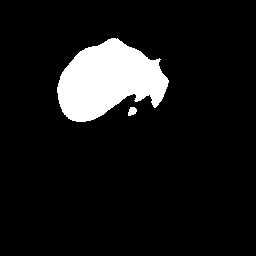}
    \end{minipage}
    \begin{minipage}{\myvaluet\textwidth}
        \centering
        \includegraphics[height=\textwidth, width=\textwidth]{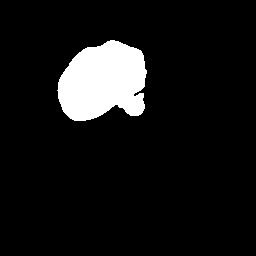}
    \end{minipage}
\end{minipage}

\vspace{2pt}
\begin{minipage}{0.04\textwidth}
    \centering
    \rotatebox{90}{\scriptsize{BrEaST}}
\end{minipage}
\begin{minipage}{0.92\textwidth}
    \begin{minipage}{\myvaluet\textwidth}
        \centering
        \includegraphics[height=\textwidth, width=\textwidth]{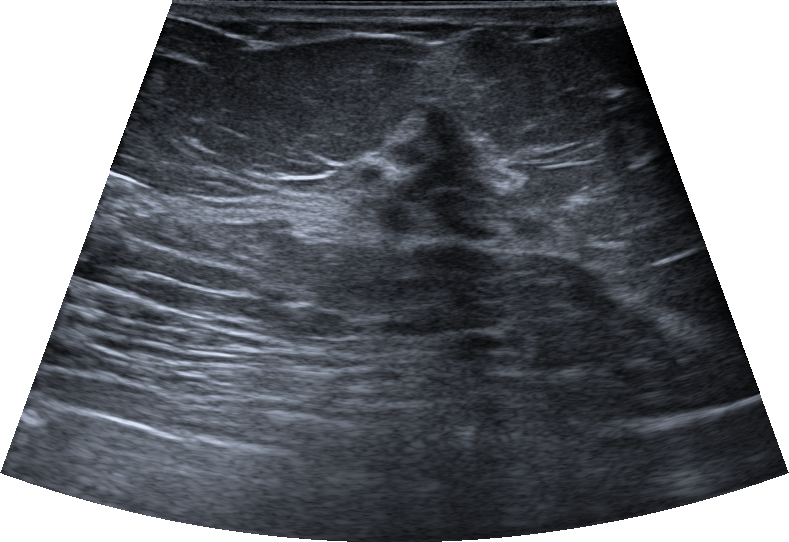}
    \end{minipage}
    \begin{minipage}{\myvaluet\textwidth}
        \centering
        \includegraphics[height=\textwidth, width=\textwidth]{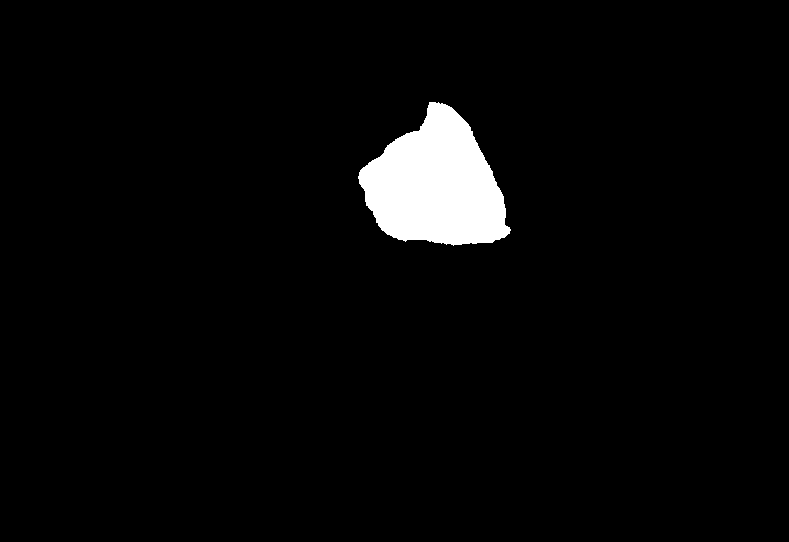}
    \end{minipage}
    \begin{minipage}{\myvaluet\textwidth}
        \centering
        \includegraphics[height=\textwidth, width=\textwidth]{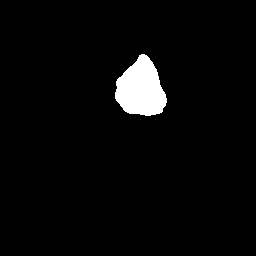}
    \end{minipage}
    \begin{minipage}{\myvaluet\textwidth}
        \centering
        \includegraphics[height=\textwidth, width=\textwidth]{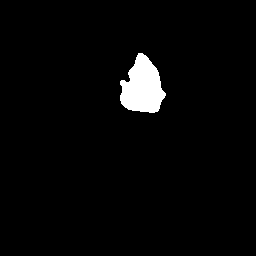}
    \end{minipage}
    \begin{minipage}{\myvaluet\textwidth}
        \centering
        \includegraphics[height=\textwidth, width=\textwidth]{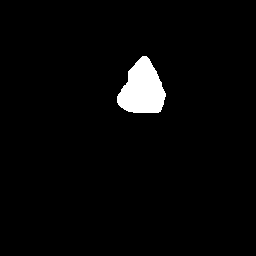}
    \end{minipage}
    \begin{minipage}{\myvaluet\textwidth}
        \centering
        \includegraphics[height=\textwidth, width=\textwidth]{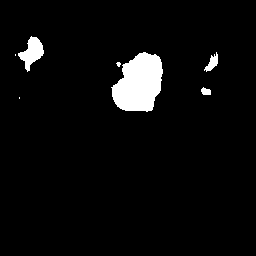}
    \end{minipage}
    \begin{minipage}{\myvaluet\textwidth}
        \centering
        \includegraphics[height=\textwidth, width=\textwidth]{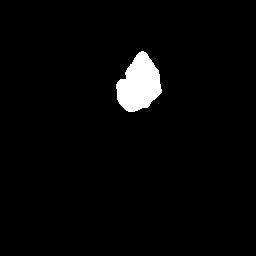}
    \end{minipage}
    \begin{minipage}{\myvaluet\textwidth}
        \centering
        \includegraphics[height=\textwidth, width=\textwidth]{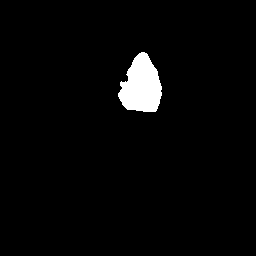}
    \end{minipage}
    \begin{minipage}{\myvaluet\textwidth}
        \centering
        \includegraphics[height=\textwidth, width=\textwidth]{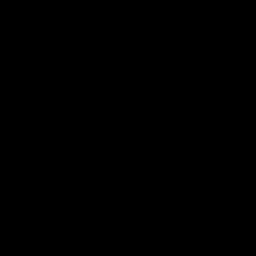}
    \end{minipage}
    \begin{minipage}{\myvaluet\textwidth}
        \centering
        \includegraphics[height=\textwidth, width=\textwidth]{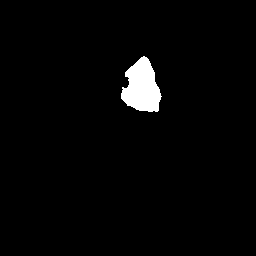}
    \end{minipage}
    \begin{minipage}{\myvaluet\textwidth}
        \centering
        \includegraphics[height=\textwidth, width=\textwidth]{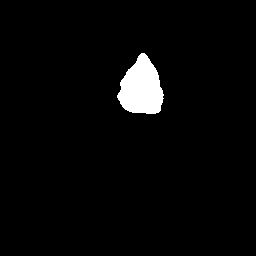}
    \end{minipage}
    \begin{minipage}{\myvaluet\textwidth}
        \centering
        \includegraphics[height=\textwidth, width=\textwidth]{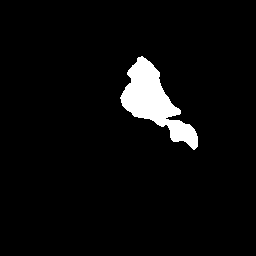}
    \end{minipage}
    \begin{minipage}{\myvaluet\textwidth}
        \centering
        \includegraphics[height=\textwidth, width=\textwidth]{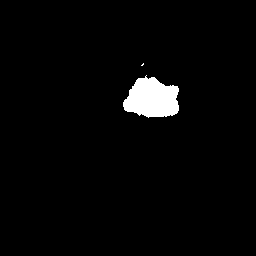}
    \end{minipage}
    \begin{minipage}{\myvaluet\textwidth}
        \centering
        \includegraphics[height=\textwidth, width=\textwidth]{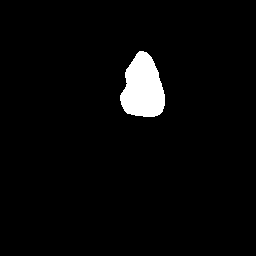}
    \end{minipage}
\end{minipage}

\vspace{2pt}
\begin{minipage}{0.04\textwidth}
    \centering
    \rotatebox{90}{\scriptsize{BUSBRA}}
\end{minipage}
\begin{minipage}{0.92\textwidth}
    \begin{minipage}{\myvaluet\textwidth}
        \centering
        \includegraphics[height=\textwidth, width=\textwidth]{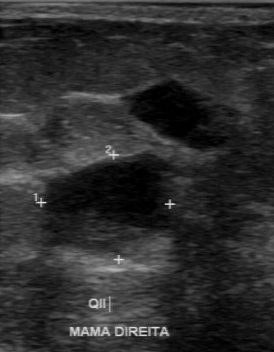}
    \end{minipage}
    \begin{minipage}{\myvaluet\textwidth}
        \centering
        \includegraphics[height=\textwidth, width=\textwidth]{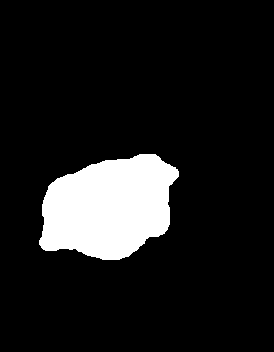}
    \end{minipage}
    \begin{minipage}{\myvaluet\textwidth}
        \centering
        \includegraphics[height=\textwidth, width=\textwidth]{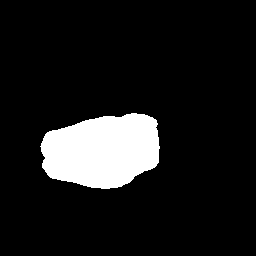}
    \end{minipage}
    \begin{minipage}{\myvaluet\textwidth}
        \centering
        \includegraphics[height=\textwidth, width=\textwidth]{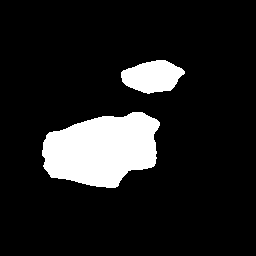}
    \end{minipage}
    \begin{minipage}{\myvaluet\textwidth}
        \centering
        \includegraphics[height=\textwidth, width=\textwidth]{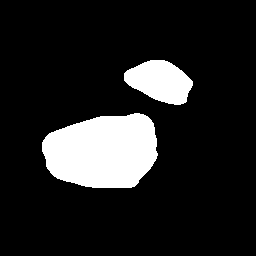}
    \end{minipage}
    \begin{minipage}{\myvaluet\textwidth}
        \centering
        \includegraphics[height=\textwidth, width=\textwidth]{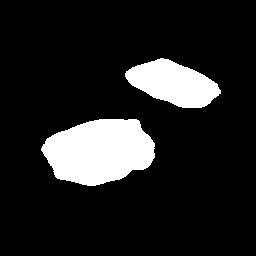}
    \end{minipage}
    \begin{minipage}{\myvaluet\textwidth}
        \centering
        \includegraphics[height=\textwidth, width=\textwidth]{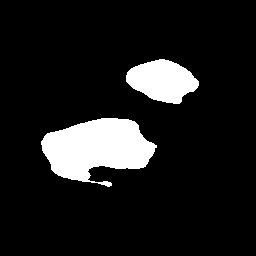}
    \end{minipage}
    \begin{minipage}{\myvaluet\textwidth}
        \centering
        \includegraphics[height=\textwidth, width=\textwidth]{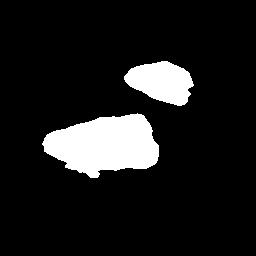}
    \end{minipage}
    \begin{minipage}{\myvaluet\textwidth}
        \centering
        \includegraphics[height=\textwidth, width=\textwidth]{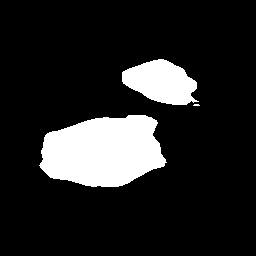}
    \end{minipage}
    \begin{minipage}{\myvaluet\textwidth}
        \centering
        \includegraphics[height=\textwidth, width=\textwidth]{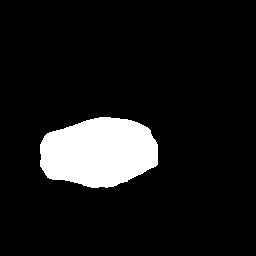}
    \end{minipage}
    \begin{minipage}{\myvaluet\textwidth}
        \centering
        \includegraphics[height=\textwidth, width=\textwidth]{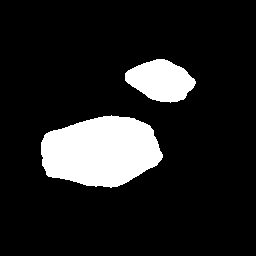}
    \end{minipage}
    \begin{minipage}{\myvaluet\textwidth}
        \centering
        \includegraphics[height=\textwidth, width=\textwidth]{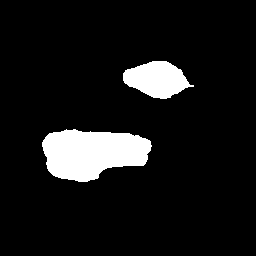}
    \end{minipage}
    \begin{minipage}{\myvaluet\textwidth}
        \centering
        \includegraphics[height=\textwidth, width=\textwidth]{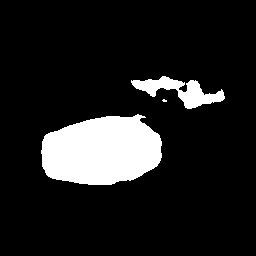}
    \end{minipage}
    \begin{minipage}{\myvaluet\textwidth}
        \centering
        \includegraphics[height=\textwidth, width=\textwidth]{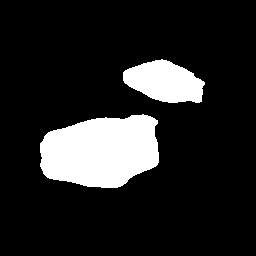}
    \end{minipage}
\end{minipage}
\caption{The qualitative comparison between our model and other models regarding \textcolor{red}{Medium Sized Breast Lesions}. On medium-sized breast lesions, all models tend to provide consistently better segmentation results. However, some models, such as UNet, LF-UNet, are more likely to experience segmentation failures. Overall, we believe our model can be a novel segmentation framework with promising performance.}
\label{fig:ins_med_cmp_sota}
\end{figure*}

\begin{figure*}[htbp]
\centering
\begin{minipage}{0.04\textwidth}
    \centering
    \rotatebox{90}{\scriptsize{Small}}
\end{minipage}
\begin{minipage}{0.92\textwidth}
    \begin{minipage}{\myvaluet\textwidth}
        \centering
        \textbf{\scriptsize Images}\\[1ex]
        \includegraphics[height=\textwidth, width=\textwidth]{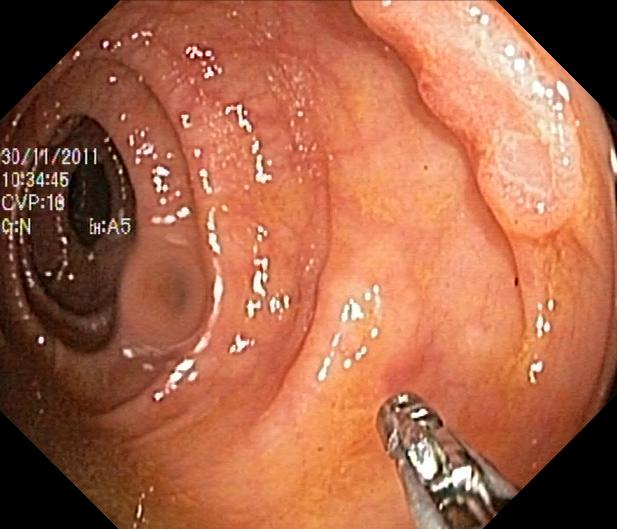}
    \end{minipage}
    \begin{minipage}{\myvaluet\textwidth}
        \centering
        \textbf{\scriptsize GTs}\\[1ex]
        \includegraphics[height=\textwidth, width=\textwidth]{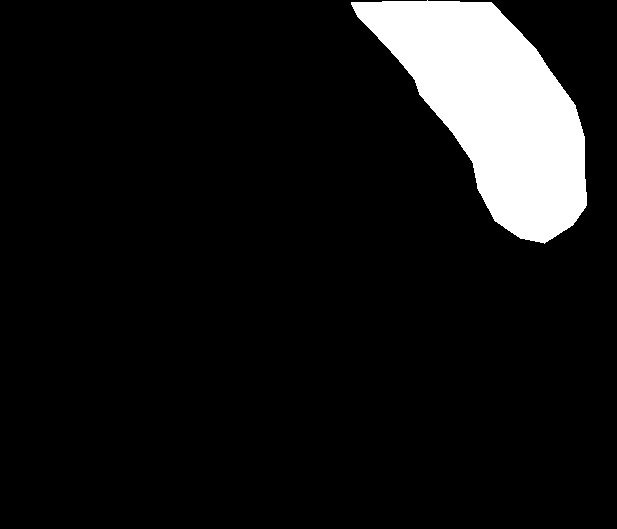}
    \end{minipage}
    \begin{minipage}{\myvaluet\textwidth}
        \centering
        \textbf{\tiny{SimpleESKNet}}\\[1ex]
        \includegraphics[height=\textwidth, width=\textwidth]{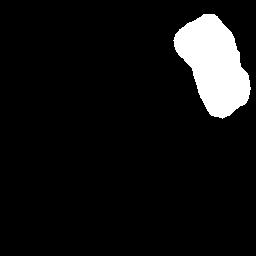}
    \end{minipage}
    \begin{minipage}{\myvaluet\textwidth}
        \centering
        \textbf{\scriptsize MALUNet}\\[1ex]
        \includegraphics[height=\textwidth, width=\textwidth]{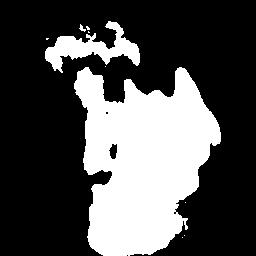}
    \end{minipage}
    \begin{minipage}{\myvaluet\textwidth}
        \centering
        \textbf{\scriptsize UNext$_{s}$}\\[1ex]
        \includegraphics[height=\textwidth, width=\textwidth]{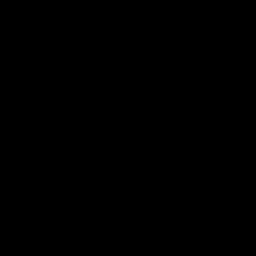}
    \end{minipage}
    \begin{minipage}{\myvaluet\textwidth}
        \centering
        \textbf{\scriptsize LF-UNet}\\[1ex]
        \includegraphics[height=\textwidth, width=\textwidth]{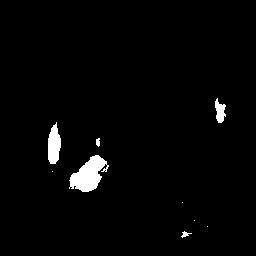}
    \end{minipage}
    \begin{minipage}{\myvaluet\textwidth}
        \centering
        \textbf{\scriptsize LBUNet}\\[1ex]
        \includegraphics[height=\textwidth, width=\textwidth]{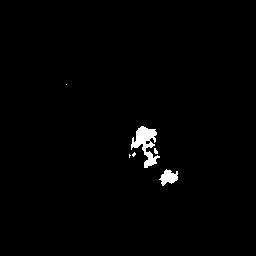}
    \end{minipage}
    \begin{minipage}{\myvaluet\textwidth}
        \centering
        \textbf{\tiny{UltraVMUNet}}\\[1ex]
        \includegraphics[height=\textwidth, width=\textwidth]{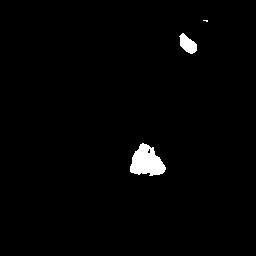}
    \end{minipage}
    \begin{minipage}{\myvaluet\textwidth}
        \centering
        \textbf{\scriptsize UNet}\\[1ex]
        \includegraphics[height=\textwidth, width=\textwidth]{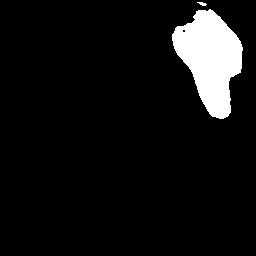}
    \end{minipage}
    \begin{minipage}{\myvaluet\textwidth}
        \centering
        \textbf{\scriptsize TransUNet}\\[1ex]
        \includegraphics[height=\textwidth, width=\textwidth]{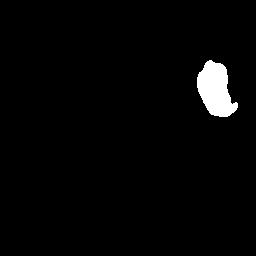}
    \end{minipage}
    \begin{minipage}{\myvaluet\textwidth}
        \centering
        \textbf{\scriptsize MDA-Net}\\[1ex]
        \includegraphics[height=\textwidth, width=\textwidth]{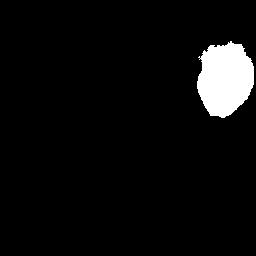}
    \end{minipage}
    \begin{minipage}{\myvaluet\textwidth}
        \centering
        \textbf{\scriptsize UNetv2}\\[1ex]
        \includegraphics[height=\textwidth, width=\textwidth]{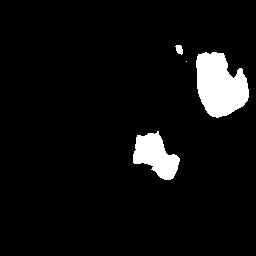}
    \end{minipage}
    \begin{minipage}{\myvaluet\textwidth}
        \centering
        \textbf{\tiny{Tiny-UNet}}\\[1ex]
        \includegraphics[height=\textwidth, width=\textwidth]{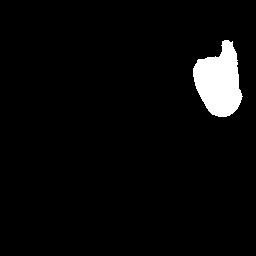}
    \end{minipage}
    \begin{minipage}{\myvaluet\textwidth}
        \centering
        \textbf{\scriptsize ESKNet}\\[1ex]
        \includegraphics[height=\textwidth, width=\textwidth]{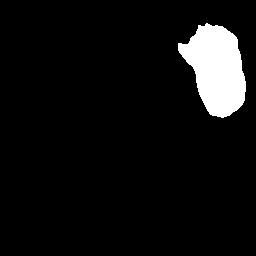}
    \end{minipage}
\end{minipage}

\vspace{2pt}
\begin{minipage}{0.04\textwidth}
    \centering
    \rotatebox{90}{\scriptsize{~}}
\end{minipage}
\begin{minipage}{0.92\textwidth}
    \begin{minipage}{\myvaluet\textwidth}
        \centering
        \includegraphics[height=\textwidth, width=\textwidth]{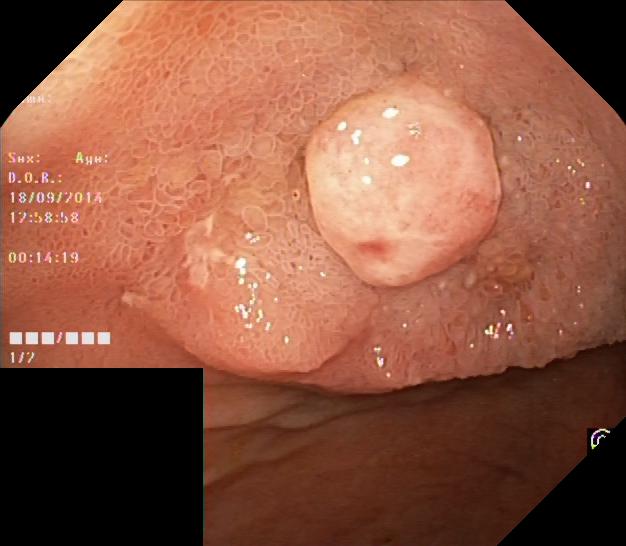}
    \end{minipage}
    \begin{minipage}{\myvaluet\textwidth}
        \centering
        \includegraphics[height=\textwidth, width=\textwidth]{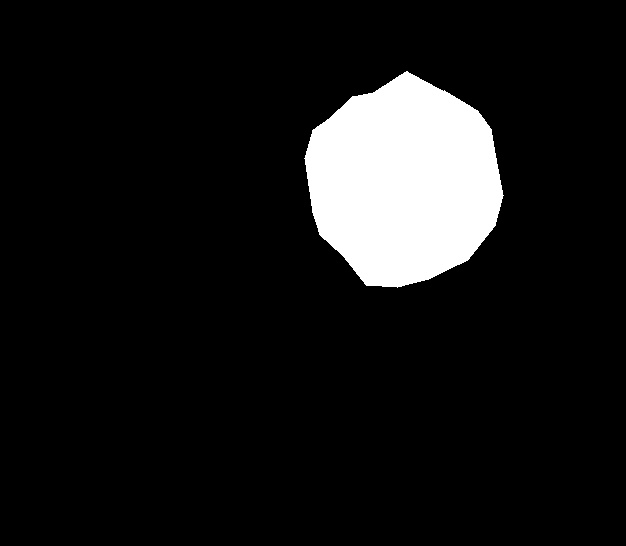}
    \end{minipage}
    \begin{minipage}{\myvaluet\textwidth}
        \centering
        \includegraphics[height=\textwidth, width=\textwidth]{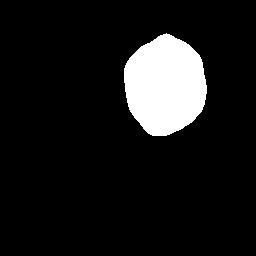}
    \end{minipage}
    \begin{minipage}{\myvaluet\textwidth}
        \centering
        \includegraphics[height=\textwidth, width=\textwidth]{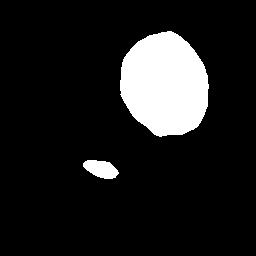}
    \end{minipage}
    \begin{minipage}{\myvaluet\textwidth}
        \centering
        \includegraphics[height=\textwidth, width=\textwidth]{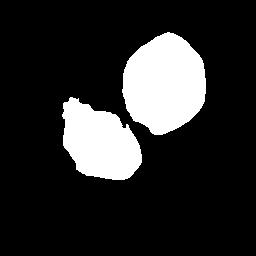}
    \end{minipage}
    \begin{minipage}{\myvaluet\textwidth}
        \centering
        \includegraphics[height=\textwidth, width=\textwidth]{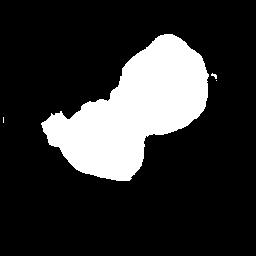}
    \end{minipage}
    \begin{minipage}{\myvaluet\textwidth}
        \centering
        \includegraphics[height=\textwidth, width=\textwidth]{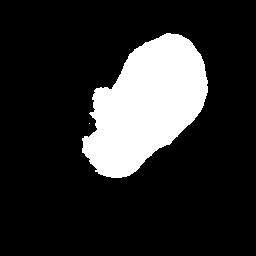}
    \end{minipage}
    \begin{minipage}{\myvaluet\textwidth}
        \centering
        \includegraphics[height=\textwidth, width=\textwidth]{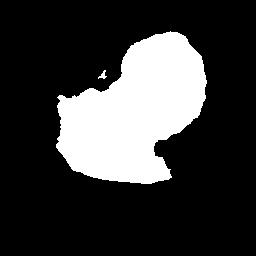}
    \end{minipage}
    \begin{minipage}{\myvaluet\textwidth}
        \centering
        \includegraphics[height=\textwidth, width=\textwidth]{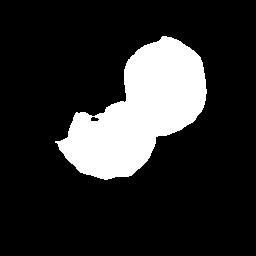}
    \end{minipage}
    \begin{minipage}{\myvaluet\textwidth}
        \centering
        \includegraphics[height=\textwidth, width=\textwidth]{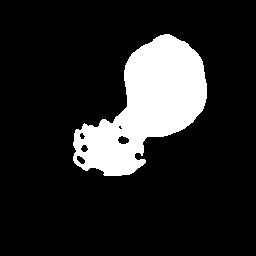}
    \end{minipage}
    \begin{minipage}{\myvaluet\textwidth}
        \centering
        \includegraphics[height=\textwidth, width=\textwidth]{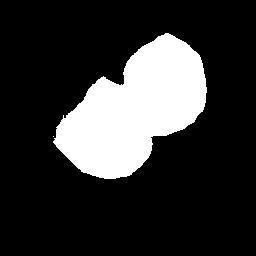}
    \end{minipage}
    \begin{minipage}{\myvaluet\textwidth}
        \centering
        \includegraphics[height=\textwidth, width=\textwidth]{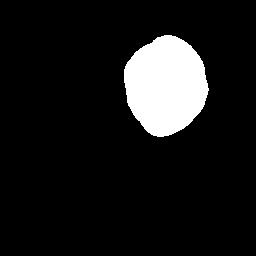}
    \end{minipage}
    \begin{minipage}{\myvaluet\textwidth}
        \centering
        \includegraphics[height=\textwidth, width=\textwidth]{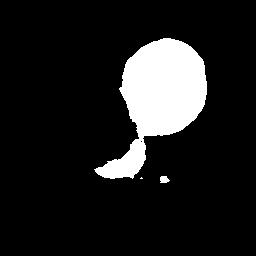}
    \end{minipage}
    \begin{minipage}{\myvaluet\textwidth}
        \centering
        \includegraphics[height=\textwidth, width=\textwidth]{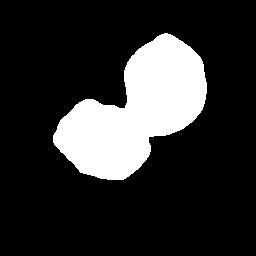}
    \end{minipage}
\end{minipage}

\vspace{2pt}
\begin{minipage}{0.04\textwidth}
    \centering
    \rotatebox{90}{\scriptsize{Medium}}
\end{minipage}
\begin{minipage}{0.92\textwidth}
    \begin{minipage}{\myvaluet\textwidth}
        \centering
        \includegraphics[height=\textwidth, width=\textwidth]{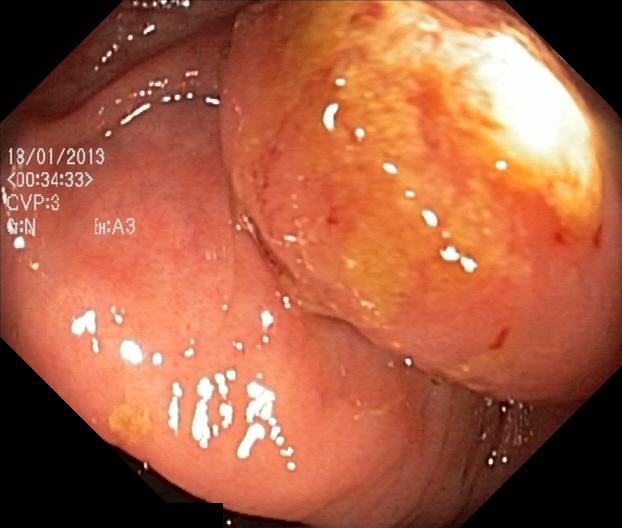}
    \end{minipage}
    \begin{minipage}{\myvaluet\textwidth}
        \centering
        \includegraphics[height=\textwidth, width=\textwidth]{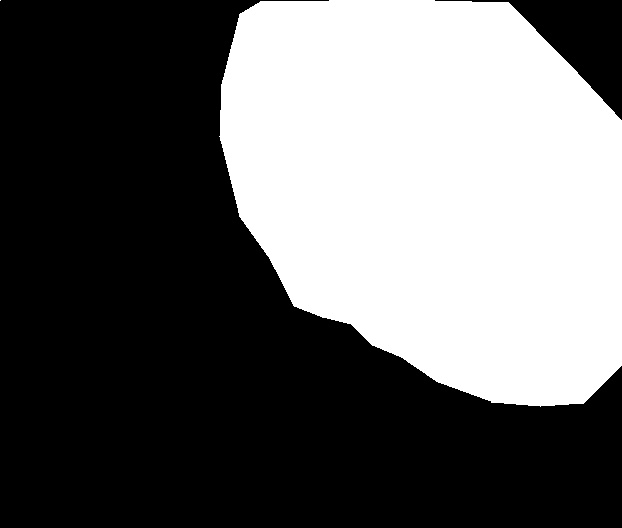}
    \end{minipage}
    \begin{minipage}{\myvaluet\textwidth}
        \centering
        \includegraphics[height=\textwidth, width=\textwidth]{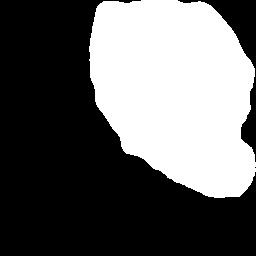}
    \end{minipage}
    \begin{minipage}{\myvaluet\textwidth}
        \centering
        \includegraphics[height=\textwidth, width=\textwidth]{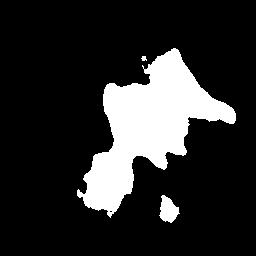}
    \end{minipage}
    \begin{minipage}{\myvaluet\textwidth}
        \centering
        \includegraphics[height=\textwidth, width=\textwidth]{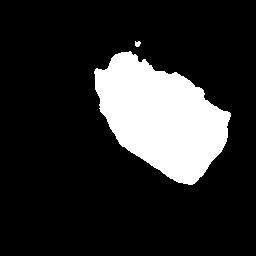}
    \end{minipage}
    \begin{minipage}{\myvaluet\textwidth}
        \centering
        \includegraphics[height=\textwidth, width=\textwidth]{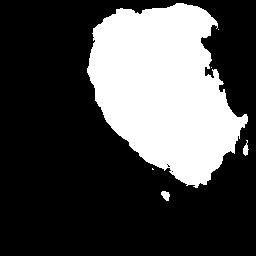}
    \end{minipage}
    \begin{minipage}{\myvaluet\textwidth}
        \centering
        \includegraphics[height=\textwidth, width=\textwidth]{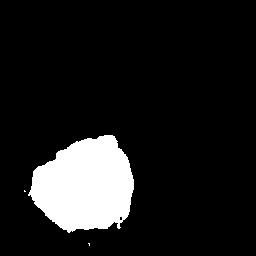}
    \end{minipage}
    \begin{minipage}{\myvaluet\textwidth}
        \centering
        \includegraphics[height=\textwidth, width=\textwidth]{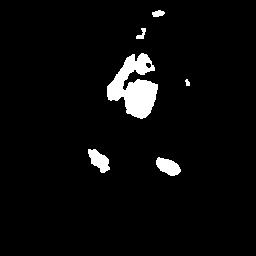}
    \end{minipage}
    \begin{minipage}{\myvaluet\textwidth}
        \centering
        \includegraphics[height=\textwidth, width=\textwidth]{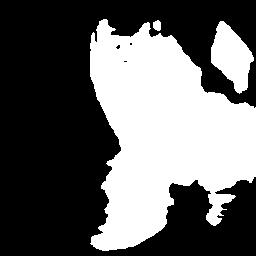}
    \end{minipage}
    \begin{minipage}{\myvaluet\textwidth}
        \centering
        \includegraphics[height=\textwidth, width=\textwidth]{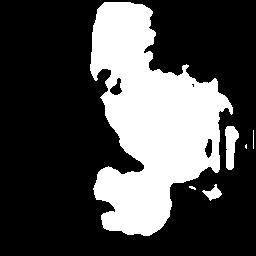}
    \end{minipage}
    \begin{minipage}{\myvaluet\textwidth}
        \centering
        \includegraphics[height=\textwidth, width=\textwidth]{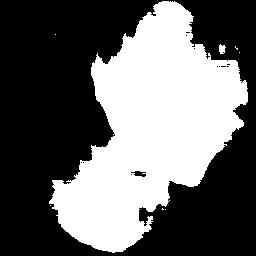}
    \end{minipage}
    \begin{minipage}{\myvaluet\textwidth}
        \centering
        \includegraphics[height=\textwidth, width=\textwidth]{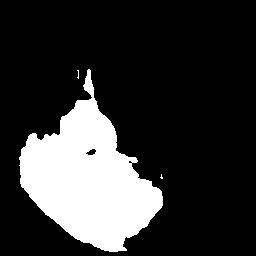}
    \end{minipage}
    \begin{minipage}{\myvaluet\textwidth}
        \centering
        \includegraphics[height=\textwidth, width=\textwidth]{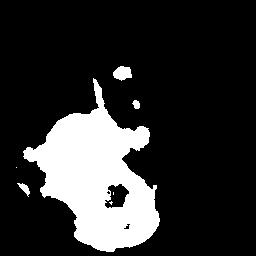}
    \end{minipage}
    \begin{minipage}{\myvaluet\textwidth}
        \centering
        \includegraphics[height=\textwidth, width=\textwidth]{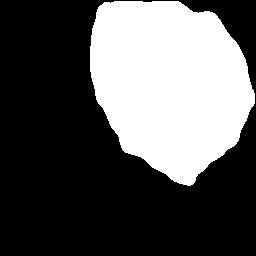}
    \end{minipage}
\end{minipage}

\vspace{2pt}
\begin{minipage}{0.04\textwidth}
    \centering
    \rotatebox{90}{\scriptsize{}}
\end{minipage}
\begin{minipage}{0.92\textwidth}
    \begin{minipage}{\myvaluet\textwidth}
        \centering
        \includegraphics[height=\textwidth, width=\textwidth]{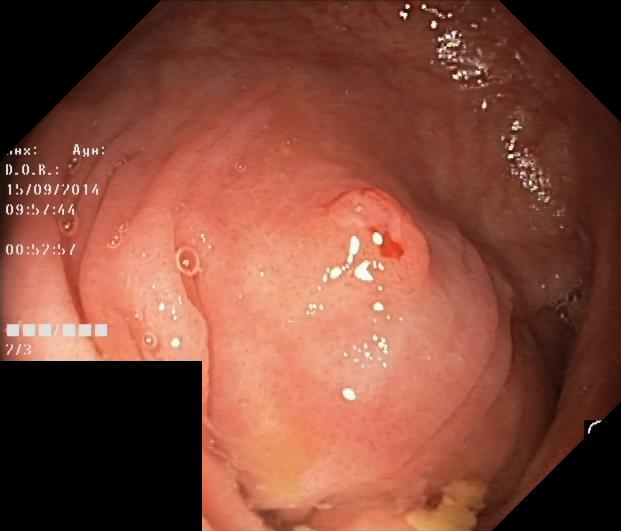}
    \end{minipage}
    \begin{minipage}{\myvaluet\textwidth}
        \centering
        \includegraphics[height=\textwidth, width=\textwidth]{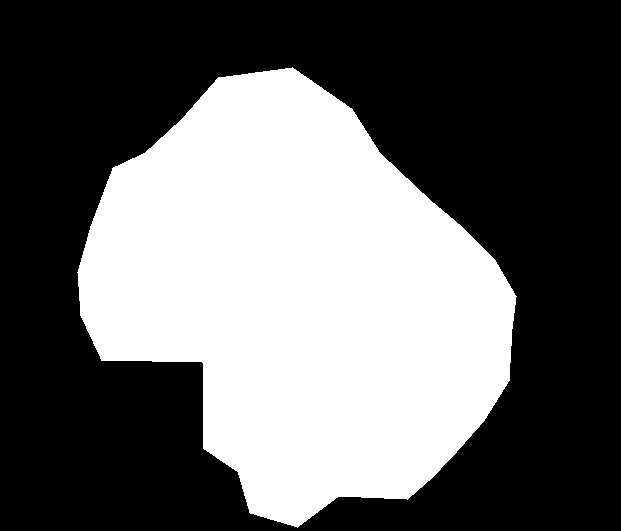}
    \end{minipage}
    \begin{minipage}{\myvaluet\textwidth}
        \centering
        \includegraphics[height=\textwidth, width=\textwidth]{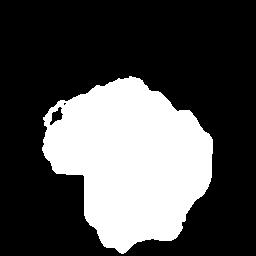}
    \end{minipage}
    \begin{minipage}{\myvaluet\textwidth}
        \centering
        \includegraphics[height=\textwidth, width=\textwidth]{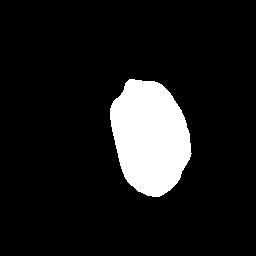}
    \end{minipage}
    \begin{minipage}{\myvaluet\textwidth}
        \centering
        \includegraphics[height=\textwidth, width=\textwidth]{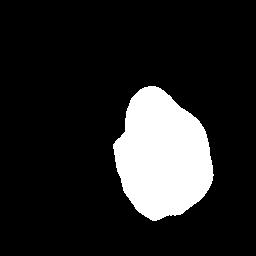}
    \end{minipage}
    \begin{minipage}{\myvaluet\textwidth}
        \centering
        \includegraphics[height=\textwidth, width=\textwidth]{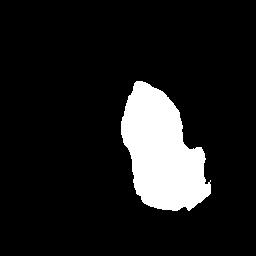}
    \end{minipage}
    \begin{minipage}{\myvaluet\textwidth}
        \centering
        \includegraphics[height=\textwidth, width=\textwidth]{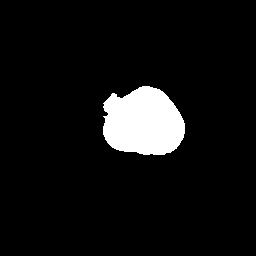}
    \end{minipage}
    \begin{minipage}{\myvaluet\textwidth}
        \centering
        \includegraphics[height=\textwidth, width=\textwidth]{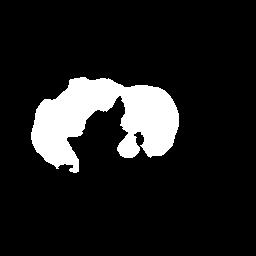}
    \end{minipage}
    \begin{minipage}{\myvaluet\textwidth}
        \centering
        \includegraphics[height=\textwidth, width=\textwidth]{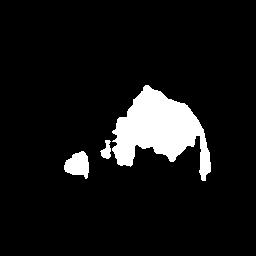}
    \end{minipage}
    \begin{minipage}{\myvaluet\textwidth}
        \centering
        \includegraphics[height=\textwidth, width=\textwidth]{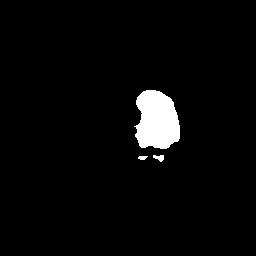}
    \end{minipage}
    \begin{minipage}{\myvaluet\textwidth}
        \centering
        \includegraphics[height=\textwidth, width=\textwidth]{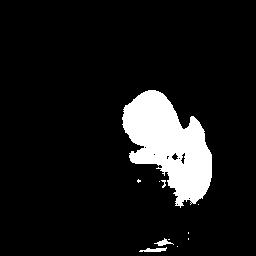}
    \end{minipage}
    \begin{minipage}{\myvaluet\textwidth}
        \centering
        \includegraphics[height=\textwidth, width=\textwidth]{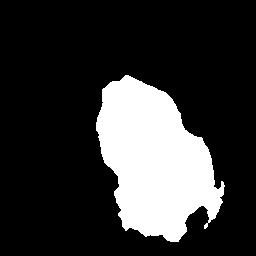}
    \end{minipage}
    \begin{minipage}{\myvaluet\textwidth}
        \centering
        \includegraphics[height=\textwidth, width=\textwidth]{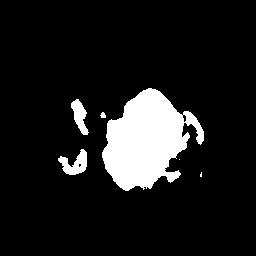}
    \end{minipage}
    \begin{minipage}{\myvaluet\textwidth}
        \centering
        \includegraphics[height=\textwidth, width=\textwidth]{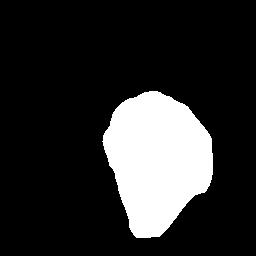}
    \end{minipage}
\end{minipage}

\vspace{2pt}
\begin{minipage}{0.04\textwidth}
    \centering
    \rotatebox{90}{\scriptsize{Large}}
\end{minipage}
\begin{minipage}{0.92\textwidth}
    \begin{minipage}{\myvaluet\textwidth}
        \centering
        \includegraphics[height=\textwidth, width=\textwidth]{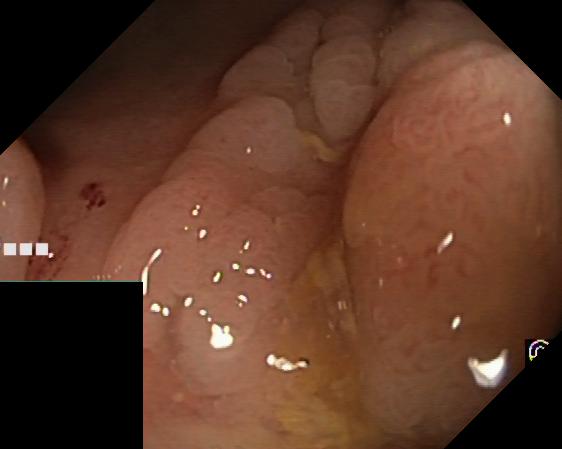}
    \end{minipage}
    \begin{minipage}{\myvaluet\textwidth}
        \centering
        \includegraphics[height=\textwidth, width=\textwidth]{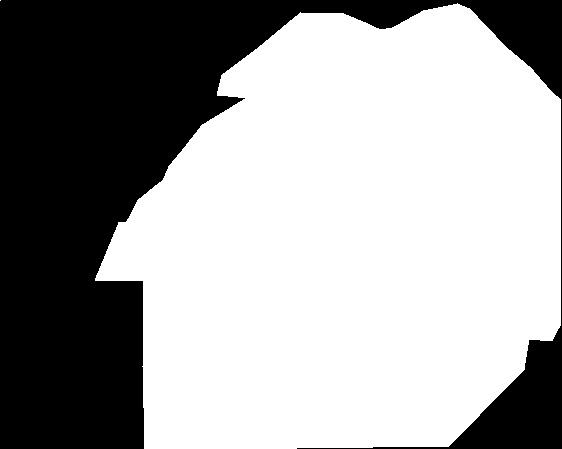}
    \end{minipage}
    \begin{minipage}{\myvaluet\textwidth}
        \centering
        \includegraphics[height=\textwidth, width=\textwidth]{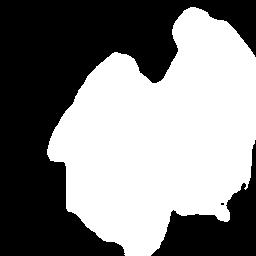}
    \end{minipage}
    \begin{minipage}{\myvaluet\textwidth}
        \centering
        \includegraphics[height=\textwidth, width=\textwidth]{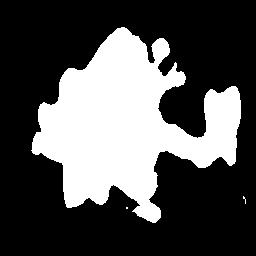}
    \end{minipage}
    \begin{minipage}{\myvaluet\textwidth}
        \centering
        \includegraphics[height=\textwidth, width=\textwidth]{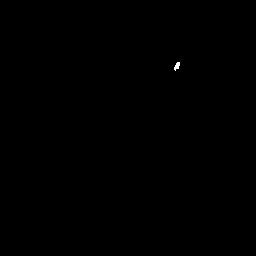}
    \end{minipage}
    \begin{minipage}{\myvaluet\textwidth}
        \centering
        \includegraphics[height=\textwidth, width=\textwidth]{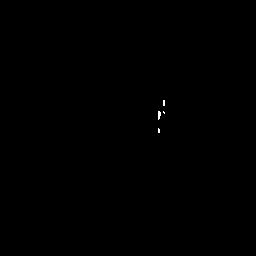}
    \end{minipage}
    \begin{minipage}{\myvaluet\textwidth}
        \centering
        \includegraphics[height=\textwidth, width=\textwidth]{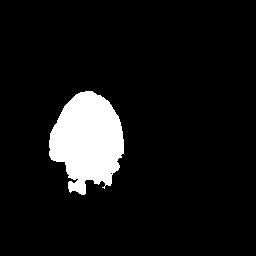}
    \end{minipage}
    \begin{minipage}{\myvaluet\textwidth}
        \centering
        \includegraphics[height=\textwidth, width=\textwidth]{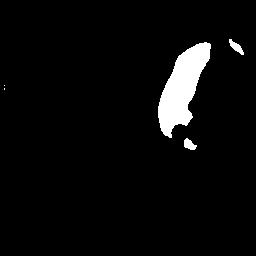}
    \end{minipage}
    \begin{minipage}{\myvaluet\textwidth}
        \centering
        \includegraphics[height=\textwidth, width=\textwidth]{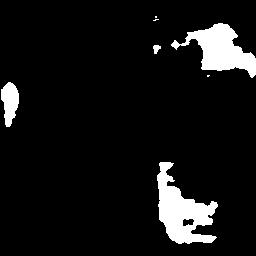}
    \end{minipage}
    \begin{minipage}{\myvaluet\textwidth}
        \centering
        \includegraphics[height=\textwidth, width=\textwidth]{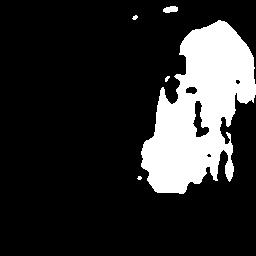}
    \end{minipage}
    \begin{minipage}{\myvaluet\textwidth}
        \centering
        \includegraphics[height=\textwidth, width=\textwidth]{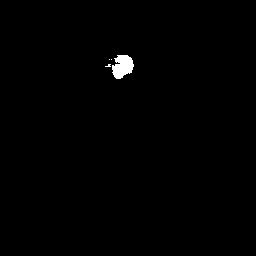}
    \end{minipage}
    \begin{minipage}{\myvaluet\textwidth}
        \centering
        \includegraphics[height=\textwidth, width=\textwidth]{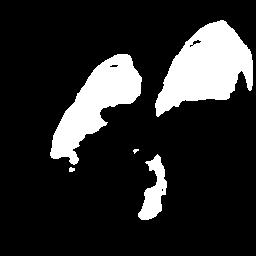}
    \end{minipage}
    \begin{minipage}{\myvaluet\textwidth}
        \centering
        \includegraphics[height=\textwidth, width=\textwidth]{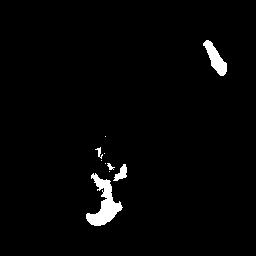}
    \end{minipage}
    \begin{minipage}{\myvaluet\textwidth}
        \centering
        \includegraphics[height=\textwidth, width=\textwidth]{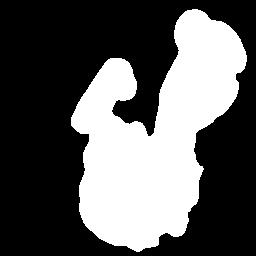}
    \end{minipage}
\end{minipage}

\vspace{2pt}
\begin{minipage}{0.04\textwidth}
    \centering
    \rotatebox{90}{\scriptsize{~}}
\end{minipage}
\begin{minipage}{0.92\textwidth}
    \begin{minipage}{\myvaluet\textwidth}
        \centering
        \includegraphics[height=\textwidth, width=\textwidth]{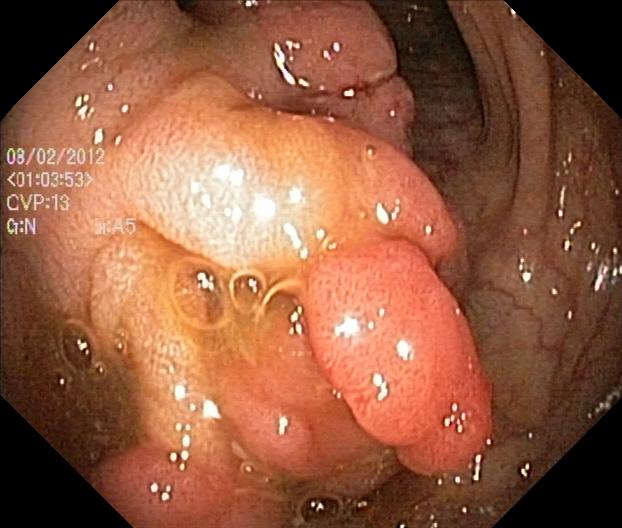}
    \end{minipage}
    \begin{minipage}{\myvaluet\textwidth}
        \centering
        \includegraphics[height=\textwidth, width=\textwidth]{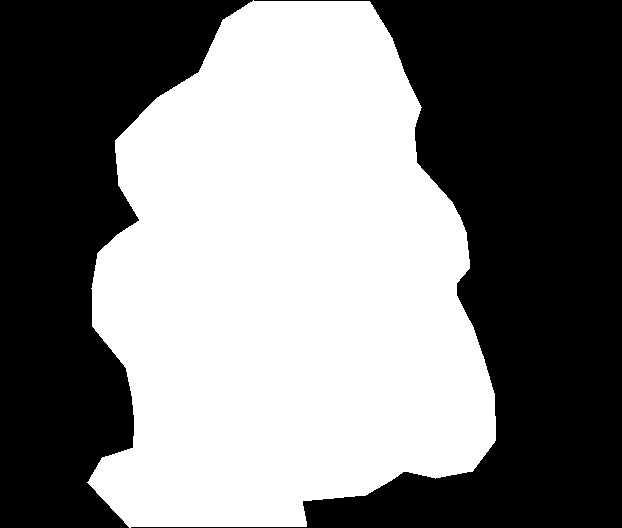}
    \end{minipage}
    \begin{minipage}{\myvaluet\textwidth}
        \centering
        \includegraphics[height=\textwidth, width=\textwidth]{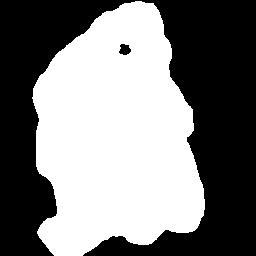}
    \end{minipage}
    \begin{minipage}{\myvaluet\textwidth}
        \centering
        \includegraphics[height=\textwidth, width=\textwidth]{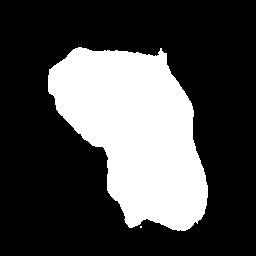}
    \end{minipage}
    \begin{minipage}{\myvaluet\textwidth}
        \centering
        \includegraphics[height=\textwidth, width=\textwidth]{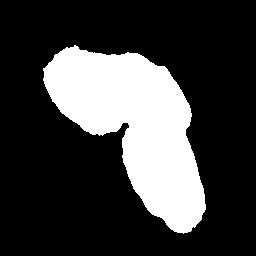}
    \end{minipage}
    \begin{minipage}{\myvaluet\textwidth}
        \centering
        \includegraphics[height=\textwidth, width=\textwidth]{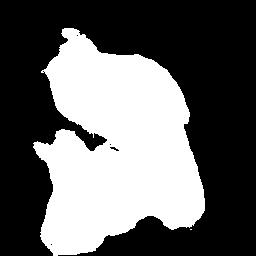}
    \end{minipage}
    \begin{minipage}{\myvaluet\textwidth}
        \centering
        \includegraphics[height=\textwidth, width=\textwidth]{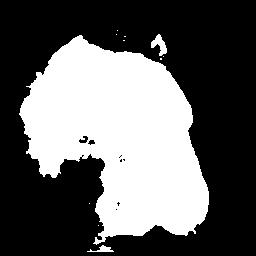}
    \end{minipage}
    \begin{minipage}{\myvaluet\textwidth}
        \centering
        \includegraphics[height=\textwidth, width=\textwidth]{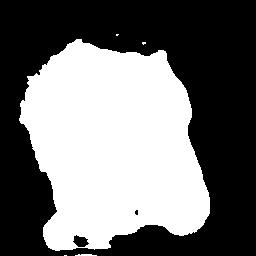}
    \end{minipage}
    \begin{minipage}{\myvaluet\textwidth}
        \centering
        \includegraphics[height=\textwidth, width=\textwidth]{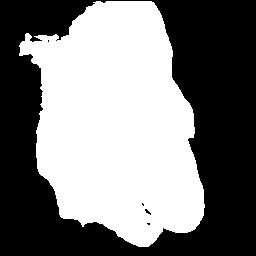}
    \end{minipage}
    \begin{minipage}{\myvaluet\textwidth}
        \centering
        \includegraphics[height=\textwidth, width=\textwidth]{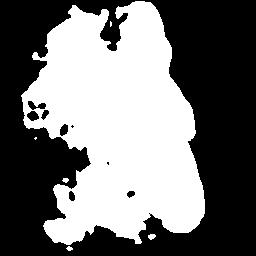}
    \end{minipage}
    \begin{minipage}{\myvaluet\textwidth}
        \centering
        \includegraphics[height=\textwidth, width=\textwidth]{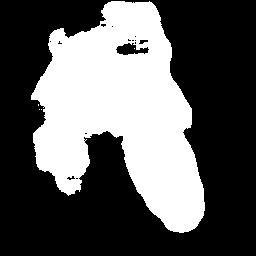}
    \end{minipage}
    \begin{minipage}{\myvaluet\textwidth}
        \centering
        \includegraphics[height=\textwidth, width=\textwidth]{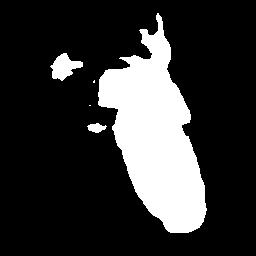}
    \end{minipage}
    \begin{minipage}{\myvaluet\textwidth}
        \centering
        \includegraphics[height=\textwidth, width=\textwidth]{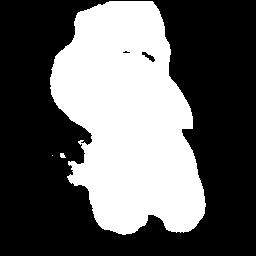}
    \end{minipage}
    \begin{minipage}{\myvaluet\textwidth}
        \centering
        \includegraphics[height=\textwidth, width=\textwidth]{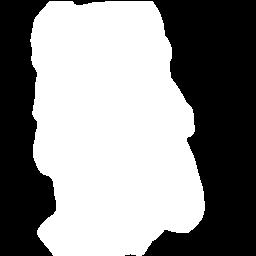}
    \end{minipage}
\end{minipage}
\caption{The qualitative comparison between our model and other models regarding \textcolor{red}{vaired-size gastrointestinal polyp}. As seen, the performance of existing approaches is severely compromised by their tendency to either over-segment or under-segment targets while our models. In contrast, our model delivers segmentation results that closely align with ground truth annotations, demonstrating superior boundary adherence.}
\label{fig:ins_kvasir_cmp_sota}
\end{figure*}

\textbf{Results on KVASIR-SEG} We consider it a desirable nature of a semantic model that can achieve high performance on limited training data. Therefore, to further validate the high performance of the developed models, we then train and validate all models mentioned above on the KVASIR-SEG dataset. As seen from Table \ref{tab:mc_mbd}, the experimental results here indicate that our models can still perform well, though the data size is small. Throughout the lightweight models, all models are experiencing under-performance where the mDice values barely exceed 70$\%$. However, our models showed superiority in terms of segmentation performance, where $SimpleESKNet_{16}^{1+AFF}$ reached 82.67$\%$ mDice and 70.87$\%$ mIoU, respectively. When it comes to relatively larger models, ESKNet becomes the best one with a mDice and mIoU of 86.60 $\%$ and 76.67 $\%$, followed by our models, which suggests that our models can still achieve high segmentation performance though with limited data. Some segmentation results concerning varied lesions are given in Fig. \ref{fig:ins_kvasir_cmp_sota}, where our best model provided accurate segmentation results.

\textbf{Results on ISIC2017 and ISIC2018} Previous experiments have demonstrated the superiority of our developed models on relatively small-scale heterogeneous and homogeneous datasets. However, we reckon it necessary to examine the learning capacity of the developed models on large-scale datasets and diverse tasks. Consequently, we validated previous models on ISIC2017 and ISIC2018 for skin lesion segmentation. In the lightweight models,  our model again becomes the best model across the two datasets and even surpasses some heavy models, such as UNet, MDA-Net, UNetv2, and Tiny-UNet. This exhibits the possibilities for lightweight models to achieve high performance under a specific segmentation task. Nevertheless, our lightweight model remains competitive because it performed better than models such as MALUNet and LF-UNet. As for the large models, TransUNet has become the best model thanks to the intrinsic advantage of the transformer on large-scale datasets. Nevertheless, our models remain competitive given the model size and the performance. The experimental results here also indicate the generality of our models, which can consistently show promising performance across varied dataset scales and segmentation tasks.

\begin{figure*}[htbp]
\centering
\begin{minipage}{0.04\textwidth}
    \centering
    \rotatebox{90}{\scriptsize{ISIC2017}}
\end{minipage}
\begin{minipage}{0.92\textwidth}
    \begin{minipage}{\myvaluet\textwidth}
        \centering
        \textbf{\scriptsize Images}\\[1ex]
        \includegraphics[height=\textwidth, width=\textwidth]{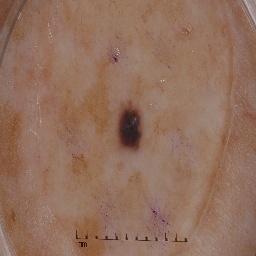}
    \end{minipage}
    \begin{minipage}{\myvaluet\textwidth}
        \centering
        \textbf{\scriptsize GTs}\\[1ex]
        \includegraphics[height=\textwidth, width=\textwidth]{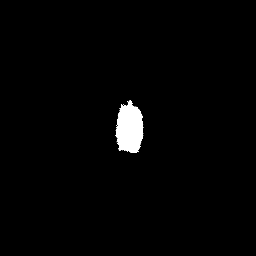}
    \end{minipage}
    \begin{minipage}{\myvaluet\textwidth}
        \centering
        \textbf{\tiny{SimpleESKNet}}\\[1ex]
        \includegraphics[height=\textwidth, width=\textwidth]{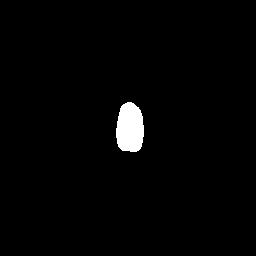}
    \end{minipage}
    \begin{minipage}{\myvaluet\textwidth}
        \centering
        \textbf{\scriptsize MALUNet}\\[1ex]
        \includegraphics[height=\textwidth, width=\textwidth]{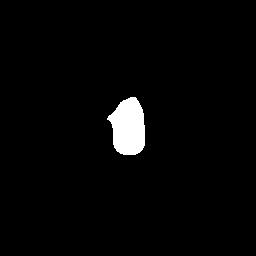}
    \end{minipage}
    \begin{minipage}{\myvaluet\textwidth}
        \centering
        \textbf{\scriptsize UNext$_{s}$}\\[1ex]
        \includegraphics[height=\textwidth, width=\textwidth]{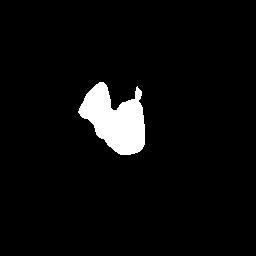}
    \end{minipage}
    \begin{minipage}{\myvaluet\textwidth}
        \centering
        \textbf{\scriptsize LF-UNet}\\[1ex]
        \includegraphics[height=\textwidth, width=\textwidth]{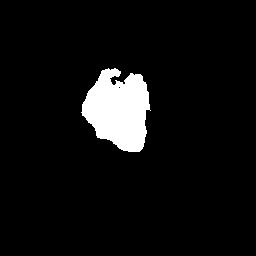}
    \end{minipage}
    \begin{minipage}{\myvaluet\textwidth}
        \centering
        \textbf{\scriptsize LBUNet}\\[1ex]
        \includegraphics[height=\textwidth, width=\textwidth]{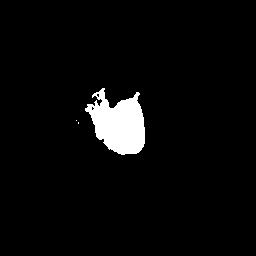}
    \end{minipage}
    \begin{minipage}{\myvaluet\textwidth}
        \centering
        \textbf{\tiny{UltraVMUNet}}\\[1ex]
        \includegraphics[height=\textwidth, width=\textwidth]{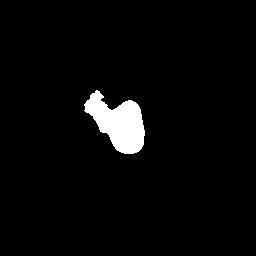}
    \end{minipage}
    \begin{minipage}{\myvaluet\textwidth}
        \centering
        \textbf{\scriptsize UNet}\\[1ex]
        \includegraphics[height=\textwidth, width=\textwidth]{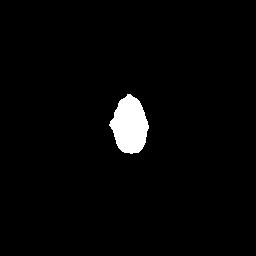}
    \end{minipage}
    \begin{minipage}{\myvaluet\textwidth}
        \centering
        \textbf{\scriptsize TransUNet}\\[1ex]
        \includegraphics[height=\textwidth, width=\textwidth]{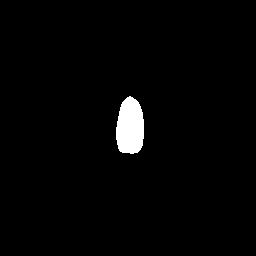}
    \end{minipage}
    \begin{minipage}{\myvaluet\textwidth}
        \centering
        \textbf{\scriptsize MDA-Net}\\[1ex]
        \includegraphics[height=\textwidth, width=\textwidth]{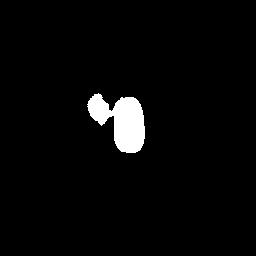}
    \end{minipage}
    \begin{minipage}{\myvaluet\textwidth}
        \centering
        \textbf{\scriptsize UNetv2}\\[1ex]
        \includegraphics[height=\textwidth, width=\textwidth]{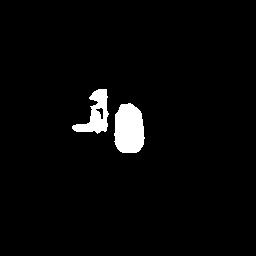}
    \end{minipage}
    \begin{minipage}{\myvaluet\textwidth}
        \centering
        \textbf{\tiny{Tiny-UNet}}\\[1ex]
        \includegraphics[height=\textwidth, width=\textwidth]{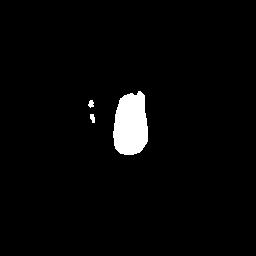}
    \end{minipage}
    \begin{minipage}{\myvaluet\textwidth}
        \centering
        \textbf{\scriptsize ESKNet}\\[1ex]
        \includegraphics[height=\textwidth, width=\textwidth]{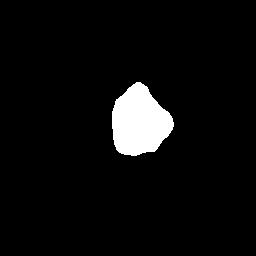}
    \end{minipage}
\end{minipage}

\vspace{2pt}
\begin{minipage}{0.04\textwidth}
    \centering
    \rotatebox{90}{\scriptsize{~}}
\end{minipage}
\begin{minipage}{0.92\textwidth}
    \begin{minipage}{\myvaluet\textwidth}
        \centering
        \includegraphics[height=\textwidth, width=\textwidth]{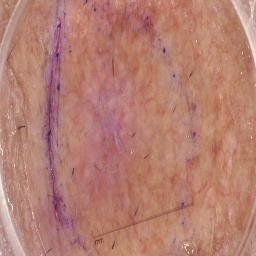}
    \end{minipage}
    \begin{minipage}{\myvaluet\textwidth}
        \centering
        \includegraphics[height=\textwidth, width=\textwidth]{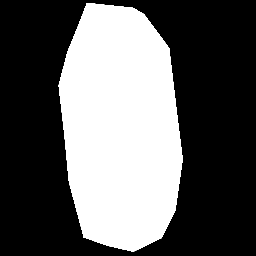}
    \end{minipage}
    \begin{minipage}{\myvaluet\textwidth}
        \centering
        \includegraphics[height=\textwidth, width=\textwidth]{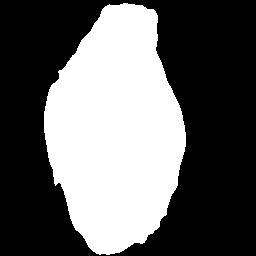}
    \end{minipage}
    \begin{minipage}{\myvaluet\textwidth}
        \centering
        \includegraphics[height=\textwidth, width=\textwidth]{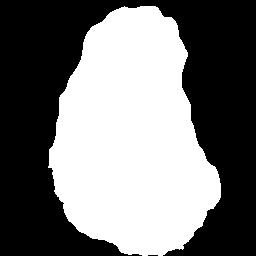}
    \end{minipage}
    \begin{minipage}{\myvaluet\textwidth}
        \centering
        \includegraphics[height=\textwidth, width=\textwidth]{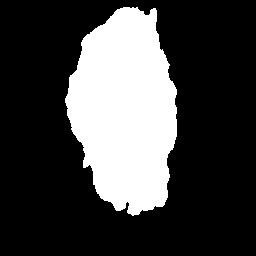}
    \end{minipage}
    \begin{minipage}{\myvaluet\textwidth}
        \centering
        \includegraphics[height=\textwidth, width=\textwidth]{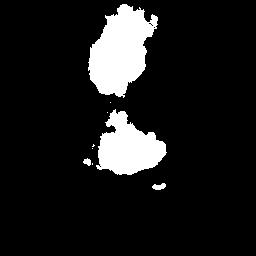}
    \end{minipage}
    \begin{minipage}{\myvaluet\textwidth}
        \centering
        \includegraphics[height=\textwidth, width=\textwidth]{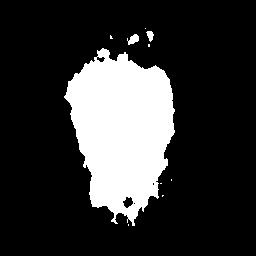}
    \end{minipage}
    \begin{minipage}{\myvaluet\textwidth}
        \centering
        \includegraphics[height=\textwidth, width=\textwidth]{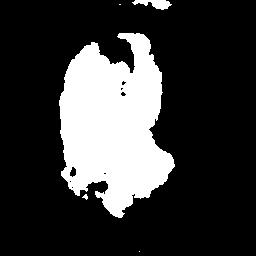}
    \end{minipage}
    \begin{minipage}{\myvaluet\textwidth}
        \centering
        \includegraphics[height=\textwidth, width=\textwidth]{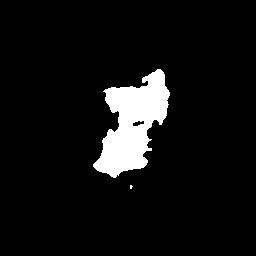}
    \end{minipage}
    \begin{minipage}{\myvaluet\textwidth}
        \centering
        \includegraphics[height=\textwidth, width=\textwidth]{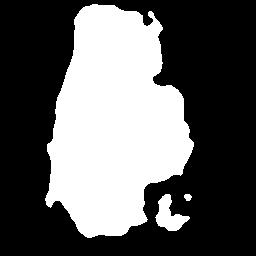}
    \end{minipage}
    \begin{minipage}{\myvaluet\textwidth}
        \centering
        \includegraphics[height=\textwidth, width=\textwidth]{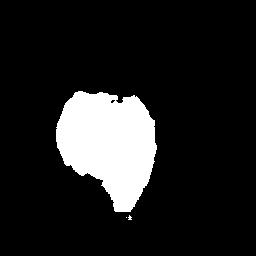}
    \end{minipage}
    \begin{minipage}{\myvaluet\textwidth}
        \centering
        \includegraphics[height=\textwidth, width=\textwidth]{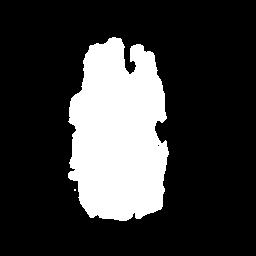}
    \end{minipage}
    \begin{minipage}{\myvaluet\textwidth}
        \centering
        \includegraphics[height=\textwidth, width=\textwidth]{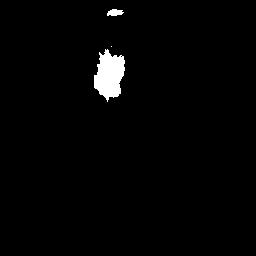}
    \end{minipage}
    \begin{minipage}{\myvaluet\textwidth}
        \centering
        \includegraphics[height=\textwidth, width=\textwidth]{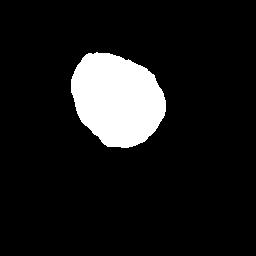}
    \end{minipage}
\end{minipage}

\vspace{2pt}
\begin{minipage}{0.04\textwidth}
    \centering
    \rotatebox{90}{\scriptsize{~}}
\end{minipage}
\begin{minipage}{0.92\textwidth}
    \begin{minipage}{\myvaluet\textwidth}
        \centering
        \includegraphics[height=\textwidth, width=\textwidth]{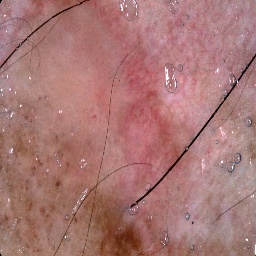}
    \end{minipage}
    \begin{minipage}{\myvaluet\textwidth}
        \centering
        \includegraphics[height=\textwidth, width=\textwidth]{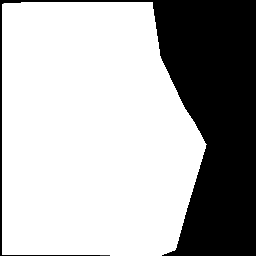}
    \end{minipage}
    \begin{minipage}{\myvaluet\textwidth}
        \centering
        \includegraphics[height=\textwidth, width=\textwidth]{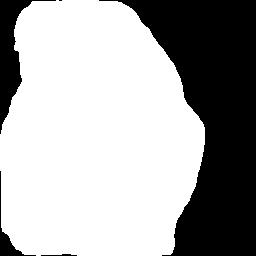}
    \end{minipage}
    \begin{minipage}{\myvaluet\textwidth}
        \centering
        \includegraphics[height=\textwidth, width=\textwidth]{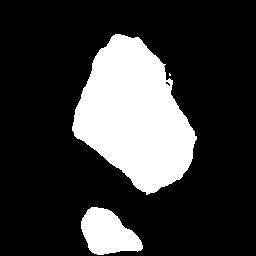}
    \end{minipage}
    \begin{minipage}{\myvaluet\textwidth}
        \centering
        \includegraphics[height=\textwidth, width=\textwidth]{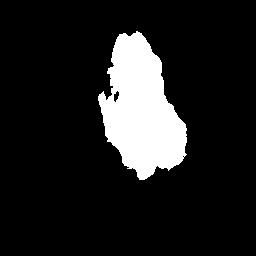}
    \end{minipage}
    \begin{minipage}{\myvaluet\textwidth}
        \centering
        \includegraphics[height=\textwidth, width=\textwidth]{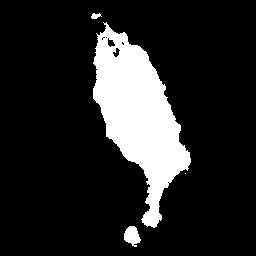}
    \end{minipage}
    \begin{minipage}{\myvaluet\textwidth}
        \centering
        \includegraphics[height=\textwidth, width=\textwidth]{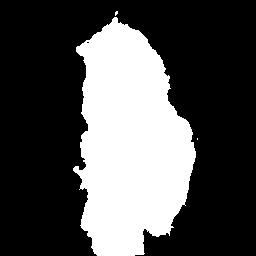}
    \end{minipage}
    \begin{minipage}{\myvaluet\textwidth}
        \centering
        \includegraphics[height=\textwidth, width=\textwidth]{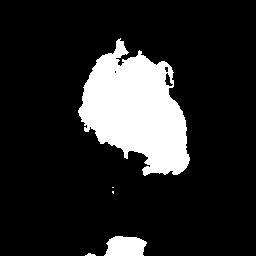}
    \end{minipage}
    \begin{minipage}{\myvaluet\textwidth}
        \centering
        \includegraphics[height=\textwidth, width=\textwidth]{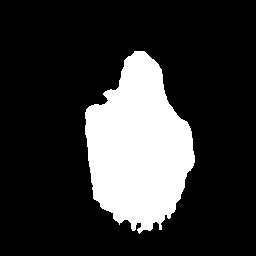}
    \end{minipage}
    \begin{minipage}{\myvaluet\textwidth}
        \centering
        \includegraphics[height=\textwidth, width=\textwidth]{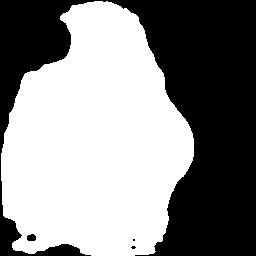}
    \end{minipage}
    \begin{minipage}{\myvaluet\textwidth}
        \centering
        \includegraphics[height=\textwidth, width=\textwidth]{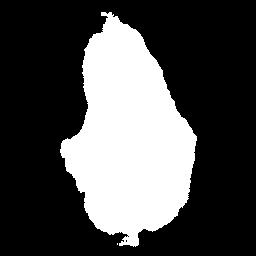}
    \end{minipage}
    \begin{minipage}{\myvaluet\textwidth}
        \centering
        \includegraphics[height=\textwidth, width=\textwidth]{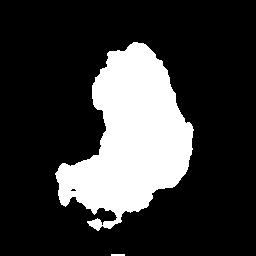}
    \end{minipage}
    \begin{minipage}{\myvaluet\textwidth}
        \centering
        \includegraphics[height=\textwidth, width=\textwidth]{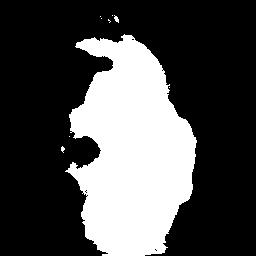}
    \end{minipage}
    \begin{minipage}{\myvaluet\textwidth}
        \centering
        \includegraphics[height=\textwidth, width=\textwidth]{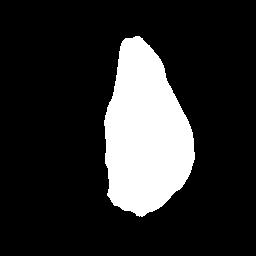}
    \end{minipage}
\end{minipage}

\vspace{2pt}
\begin{minipage}{0.04\textwidth}
    \centering
    \rotatebox{90}{\scriptsize{ISIC2018}}
\end{minipage}
\begin{minipage}{0.92\textwidth}
    \begin{minipage}{\myvaluet\textwidth}
        \centering
        \includegraphics[height=\textwidth, width=\textwidth]{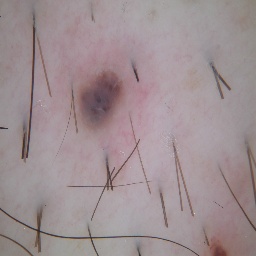}
    \end{minipage}
    \begin{minipage}{\myvaluet\textwidth}
        \centering
        \includegraphics[height=\textwidth, width=\textwidth]{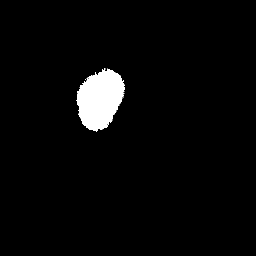}
    \end{minipage}
    \begin{minipage}{\myvaluet\textwidth}
        \centering
        \includegraphics[height=\textwidth, width=\textwidth]{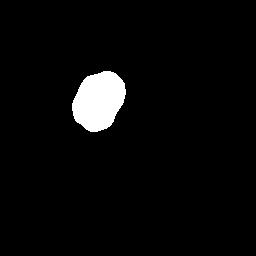}
    \end{minipage}
    \begin{minipage}{\myvaluet\textwidth}
        \centering
        \includegraphics[height=\textwidth, width=\textwidth]{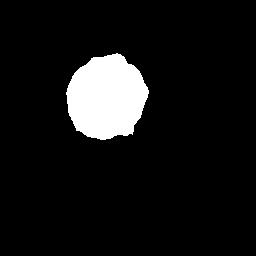}
    \end{minipage}
    \begin{minipage}{\myvaluet\textwidth}
        \centering
        \includegraphics[height=\textwidth, width=\textwidth]{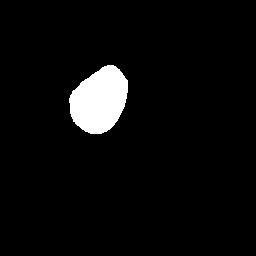}
    \end{minipage}
    \begin{minipage}{\myvaluet\textwidth}
        \centering
        \includegraphics[height=\textwidth, width=\textwidth]{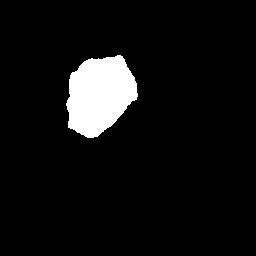}
    \end{minipage}
    \begin{minipage}{\myvaluet\textwidth}
        \centering
        \includegraphics[height=\textwidth, width=\textwidth]{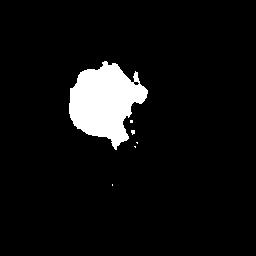}
    \end{minipage}
    \begin{minipage}{\myvaluet\textwidth}
        \centering
        \includegraphics[height=\textwidth, width=\textwidth]{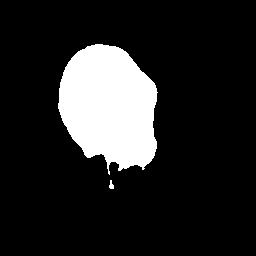}
    \end{minipage}
    \begin{minipage}{\myvaluet\textwidth}
        \centering
        \includegraphics[height=\textwidth, width=\textwidth]{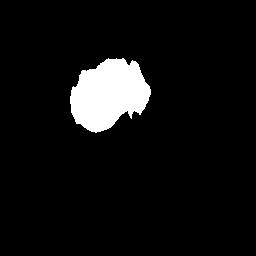}
    \end{minipage}
    \begin{minipage}{\myvaluet\textwidth}
        \centering
        \includegraphics[height=\textwidth, width=\textwidth]{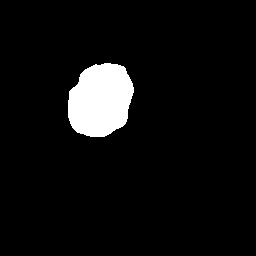}
    \end{minipage}
    \begin{minipage}{\myvaluet\textwidth}
        \centering
        \includegraphics[height=\textwidth, width=\textwidth]{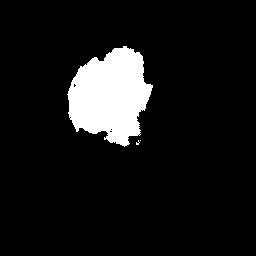}
    \end{minipage}
    \begin{minipage}{\myvaluet\textwidth}
        \centering
        \includegraphics[height=\textwidth, width=\textwidth]{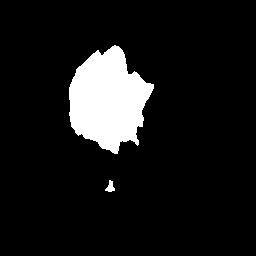}
    \end{minipage}
    \begin{minipage}{\myvaluet\textwidth}
        \centering
        \includegraphics[height=\textwidth, width=\textwidth]{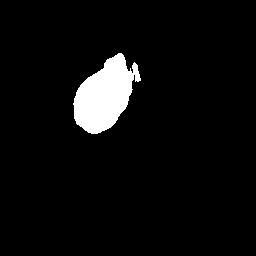}
    \end{minipage}
    \begin{minipage}{\myvaluet\textwidth}
        \centering
        \includegraphics[height=\textwidth, width=\textwidth]{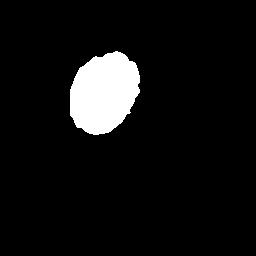}
    \end{minipage}
\end{minipage}

\vspace{2pt}
\begin{minipage}{0.04\textwidth}
    \centering
    \rotatebox{90}{\scriptsize{~}}
\end{minipage}
\begin{minipage}{0.92\textwidth}
    \begin{minipage}{\myvaluet\textwidth}
        \centering
        \includegraphics[height=\textwidth, width=\textwidth]{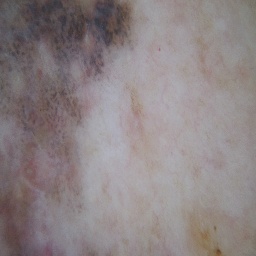}
    \end{minipage}
    \begin{minipage}{\myvaluet\textwidth}
        \centering
        \includegraphics[height=\textwidth, width=\textwidth]{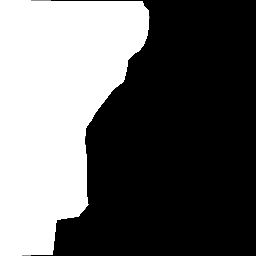}
    \end{minipage}
    \begin{minipage}{\myvaluet\textwidth}
        \centering
        \includegraphics[height=\textwidth, width=\textwidth]{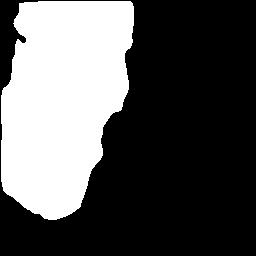}
    \end{minipage}
    \begin{minipage}{\myvaluet\textwidth}
        \centering
        \includegraphics[height=\textwidth, width=\textwidth]{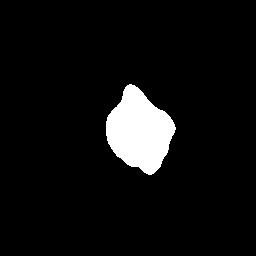}
    \end{minipage}
    \begin{minipage}{\myvaluet\textwidth}
        \centering
        \includegraphics[height=\textwidth, width=\textwidth]{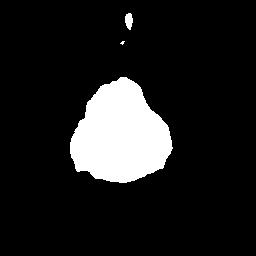}
    \end{minipage}
    \begin{minipage}{\myvaluet\textwidth}
        \centering
        \includegraphics[height=\textwidth, width=\textwidth]{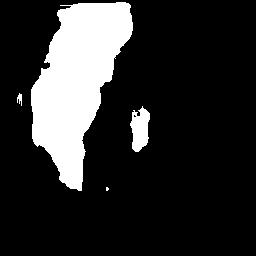}
    \end{minipage}
    \begin{minipage}{\myvaluet\textwidth}
        \centering
        \includegraphics[height=\textwidth, width=\textwidth]{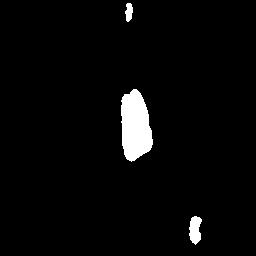}
    \end{minipage}
    \begin{minipage}{\myvaluet\textwidth}
        \centering
        \includegraphics[height=\textwidth, width=\textwidth]{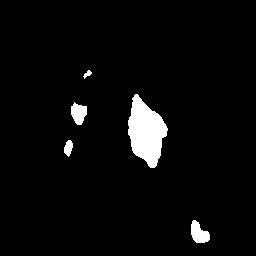}
    \end{minipage}
    \begin{minipage}{\myvaluet\textwidth}
        \centering
        \includegraphics[height=\textwidth, width=\textwidth]{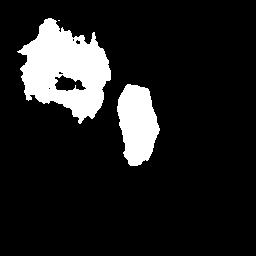}
    \end{minipage}
    \begin{minipage}{\myvaluet\textwidth}
        \centering
        \includegraphics[height=\textwidth, width=\textwidth]{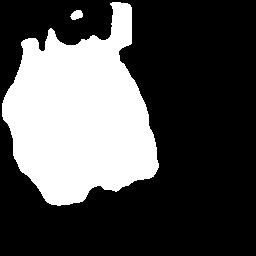}
    \end{minipage}
    \begin{minipage}{\myvaluet\textwidth}
        \centering
        \includegraphics[height=\textwidth, width=\textwidth]{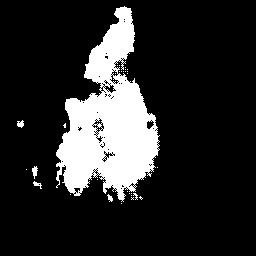}
    \end{minipage}
    \begin{minipage}{\myvaluet\textwidth}
        \centering
        \includegraphics[height=\textwidth, width=\textwidth]{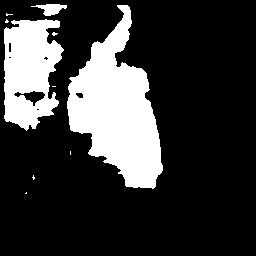}
    \end{minipage}
    \begin{minipage}{\myvaluet\textwidth}
        \centering
        \includegraphics[height=\textwidth, width=\textwidth]{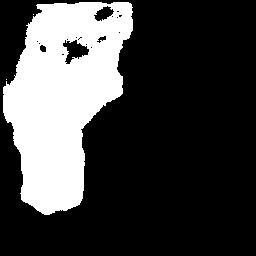}
    \end{minipage}
    \begin{minipage}{\myvaluet\textwidth}
        \centering
        \includegraphics[height=\textwidth, width=\textwidth]{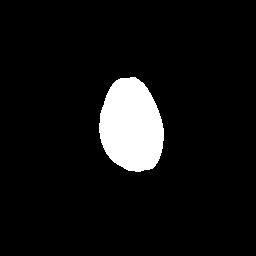}
    \end{minipage}
\end{minipage}

\vspace{2pt}
\begin{minipage}{0.04\textwidth}
    \centering
    \rotatebox{90}{\scriptsize{~}}
\end{minipage}
\begin{minipage}{0.92\textwidth}
    \begin{minipage}{\myvaluet\textwidth}
        \centering
        \includegraphics[height=\textwidth, width=\textwidth]{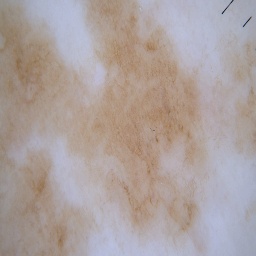}
    \end{minipage}
    \begin{minipage}{\myvaluet\textwidth}
        \centering
        \includegraphics[height=\textwidth, width=\textwidth]{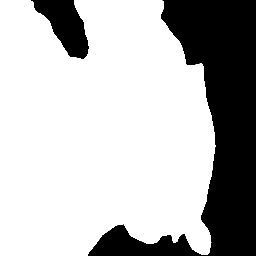}
    \end{minipage}
    \begin{minipage}{\myvaluet\textwidth}
        \centering
        \includegraphics[height=\textwidth, width=\textwidth]{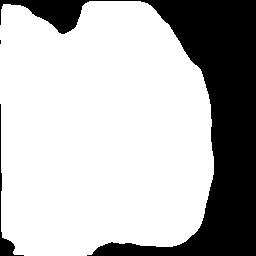}
    \end{minipage}
    \begin{minipage}{\myvaluet\textwidth}
        \centering
        \includegraphics[height=\textwidth, width=\textwidth]{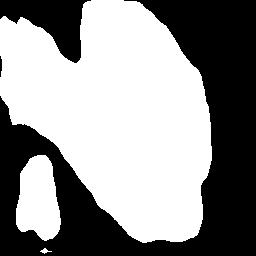}
    \end{minipage}
    \begin{minipage}{\myvaluet\textwidth}
        \centering
        \includegraphics[height=\textwidth, width=\textwidth]{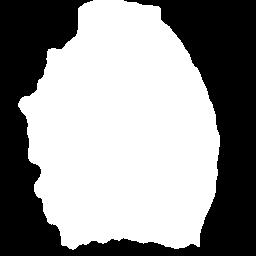}
    \end{minipage}
    \begin{minipage}{\myvaluet\textwidth}
        \centering
        \includegraphics[height=\textwidth, width=\textwidth]{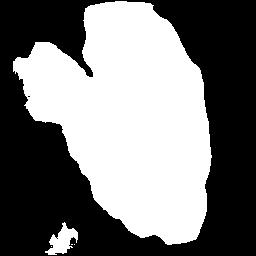}
    \end{minipage}
    \begin{minipage}{\myvaluet\textwidth}
        \centering
        \includegraphics[height=\textwidth, width=\textwidth]{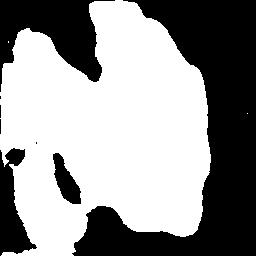}
    \end{minipage}
    \begin{minipage}{\myvaluet\textwidth}
        \centering
        \includegraphics[height=\textwidth, width=\textwidth]{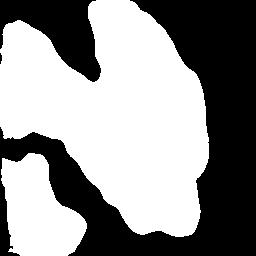}
    \end{minipage}
    \begin{minipage}{\myvaluet\textwidth}
        \centering
        \includegraphics[height=\textwidth, width=\textwidth]{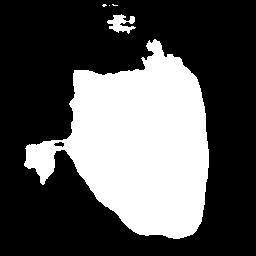}
    \end{minipage}
    \begin{minipage}{\myvaluet\textwidth}
        \centering
        \includegraphics[height=\textwidth, width=\textwidth]{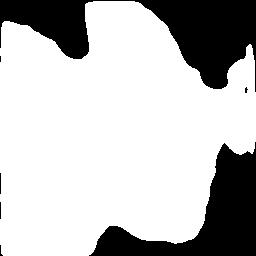}
    \end{minipage}
    \begin{minipage}{\myvaluet\textwidth}
        \centering
        \includegraphics[height=\textwidth, width=\textwidth]{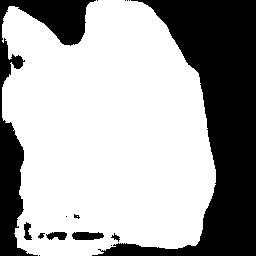}
    \end{minipage}
    \begin{minipage}{\myvaluet\textwidth}
        \centering
        \includegraphics[height=\textwidth, width=\textwidth]{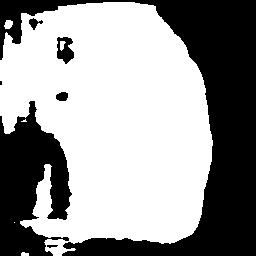}
    \end{minipage}
    \begin{minipage}{\myvaluet\textwidth}
        \centering
        \includegraphics[height=\textwidth, width=\textwidth]{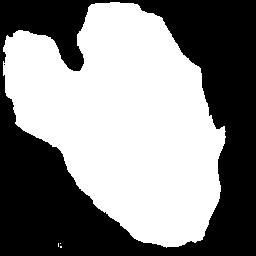}
    \end{minipage}
    \begin{minipage}{\myvaluet\textwidth}
        \centering
        \includegraphics[height=\textwidth, width=\textwidth]{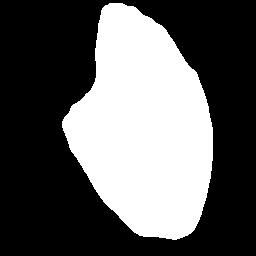}
    \end{minipage}
\end{minipage}
\caption{The qualitative comparison between our model and other models regarding \textcolor{red}{small, medium, and large skin lesions on ISIC2017 and ISIC2018}. As seen, conventional techniques suffer marked efficacy impairment from their dichotomous segmentation errors (over-/under-segmentation), while our framework produces outputs demonstrating exceptional anatomical congruence with reference standards and improved contour fidelity.}
\label{fig:ins_isic_cmp_sota}
\end{figure*}

Compared to the ESKNet, our best model is slightly inferior to it concerning endoscopic polyps. However, our best model has proven the be comparable or even better performance on the other three segmentation tasks, which also emphasizes the importance of a strong backbone. Moreover, the model size has been reduced by around 40 times instead with a 5 times computation efficiency. More importantly, the best model even defeated TransUNet on the MBD and KVASIR-SEG datasets while performing slightly worse on ISIC2017 and ISIC2018. Impressively, our best model had around 200x parameters and 8x computation efficiency than TransUNets. Therefore, we reckon that our \netname{} and the advanced models can be potential lightweight baselines for general medical image segmentation tasks. The qualitative comparisons between our models and the existing models can be seen in Fig. \ref{fig:ins_isic_cmp_sota}.

\section{Discussion}
We would like to discuss a few issues we encountered during the model design and experiment. Firstly, we note that the developed model with fewer parameters usually has higher computational costs compared to its lightweight counterpart. This is because we didn't carry out any other lightweight computing operations, while the spatial dimensions in the first stage of convolution didn't experience any dimensional reduction. Therefore, the proposed model can be further improved by introducing lightweight operations such as grouped depthwise convolution or early spatial reduction. Otherwise, simply staying with a small initial width and stagewise increasing the width would be another much more straightforward strategy. As for the feature fusion in the decoder of \netname{}, we also tried to directly add the shortcut features to the deep features. By doing so, the overall parameters can again be greatly reduced, and so can the computational cost. However, the model performance drops drastically, and the optimal fusion method remains to be explored. Alternatively, the adaptive features fusion can also be implemented through attention modules such as channel attention, spatial attention, or gated attention, and this remains another possible direction for model improvement. Moreover, we didn't deploy the deep supervision strategy like other lightweight models, and the introduction of deep supervision may help the model for faster convergence and better performance, but remains to be explored. Also, we extended our model by replacing the convolution blocks, however, more explorations can be done concerning novel computing units. Nevertheless, the experimental results have shown the simplicity, effectiveness, and extendability of the developed framework.

\section{Conclusion}
When developing our \netname{}, we worked on the vanilla U-Net and found that the key factors leading to a large-scale segmentation model are progressively increased width and feature concatenation. Oriented to lightweight model design, we, therefore, developed three straightforward strategies to reduce the model parameters to less than 1 MB while improving the performance of U-Net. Based on the developed strategies, the developed models outperform the existing cutting-edge ones across multiple public datasets with considerable parameter and computation efficiency. As a result, our models demonstrated the superiority both in terms of model size and performance. To demonstrate the easy extendability of the developed models, we simply replaced the convolution blocks with the ones in ESKNet and successfully advanced the models as well as the performance. In conclusion, we believe that the developed framework can serve as a novel benchmark for medical image segmentation while embracing friendly extension.

\section*{Data availability}
All datasets involved in this study are publicly accessible, and no potential ethical concerns might arise. Once the article has been accepted, the source code will be publicly available. 

\section*{Acknowledgments}
 This work was partially supported by the National Key Research and Development Program (2022YFF0606301), National Natural Science Foundation of China (62227808), Shenzhen Science and Technology Program (JCYJ20230807120304009, JCYJ20210324130812035, JCYJ20220530142002005, JCYJ20220530155208018), and Research Fund Project of Shenzhen Maternity and Child Healthcare Hospital (Grant numbers [FYB2022001]).

\bibliographystyle{IEEEtran}
\bibliography{reference}

\newpage

\end{document}